\documentclass{aa}

\usepackage{physics}
\usepackage{epstopdf}
\usepackage{color}
\usepackage{natbib}
\bibpunct{(}{)}{;}{a}{}{,}
\usepackage[normalem]{ulem}
\usepackage{graphicx,txfonts}
\usepackage{url}
\usepackage[colorlinks]{hyperref}
\usepackage{xspace}
\usepackage{todonotes}
\usepackage{mathtools}
\usepackage{amsmath}
\usepackage{longtable}
\usepackage{comment}
\usepackage{multirow}
\usepackage{multicol}
\usepackage{gensymb}
\usepackage[percent]{overpic}
\usepackage{filecontents}

\usepackage{subcaption}
\usepackage[normalem]{ulem}

\setcitestyle{authoryear, round, aysep={comma}}
\hypersetup{colorlinks=blue, linkcolor=blue, urlcolor=blue, citecolor=blue}

\begin{document}

\defcitealias{Spaghettichugai}{Paper II}

\title{Study of X-ray emission from the S147 nebula with SRG/eROSITA: X-ray imaging, spectral characterization, and a multiwavelength picture}

\author{Miltiadis Michailidis\inst{1}, Gerd P{\"u}hlhofer\inst{1}, Werner Becker\inst{2,3}, Michael Freyberg\inst{2}, Andrea Merloni\inst{2}, Andrea Santangelo\inst{1}, Manami Sasaki\inst{4}, Andrei Bykov\inst{5}, Nikolai Chugai\inst{6}, Eugene Churazov\inst{7,8}, Ildar Khabibullin\inst{7,9,8}, Rashid Sunyaev\inst{7,8,10}, Victor Utrobin\inst{6,11}, Igor Zinchenko\inst{12}}

 \institute{Institut f{\"u}r Astronomie und Astrophysik T{\"u}bingen, Universit{\"a}t T{\"u}bingen, Sand 1, D-72076 T{\"u}bingen, Germany
 \and Max-Planck Institut f{\"u}r extraterrestrische Physik, Giessenbachstraße, 85748 Garching, Germany
 \and Max-Planck Institut für Radioastronomie, Auf dem Hügel 69, 53121 Bonn, Germany
 \and Dr.\ Karl Remeis Observatory, Erlangen Centre for Astroparticle Physics, Friedrich-Alexander-Universit\"{a}t Erlangen-N\"{u}rnberg, Sternwartstra{\ss}e 7, 96049 Bamberg, Germany
 \and Ioffe Institute, Politekhnicheskaya st. 26, Saint Petersburg 194021, Russia
 \and Institute of Astronomy, Russian Academy of Sciences, 48 Pyatnitskaya str., Moscow 119017, Russia
 \and Space Research Institute (IKI), Profsoyuznaya 84/32, Moscow 117997, Russia
 \and Max Planck Institute for Astrophysics, Karl-Schwarzschild-Str. 1, D-85741 Garching, Germany
 \and Universitäts-Sternwarte, Fakultät für Physik, Ludwig-Maximilians-Universität München, Scheinerstr.1, 81679 München, Germany
 \and Institute for Advanced Study, Einstein Drive, Princeton, New Jersey 08540, USA
 \and NRC `Kurchatov Institute', acad. Kurchatov Square 1, Moscow 123182, Russia
 \and Institute of Applied Physics of the Russian Academy of Sciences, 46 Ul'yanov~str., Nizhny Novgorod 603950, Russia
}

\date{Received 30 January 2024 / Accepted 17 May 2024}

\abstract{
Simeis 147 (S147, G180.0-01.7, ``Spaghetti nebula'') is a supernova remnant (SNR) extensively studied across the entire electromagnetic spectrum, from radio to giga-electronvolt $\gamma$-rays, except in X-rays. Here, we report the first detection of significant X-ray emission from the entire SNR using data of the extended ROentgen Survey Imaging Telescope Array (eROSITA) onboard the Russian-German Spektrum Roentgen Gamma (SRG). The object is located at the Galactic anticenter, and its $3\degree$ size classifies it among the largest SNRs ever detected in X-rays. By employing $\sim$15 years of \textit{Fermi}-LAT data, our study confirms the association of the remnant with a spatially coincident diffuse giga-electronvolt excess, namely 4FGL J0540.3+2756e or FGES J0537.6+2751. The X-ray emission is purely thermal, exhibiting strong O, Ne, and Mg lines; whereas it lacks heavier-Z elements. The emission is mainly confined to the 0.5-1.0 keV band; no significant emission is detected above 2.0 keV. Both a collisional plasma model in equilibrium and a model of nonequilibrium collisional plasma can fit the total spectrum. While the equilibrium model -- though statistically disfavored -- cannot be excluded by X-ray fitting, only the absorption column of the nonequilibrium model is consistent with expectations derived from optical extinction data. Adopting an expansion in a homogeneous medium of typical interstellar medium (ISM) density, the general SNR properties are broadly consistent with an expansion model that yields an estimated age of $\sim0.66-2\times10^{5}$ yr, that is a rather old age. The preference for an X-ray-emitting plasma in nonequilibrium, however, adds to the observational evidence that favors a substantially younger age. In a companion paper, we explore an SNR-in-cavity scenario, resulting in a much younger age that alleviates some of the inconsistencies of the old-age scenario.
}
 

\titlerunning{Detection of X-rays and study of the Spaghetti nebula with SRG/eROSITA}
\authorrunning{M. Michailidis}
\maketitle

\section{Introduction}\label{sec:intro}

Supernovae (SNe) rank among the most intense events in a galaxy in terms of the amount of energy they release.
SNe are short-timescale events, observable only in a period of months to a few years maximum. On the contrary, their remnants can be observable up to more than $10^5$ years until they merge with the interstellar medium (ISM) and are dissolved. With an average rate of 1-2 SN per century \citep{10.1111/j.1365-2966.2008.14045.x}, a total of $\sim1200$ supernova remnants (SNRs) are expected to be observable in the Milky Way (MW) at any given moment. Despite the rough estimation of a few thousand SNe, up-to-date findings result in a much lower number of identified SNRs, $\sim$ 300 (Green catalog \citep{2019JApA...40...36G}, SNRcat\footnote{\url{http://snrcat.physics.umanitoba.ca/}} \citep{2012AdSpR..49.1313F}), with the majority of those discovered in radio.

The apparent size of evolved SNRs can vary based on their distance. In close proximity to Earth (hundreds of parsecs), they can reach degree-scale sizes. Current imaging X-ray instruments have a limited field of view (FoV), making them difficult to study in X-rays. In many cases, imaging survey data is the only option. In this respect, the extended ROentgen Survey Imaging Telescope Array (eROSITA), a wide-angle grazing-incident X-ray telescope providing all-sky survey data, \citet{2021A&A...647A...1P}), offers a unique opportunity to study such objects with unprecedented sensitivity. In this category, S147 is a classic example. In this paper, we report the first firm detection of X-rays from the entire S147 SNR using eROSITA all-sky survey (eRASS, \citet{2023Merloni}) data from the first four completed eROSITA all-sky surveys, eRASS:4. We present a thorough examination of the X-ray and multiwavelength properties of the object.

The object S147, listed in the atlas of optical nebulosities \citep{1952IzKry...9...52G}, was originally discussed, among other nebulosities, as an SNR candidate by \citet{1958RvMP...30.1048M}. A first radio map of the SNR was reported by \citet{1969ApJ...155...67D}, revealing the shell-like morphology of the source. More recent radio observations confirm an almost circular shell-type morphology of the source with a diameter of $\sim200'$ \citep{2002nsps.conf....1R,2003A&A...408..961R,xiao2008radio}. The radio emission originates from synchrotron radiation with a rarely observed spectral break at higher frequencies \citep{xiao2008radio}. In 1973, the first optical (H$\alpha$ emission) detailed study of the SNR was reported, showing an SNR with well-defined, delicate, long filaments \citep{1973ApJS...26...19V}. A good spatial correlation between radio synchrotron and optical emission is observed across the entire remnant.
A particular ring-like structure located at the eastern edge of the SNR, present both in radio synchrotron and H$\alpha$ data, stands out as a peculiar feature indicating a lopsided morphology that we describe as an "ear-type" morphology due to the presence of an outer shell that appears to be extended further to the east compared to the rest of the multi-shell structures of the remnant. The latter structural feature is significantly present in the optical and of a fainter nature in radio observations. This feature is also apparent in the multiband images of \citet{2018RAA....18..111R}. A void-type structure, toward the east of the center of the SNR, is a common feature in both wavebands.

Depending on the individual method utilized, SNR distance estimates between 0.6 and 1.9 kpc have been computed \citep{1976MNRAS.174..267C,1979ApJ...229..147K,1980PASJ...32....1S,1980A&A....92..225K,1985ApJ...292...29F,2003ApJ...593L..31K,2004SerAJ.168...55G,2004A&A...426..555S,Ng_2007,Chatterjee_2009,10.1093/mnras/stv124,2017MNRAS.472.3924C,2017A&A...606A..14B}. 
Earlier estimates that are based on the $\Sigma$-D relation and make use of the surface brightness of the SNR and high-velocity gas tend to emerge toward smaller distances $<1.1$~kpc \citep{1976MNRAS.174..267C,1979ApJ...229..147K,1980A&A....92..225K,2004SerAJ.168...55G,2004A&A...426..555S}, except for that of \citet{1980PASJ...32....1S}, which reported a $1.6\pm0.3$~kpc distance using the $\Sigma$-D relation.  
More recent studies employed the associated runaway star, associated pulsar, and the S147 dust cloud, and they converge toward a distance of 1.2-1.4~kpc \citep{Ng_2007,Chatterjee_2009,10.1093/mnras/stv124,2017MNRAS.472.3924C,2017A&A...606A..14B}. \citet{2024arXiv240313892K} used an alternative strategy of
examining the appearance of high-velocity CaII or NaI absorption lines in
hot stars to derive the remnant's distance, which yielded a $1.37^{+0.10}_{-0.07}$~kpc distance estimate. This result is found to be in excellent agreement with previous estimates.
Therefore, a distance of $\sim1.3$~kpc is adopted as a central fiducial value for subsequent calculations throughout the rest of this paper, unless otherwise noted.

Falling near the Galactic anticenter, morphologically similar to the Vela SNR, and having an approximate angular size of 3$\degree$, S147 is among the largest SNRs ever detected in X-rays. With the the ROentgen SATellite (ROSAT) all-sky survey (RASS) \citep{1999A&A...349..389V}, before eRASS the most sensitive available survey in this energy range, X-ray emission from the location of S147 had already been observed, but with neither a clear shell-type morphology (mostly a patchy appearance) nor enough X-ray data to exploit and analyze \citep{1995IAUC.6187....2S, 1996rftu.proc..213A}. The European X-ray Observatory Satellite (EXOSAT) was the first instrument to attempt to detect X-ray emission from the prominent, bright-in-radio, southern rim of the SNR before ROSAT. No X-ray emission at low energies $< 2$~keV - the energy range in which the EXOSAT Low Energy (LE) telescopes operated - was detected. 
In particular, X-ray flux measurements from the southern, patchy structure of the SNR resulted only in strong constraints on expansion parameters of the remnant (i.e., interstellar density and explosion energy), assuming that the SNR itself is located at nearby ($\sim$ 1.1 kpc) \citep{1990A&A...227..183S}. Conversely, X-ray emission in the $2-6$~keV energy band was detected by the Medium Energy Instrument (ME). However, its origin was likely attributed to the transient pulsar A0535+26 mainly due to its relatively hard spectrum \citep{1990A&A...227..183S}. The latter assertion is confirmed in this work since no emission from the SNR is detected with eROSITA above 2~keV.

A significant number of X-ray sources fall in the sky region where the SNR is located. Among those, the bright, high-mass X-ray binary (HMXB) 1A 0535+262, located at the south-eastern edge of the SNR and the pulsar J0538+2817, located  $40'$ to the west of its center, stand out. While 1A\,0535+262 is not associated with the SNR, various observations toward the pulsar suggest a plausible association with the remnant. Both the kinematic age ($\sim30$ kyr) and distance estimates, $1.33^{+0.22}_{-0.16}$ kpc based on parallax \citep{Chatterjee_2009} and 1.2 kpc based on dispersion measurements (DM) \citep{2003ApJ...593L..31K}, are broadly consistent with those of the SNR. The pulsar associated with the remnant was first detected in radio \citep{1996ApJ...468L..55A} and later in X-ray observations with \textit{XMM-Newton} and \textit{Chandra} \citep{2003ApJ...585L..41R}. The X-ray observations revealed extended emission of a faint pulsar-wind nebula (PWN G179.72-1.69) accompanying the pulsar. The runaway star HD 37424 with a spectral type of B5.0V$\pm$0.5 is considered to be the pre-supernova binary companion to the progenitor of the pulsar\,J0538+2817 \citep{10.1093/mnras/stv124,2017A&A...606A..14B,2021MNRAS.507.5832K}.
\begin{table*}
\centering
\caption{X-ray observations analyzed in this work.}
\renewcommand{\arraystretch}{1.7}
\begin{minipage}{20cm}
\begin{tabular}
{p{3.0cm} p{4.0cm} p{2.0cm} p{3.5cm} p{2.5cm} p{1.5cm}}
\hline
Instrument & ObsID& Year& Mode&Exposure (ks) &Pointing  \\ \hline
eROSITA & eRASS:4 (088063,085063& 2019-2021&survey&14.5\footnote{Total on-source exposure time.}&- \\ 
&084060, 087060, 082063)&&&&\\\hline
MOS1, MOS2, PN& 0693270301& 2013& Full Frame (x3)&37.5/37.5/33.6&South East\\ 
MOS1, MOS2, PN&0693270401 & 2013 & Full Frame (x3)&26.7/26.7/22.7&South  \\ \hline
ROSAT & RASS (931214)&1990&survey&17.2\footnote{Live time, on time.}& - \\ 
\hline
\label{TAB00}
\end{tabular}
\end{minipage}
\tablefoot{eROSITA, \textit{XMM-Newton} (MOS1, MOS2, PN), and ROSAT observations in the direction of the remnant. The pointing column describes the position of the \textit{XMM-Newton} observations with respect to the remnant's center.}
\end{table*}

The SNR S147 is usually considered to be one of the most evolved Galactic SNRs, with an estimated age of $\gtrsim10^{2}$ kyr. Therefore, it has been claimed to be the oldest SNR ever detected in X-rays \citep{MPEproceedings}. The age estimate is derived from the shock velocity measurements of $\sim80-120\mathrm{km~s^{-1}}$ \citep{1979ApJ...229..147K,10.1093/mnras/195.3.485}. At the same time, S147 is part of the "special group" of the 70 out of the 294 known SNRs that have been reported to be associated or likely associated with molecular clouds (MCs) \citep{Jiang_2010,ChenB2014}. Its association with the "S147 dust cloud" is extensively discussed in \citet{2017MNRAS.472.3924C}, establishing the conditions for the presence of spatially extended giga-electronvolt $\gamma$-ray emission. The giga-electronvolt emission is found to be in good spatial correlation with the actual extent of the SNR, and it is interpreted to be of hadronic origin \citep{2012ApJ...752..135K}. An updated $\gamma$-ray spectral modeling of the source, aimed at constraining the particle acceleration parameters, was conducted in \citet{2022ApJ...924...45S}. The derived giga-electronvolt spectrum was found to be in good agreement with the one reported in \citet{2012ApJ...752..135K}. However, the particle acceleration can be modeled by various parameter sets, and thus its parameters were poorly constrained. The extended giga-electronvolt source is mainly detected in the 0.2-10~GeV energy band with \textit{Fermi}-LAT and is currently listed in the 12-year \textit{Fermi}-LAT $\gamma$-ray source catalog as 4FGL J0540.3+2756e, in the first catalog of extended sources produced using \textit{Fermi}-LAT data as FGES J0537.6+2751 and in the third catalog of high-energy \textit{Fermi}-LAT sources as 3FHL J0537.6+2751e. Morphologically, the giga-electronvolt emission seems to fit the H$\alpha$ filamentary structure better than regular geometrical shapes. The giga-electronvolt and H$\alpha$ flux correlation suggests that the $\gamma$-ray emission could possibly originate from the thin filaments observed in radio-continuum and optical data \citep{2012ApJ...752..135K}. 

The paper is organized as follows. In Section~\ref{sec:sig&samp}, we report on the eROSITA observations of the SNR alongside the data analysis of the eROSITA first four all-sky surveys (eRASS:4). Additionally, we compare the eROSITA findings to results obtained from the earlier ROSAT survey and archival \textit{XMM-Newton} observations from small portions of the remnant. We further outline the implications of the presence of dust at and around the remnant's location. A reanalysis of \textit{Fermi}-LAT $\gamma$-ray data, from the location of the remnant, is also presented. Section~\ref{section3} is dedicated to the eROSITA and \textit{XMM-Newton} spectral analysis results and its potential implications (distance and age estimates), adopting a simplified setting of an expansion in homogeneous local medium. In Section~\ref{sectionGEVPEC}, we report on the updated $\gamma$-ray spectrum of the S147 giga-electronvolt counterpart and discuss potential scenarios for the origin of the giga-electronvolt emission on the basis of its giga-electronvolt spectral energy distribution (SED). Section~\ref{section5} gives concluding remarks. In a companion paper \citep[\citetalias{Spaghettichugai} hereafter]{Spaghettichugai}, we explore the possibility that the SNR expands in a wind-blown cavity, which yields a much younger age estimate for the SNR.

\begin{figure*}[h!]

    \includegraphics[width=0.495\textwidth,clip=true, trim= 1.4cm 0.1cm 0.9cm 1.4cm]{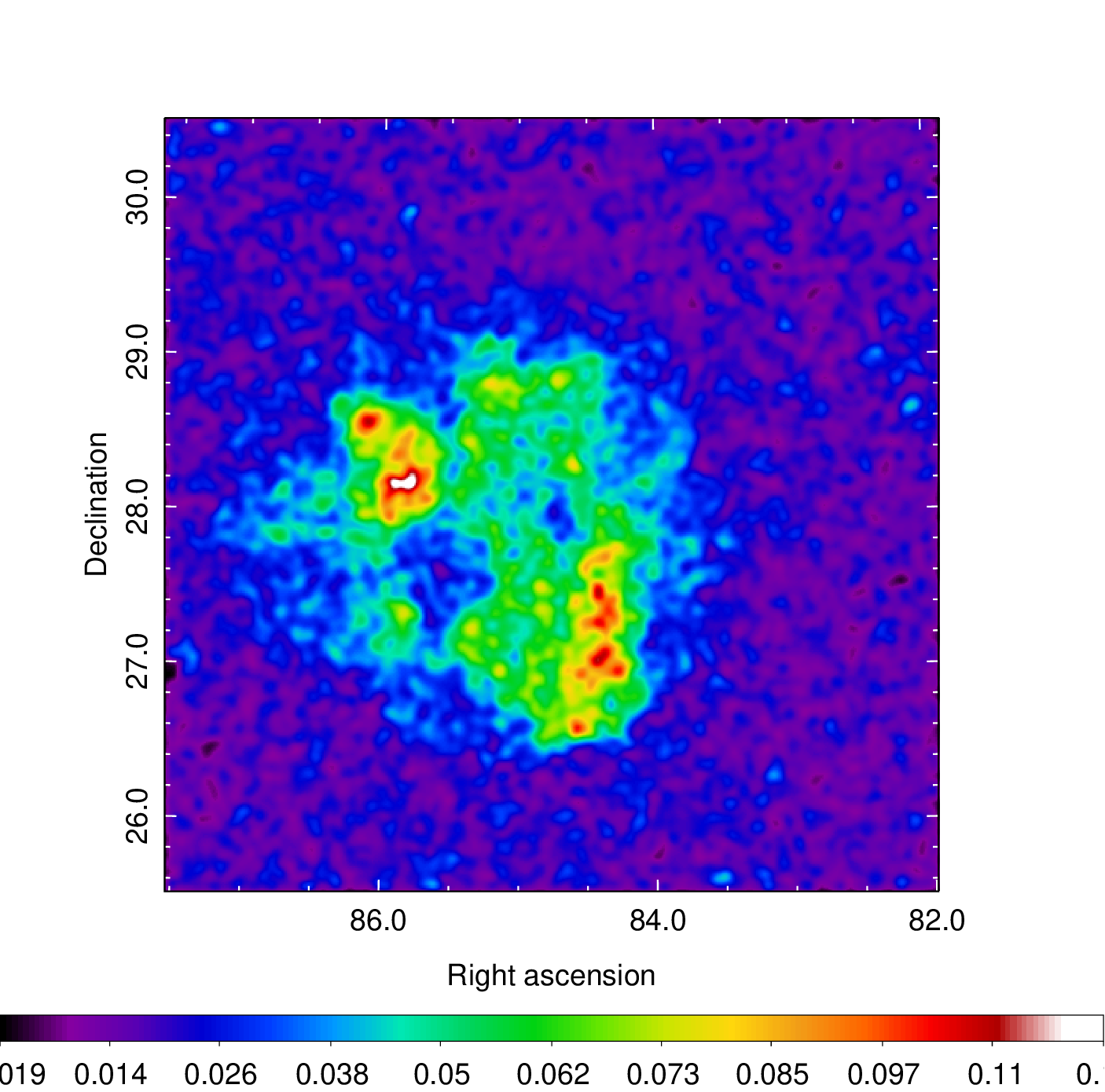}
    \includegraphics[width=0.495\textwidth,clip=true, trim= 1.4cm 0.1cm 0.9cm 1.4cm]{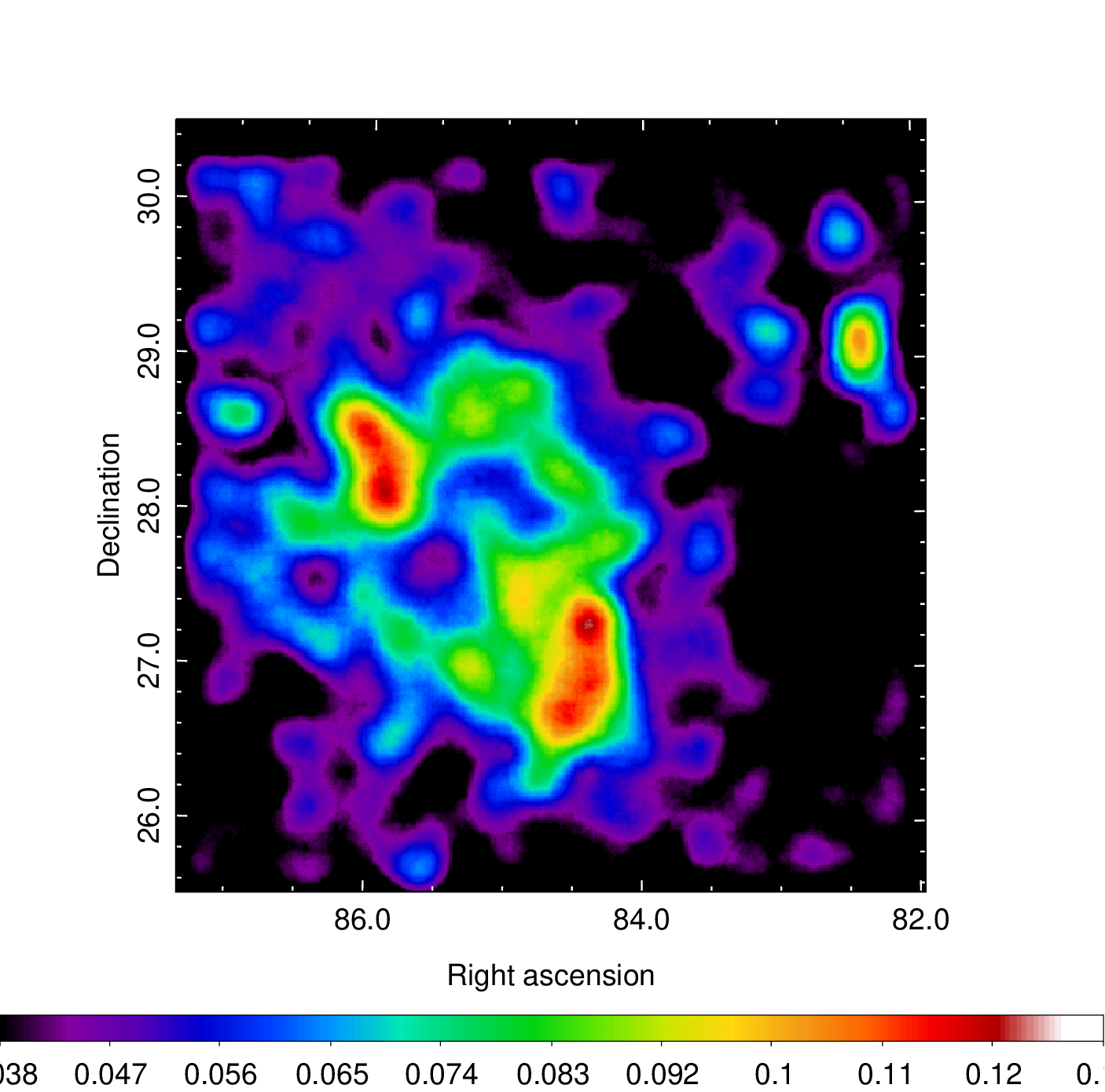}
    \caption{First detection of significant X-ray emission from the entire extension of the S147 with eROSITA and a side-by-side comparison of the eROSITA and ROSAT views of the remnant. Left panel: eRASS:4 exposure-corrected intensity sky map in the 0.5-1.0~keV energy band, in units of counts per pixel with a pixel size of $10''$. All point sources are filtered out, and the image is convolved with a $\sigma=100''$ Gaussian aimed at enhancing the visibility of the diffuse X-ray emission. 
    Right panel: ROSAT intensity sky map in the 0.4-2.4~keV energy band (medium RASS band). The ROSAT image is built on a different energy range that maximizes contrast between on-source and background regions, allowing fainter structures to be detected. The image, with a pixel size of $45''$, is convolved with a $\sigma=6.4'$ Gaussian to enhance the visibility of the diffuse emission from the location of the remnant. Except for HMXB 1A 0535+262 and the emission associated with the associated pulsar, which were treated independently and effectively masked out from the image, all remaining point sources were not removed. Their proper masking requires a substantially larger extraction radius, which heavily affects the faint diffuse emission originating from the remnant.}
    \label{01}
\end{figure*}

\section{X-ray observations and data analysis}
\label{sec:sig&samp}

The main parameters of all X-ray observations employed in this work are summarized in Table~\ref{TAB00}.
\subsection{$\textit{eROSITA data}$}\label{eROSITAanalysis}

The Russian-German Spektrum Roentgen Gamma (SRG) observatory \citep{2021A&A...656A.132S} consists of two instruments:$\textit{(i)}$ eROSITA, operating at the softer X-ray band, 0.2-8.0 keV, with unprecedented sensitivity \citep{2021A&A...647A...1P}; and $\textit{(ii)}$ the Russian X-ray telescope $\texttt{MIKHAIL PAVLINSKY}$ ART-XC, which covers the harder 4.0-30.0 keV X-ray band \citep{2021A&A...650A..42P}. In this paper, only data from the eROSITA telescopes are used. 

eROSITA \citep{2021A&A...647A...1P} consists of seven X-ray telescope modules (TM1-7), which are aligned in parallel and have identical fields of view of approximately $1^\circ$ diameter each. Each telescope holds 54 nested mirror shells. Such a system of X-ray mirrors can be roughly described by three numbers: its effective area, its vignetting function, and its point spread function (PSF). 
A preliminary analysis of the in-flight PSF calibration presented in \citet{2023Merloni} results in an $\sim30"$ average spatial resolution in survey mode.

Launched on July 13, 2019 toward the L2 Lagrangian point, eROSITA started taking regular survey data on December 13, 2019.
In January 2022, the fourth eROSITA all-sky survey was completed, and here we report on results obtained with data of eRASS1 to eRASS4, called eRASS:4. 

The German and Russian parties involved in the mission are responsible for the analysis of the data of the eastern and western Galactic hemisphere, respectively. Since S147 is a large diameter SNR located at the Galactic anticenter, half of its extension falls in the western hemisphere and half falls in the eastern hemisphere. Therefore, this work is a joint collaboration of the two parties.

We used data of the \texttt{c020} processing version. The analysis and data reduction were carried out using the $\texttt{eSASSusers\_201009}$ version \citep{2022A&A...661A...1B} of eROSITA Standard Analysis Software (eSASS). All events flagged as corrupt either individually or as part of a corrupt frame were excluded, retaining all four of the recognized legal patterns (\texttt{pattern=15}), and identifying and repairing disordered GTIs. Flare inspections, when the survey scans were passing through the S147 area, 
were also carried out with the aim of avoiding potential contamination of the data sets used for further analysis.

For ease of use, the eROSITA X-ray sky map can be divided into 4700 sky tiles (squares of $\sim3.6^o\times3.6^o$ size) that partially overlap. Each sky tile corresponds to a unique six-digit ID (the first three digits express the right ascension -RA- while the last three express the declination dec of the corresponding center of each particular sky tile). Given that S147 is an $\sim3^o$ SNR that is not particularly centered at a specific sky tile, its X-ray emission extends into more than one sky tile. Five sky tiles were individually examined for diffuse X-ray emission across the entire SNR in order to achieve complete coverage. In particular, the 088063 sky tile lies at the German half of the sky, while the other four [085063, 084060, 087060, 082063] partially fall in the German and in the Russian halves, respectively.  

\begin{figure}[h!]

    \includegraphics[width=0.51\textwidth,clip=true, trim= 0.99cm 0.1cm 0.9cm 0.9cm]{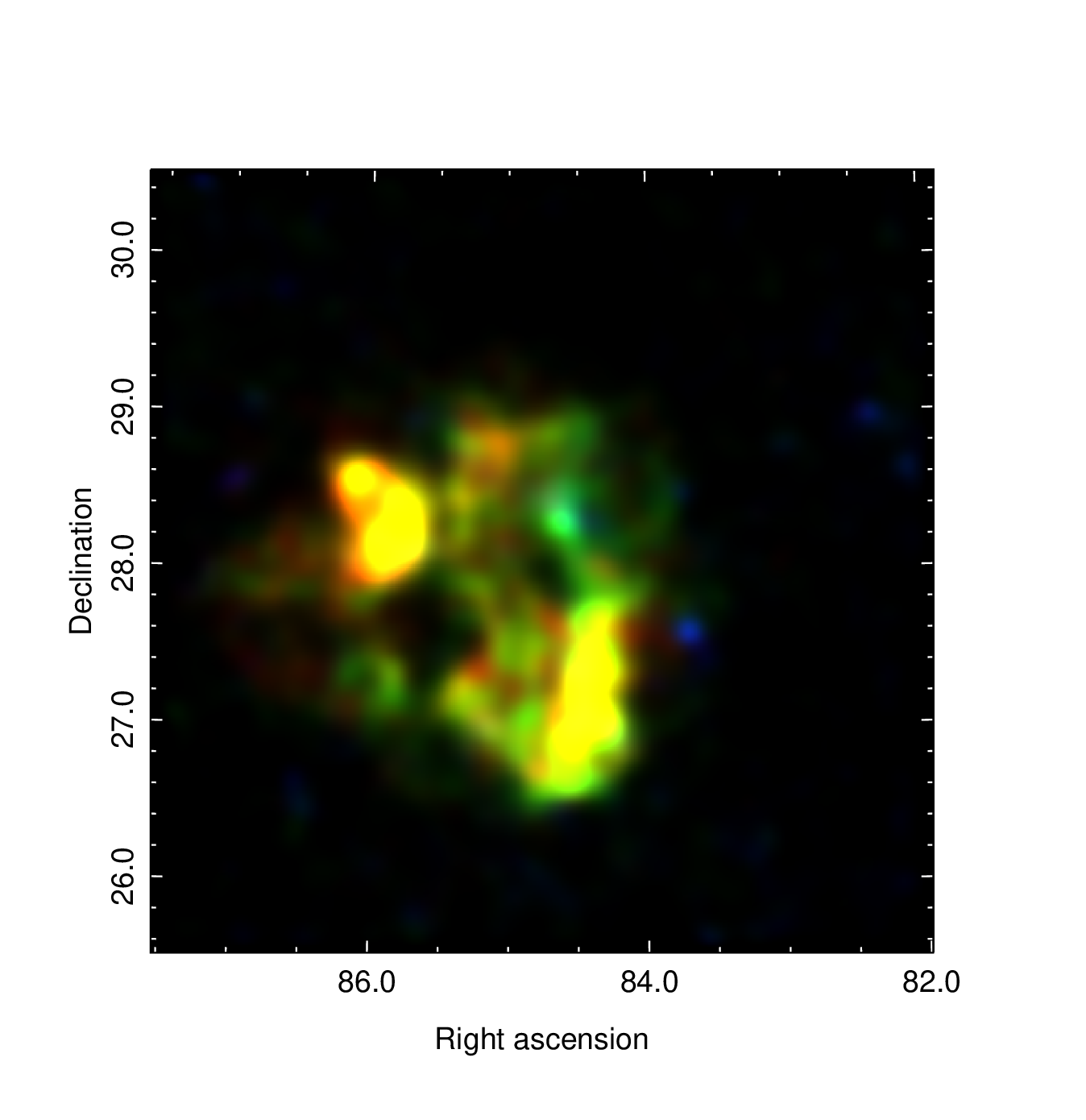}
    \caption{eRASS:4 RGB exposure-corrected intensity sky map, with energy-to-color correspondence: R: 0.3-0.6 keV, G: 0.6-1.0 keV, and B: 1.0-1.5 keV, in units of counts per pixel with a pixel size of $10''$. Similarly to Fig.~\ref{01}, all point sources are masked out, and the image is convolved with a $\sigma=2'$ Gaussian aimed at enhancing the visibility of the diffuse X-ray emission.} 
    \label{02}
\end{figure}

Figure~\ref{01} illustrates the $5\degree\times5\degree$ mosaic intensity map, in units of counts per pixel in the 0.5-1.0 keV energy range. The selection of the energy range was optimized based on the energy spectrum of the source. The image was corrected for uneven exposure by a factor ranging between 0.946 and 1.056.
From the image analysis, when exploiting different energy bands (see the corresponding RGB image of Fig.~\ref{02}), it is evident that the SNR is only detectable at very soft X-rays. The emission is mainly confined to the 0.5-1.0~keV energy band, whereas above 2.0 keV no X-ray emission originating from the SNR can be detected. In more detail, the first step of the mosaic intensity image production is the creation of the individual sky tile count maps. Using a $10''$ spatial resolution (i.e., pixel size) and utilizing the $\texttt{evtool}$ task of eSASS software, we combined the data for all four surveys and all TMs (all seven TMs were used as any uncertainty in the energy calibration resulting from the light leaks in TMs 5 and 7 \citep{2021A&A...647A...1P} do not quantitatively impact our results in the energy band of interest $>0.5~\mathrm{keV}$) to produce the corresponding count maps. To enhance the visibility of the detected diffuse emission from the source of interest and avoid likely contamination on the signal, we masked out all point sources with detection significance above $3\sigma$ using a $110''$ extraction radius. Individual brighter sources such as the HMXB 1A 0535+262 and the associated pulsar (accompanied by its faint PWN) were treated independently employing larger extraction regions to be properly masked out. Following $\texttt{evtool}$, $\texttt{expmap}$ task was employed to compute the exposure maps at that specific energy range, with an average exposure of $\mathrm{\sim500~s\cdot pixel^{-1}}$. Finally, filter-wheel-closed (FWC) data with deep and different exposures for each TM, at a scale of a few hundreds of kilo-seconds, were used for the creation of the corresponding instrumental background maps. The aforementioned maps were employed to ensure that after subtracting the average background level from the mosaic intensity map and correcting for exposure times, which might moderately vary at distinct locations across the remnant, the obtained net mosaic map is a good measure of the surface brightness of the source (i.e., no artifacts from uneven exposure are present in the mosaic intensity maps).

A three-colored image of S147, color-coded as R:0.3-0.6~keV, G:0.6-1.0~keV, and B:1.0-1.5~keV, is shown in Fig.~\ref{02}. The image reveals that the majority of the SNR's X-ray emission is soft and well-confined in the 0.5-1.0~keV band. Such a conclusion is additionally confirmed by the spectral analysis process described in Section~\ref{eROSEspectra}, where it becomes evident that only some faint X-ray diffuse emission is present above 1.0~keV, leaving the remnant totally undetected above 2.0~keV. The diffuse X-ray emission fills almost the entire remnant except for the small void structure east of the center of the remnant, which is also present in radio-continuum and H$\alpha$ data.

The peculiar morphology of S147 complicates an accurate geometrical center estimation. 
We first computed the geometrical center of the SNR in X-rays, RA: 5h40m53.0647s Dec: $27\degree$49'49.585", by fitting an annulus to the outermost parts of the S147 X-ray diffuse emission. We additionally validated the above result by applying a Minkowski tensor analysis. This analysis is used to parametrize the shape of astrophysical objects by drawing perpendicular lines to structures of our selection -based on their detection significance- and subsequently deriving those regions of the highest line density  \citep{2021A&A...653A..16C}. The above process was applied to the mosaic intensity sky map (left panel of Fig.~\ref{01}). In the absence of a strong symmetrical morphology of the remnant, three distinct regions were identified as the most probable candidates of the remnant's X-ray center.
Among those, the most probable one is found to be in agreement with the value derived above: R.A. 5h41m152s and Dec: $27\degree$44'59.532". The X-ray center appears to be moderately shifted to the east in comparison to the latest estimate based on radio synchrotron data:  RA: 5h39m00s Dec: $27\degree$50'0" \citep{2009yCat.7253....0G}. This discrepancy is likely induced by the "ear type" structure detected to the east of the remnant, which is present with a high level of significance in X-ray and H$\alpha$ data, but it appears much fainter in the radio synchrotron data sets. Thus, we speculate that the latter remnant feature might have been excluded from the geometry computation of previous works.

\subsection{\textit{ROSAT data}}\label{ROSAT}

The SNR S147 was reported as a potential X-ray emitter for the first time based on ROSAT data, although no clear shell-type morphology was reported; it instead showed a mostly
patchy appearance. In addition, the statistical quality of the X-ray data did not allow further exploitation and analysis of the remnant \citep{1995IAUC.6187....2S,1996rftu.proc..213A}. Different to eROSITA, ROSAT all-sky survey maps are divided into 1378 sky tiles (squares of $6^{o}.4\times6^{o}.4$). Therefore, the extended X-ray emission originating from the remnant is confined within a single ROSAT sky tile [ID:931214]. In order to achieve a direct comparison between eROSITA's and ROSAT's view of the remnant, we exploited the publicly available ROSAT all-sky survey data. Data were extracted from the High Energy Astrophysics Science Archive Research Center (HEASARC) \citep{bworld}). Data reduction was performed using FTOOLS \citep{cworld}. The right panel of Fig.~\ref{01} illustrates the X-ray photon emission, from the location of the remnant, in the medium ROSAT energy band [0.4-2.4 keV], using data from the ROSAT Position Sensitive Proportional Counter detector in survey mode (PSPC) \citep{2000IAUC.7432....3V}. Heavy smoothing has been applied, as specified in the caption of Fig.~\ref{01}, aiming at enhancing the diffuse X-ray emission originating from the remnant. As shown by the side-by-side comparison of the two panels of Fig.~\ref{01}, significant diffuse X-ray emission of similar morphology to the one obtained in eRASS:4 is also detected with ROSAT. It is noteworthy that the ROSAT image presented in this study is substantially improved compared to the latest ROSAT view of the remnant reported in \citet{1995IAUC.6187....2S, 1996rftu.proc..213A}. Considering the same extraction region (i.e., the extraction region used for the spectral analysis of the entire remnant, as defined in the caption of Fig.~\ref{011}) and the same energy range (i.e., 0.4-2.4 keV) for both ROSAT and eROSITA, the latter performs in an $\sim25$ times higher survey collection area at the location of the remnant. 
In particular, with eROSITA we detect a total of 78.652 counts (among which 30663 are source counts) compared to 3356 (among which 1241 are source counts) from ROSAT.

\begin{figure}[h!]

    \includegraphics[width=0.25\textwidth,clip=true, trim= 0.9cm 0.1cm 0.9cm 0.5cm]{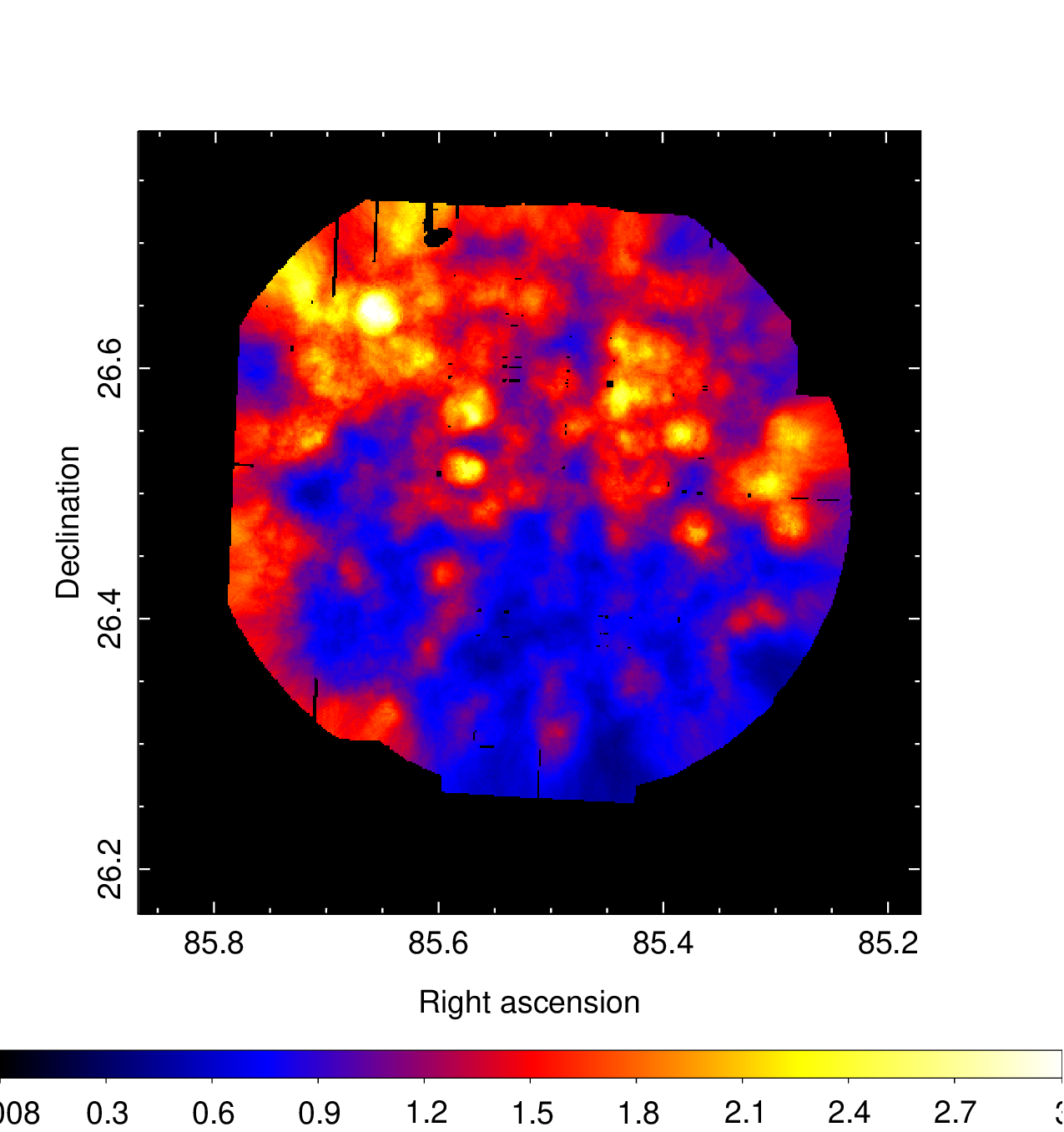}\includegraphics[width=0.25\textwidth,clip=true, trim= 0.9cm 0.1cm 0.9cm 0.5cm]{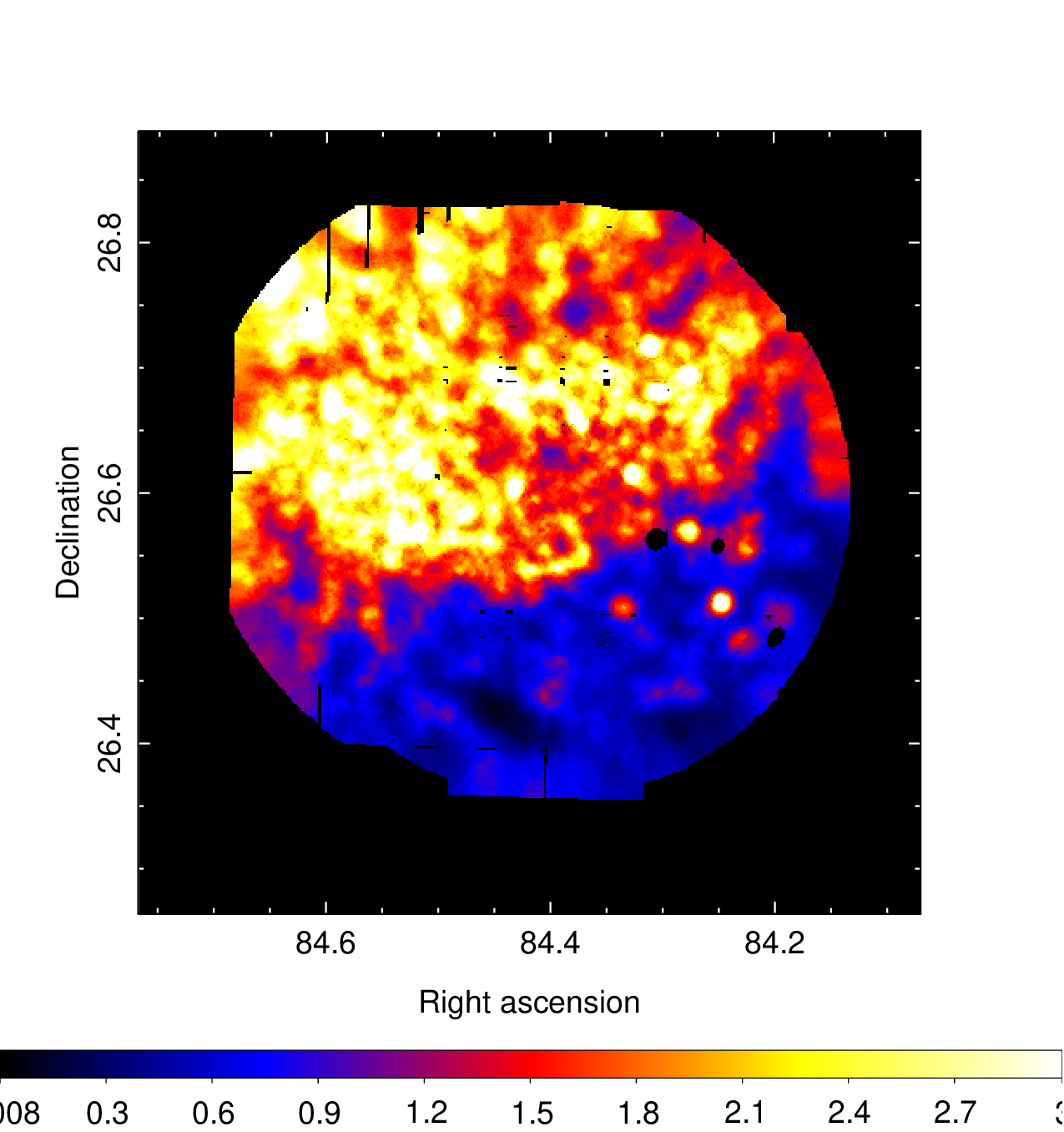}

    \caption{\textit{XMM-Newton} surface-brightness maps in the 0.5-1.0~keV energy band, in units of $\mathrm{counts~s^{-1}~deg^{-2}}$. All point sources are masked out. The maps, with $5''$ pixel sizes, are adaptively smoothed with a 50 count kernel and are vignetting-corrected. Left panel: 0693270301 \textit{XMM-Newton} pointing. Right panel: 0693270401 \textit{XMM-Newton} pointing. 
    Both \textit{XMM-Newton} pointings fall at the southern edge of the remnant, and thus its northern halves are filled with X-ray emission, whereas its southern halves can serve as background control regions.}
    \label{04}
\end{figure}
\subsection{\textit{\textit{XMM-Newton} and \textit{Chandra} data}}
\label{section4}

Individual smaller parts of S147 have been observed before with \textit{XMM-Newton} (unrelated to the team presenting the work of this paper). In particular, two dedicated \textit{XMM-Newton} pointings towards the remnant have been carried out [ID:0693270301, 0693270401]. Additionally, a third \textit{XMM-Newton} observation [ID:0674180101] toward the
HMXB 1A 0535+262 partially contains diffuse X-ray emission from the SNR. However, due to the high contamination of the signal from X-ray emission originating from the binary, we did not exploit these data any further. Given the remnant's large size, the two observations mentioned above were performed in discrete parts of the SNR to obtain a better insight into its nature. The European Photon Imaging Camera (EPIC) consists of three detectors: the \texttt{PN} camera \citep{2001A&A...365L..18S} and two \texttt{MOS} cameras \citep{2001A&A...365L..27T}. Both (yet unpublished) observations (ID: 0693270301: S147 SE, PN (extended full frame)/MOS1 (full frame)/MOS2 (full frame) exposures: 22.7/26.7/26.7 ks, PI: Jean Ballet - ID: 0693270401: S147 south, PN (extended full frame)/MOS1 (full frame)/MOS2 (full frame) exposures: 33.6/37.5/37.5 ks, PI: Jean Ballet) exhibit diffuse X-ray emission originating from the SNR. The limited field of view of \textit{XMM-Newton}, of an angular extension of $0.5\degree$, does not allow a thorough analysis of S147's X-ray morphology. Nevertheless, the high sensitivity of the instrument allows us to derive data that are complementary to eROSITA. In addition, a \textit{Chandra} observation (ID:5770), carried out with the Advanced CCD image spectrometer camera (\texttt{ACIS-I}) operating at 0.1-10 keV toward ET Tau, is positioned within the SNR extension. Due to limited statistics, we did not further exploit the available \textit{Chandra} data in the context of this work. Additional \textit{XMM-Newton} [ID:0112200401] and \textit{Chandra} [ID:5538] observations have been directed to explore the X-ray nature of the pulsar J0538+2817 associated with the remnant. The pulsar PSR J0538+2813 and its corresponding pulsar wind nebula (PWN 179.72-1.69), which has an angular size of $<10"$ (e.g., \citet{2003ApJ...585L..41R,Ng_2007}) and is not resolved in eROSITA data, are effectively excluded by our source-masking procedure (using $110"$ radius). Hence, no influence of their contribution on the spectral properties of the SNR emission presented here is expected.
A dedicated analysis of the X-ray emission properties of the region in the vicinity of  PSR J0538+2813
will be presented in a separate paper (Bykov et al, in preparation).

The eSAS software assembled with the latest calibrations was utilized to perform the X-ray data analysis of the available \textit{XMM-Newton} observations (i.e., data reduction, image production, and spectra extraction). We used the \texttt{emchain} and \texttt{epchain} eSAS tools to process observation data files. The \texttt{mos-spectra} and \texttt{pn-spectra} commands were employed to extract images and spectra of the regions of interest. The two panels of Fig.~\ref{04} illustrate the two \textit{XMM-Newton} observations toward S147 for the PN, MOS1, and MOS2 combined data (camera-CCDs found in anomalous state were excluded from the analysis). The two observations reveal a soft, mainly confined in the 0.5-1.0 keV energy band, X-ray diffuse emission component in excellent morphological agreement with the eRASS:4 data. The images were adaptively smoothed, using a smoothing kernel of 50 counts, and vignetting corrected. Point sources have been masked out to enhance the visibility of the diffuse X-ray emission from the remnant. Fig.~\ref{011} shows the eRASS:4 $5\degree\times5\degree$ intensity sky map (described in detail in Section~\ref{eROSITAanalysis}) overlaid with contours that mark the position (one level contours) of the two \textit{XMM-Newton} observations (white [ID:0693270301] and red [ID:0693270401] contours, respectively).

\begin{figure*}[h]

    \includegraphics[width=0.51\textwidth,clip=true, trim= 1.4cm 0.1cm 1.0cm 1.0cm]{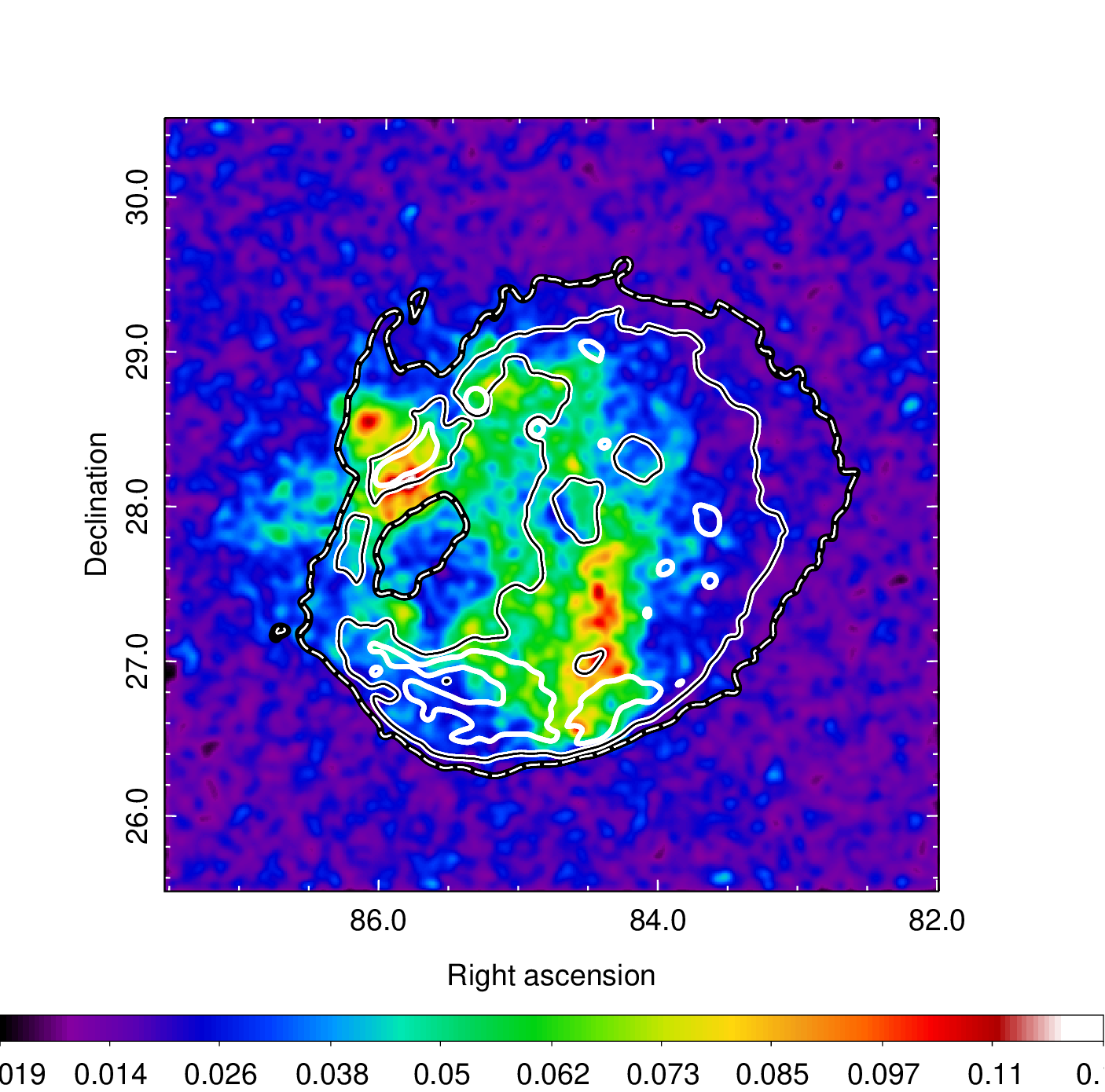}\includegraphics[width=0.51\textwidth,clip=true, trim= 1.4cm 0.1cm 1.0cm 1.0cm]{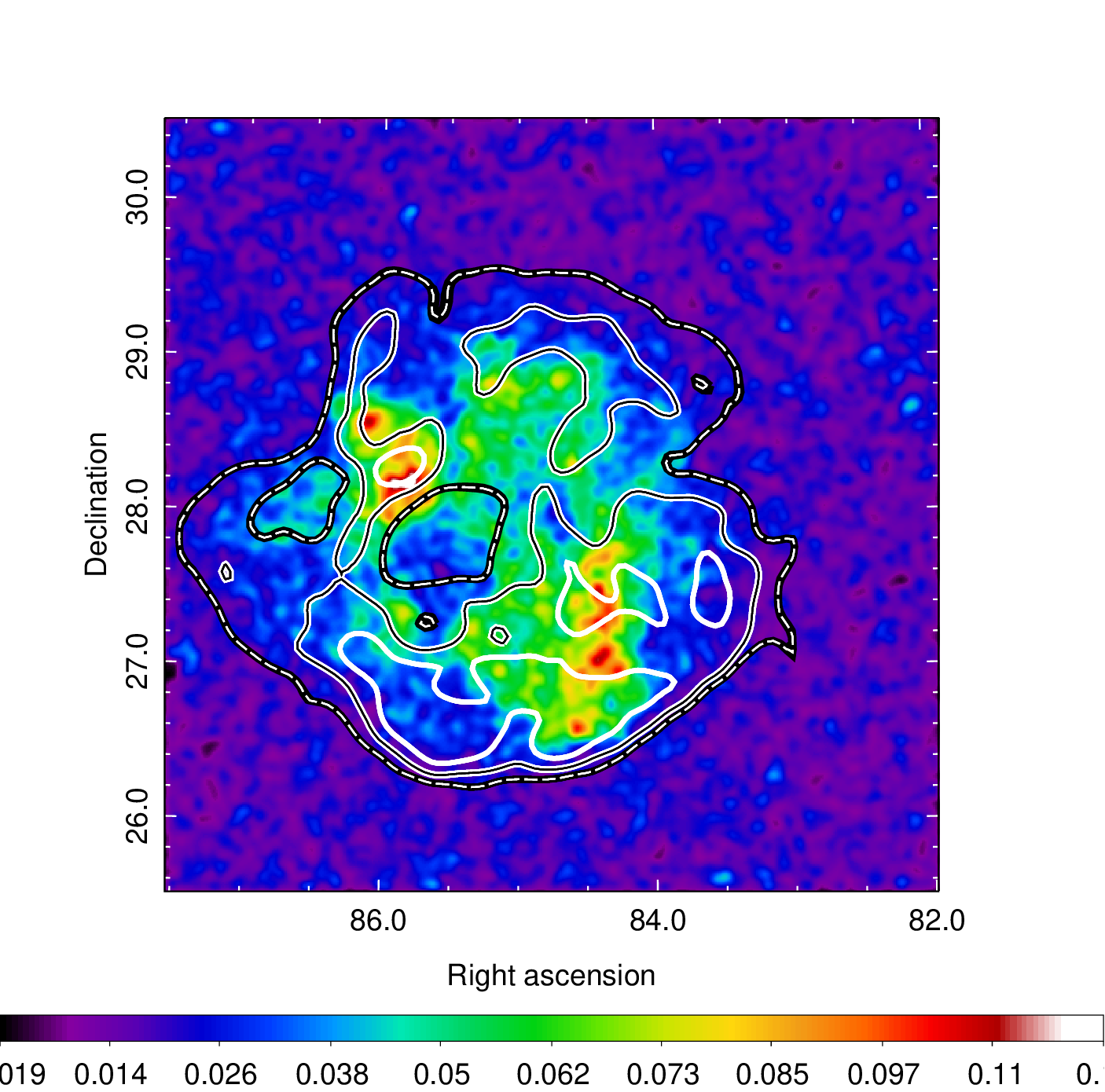}
    \caption{eRASS:4 exposure-corrected intensity sky map; the same as the one displayed on the left panel of Fig.~\ref{01}. Left panel:  Canadian Galactic Plane Survey (CGPS) data at $1.42~\mathrm{GHz }$  \citep{2003AJ....125.3145T} are overlaid as contours. Right panel: Optical H$\alpha$ data obtained from the full-sky H$\alpha$ map of $6'$ FWHM resolution \citep{2003ApJS..146..407F} are overlaid as contours. For both panels, three level contours are used from thin, dashed, white-on-thick-solid-black, to thin, solid black on thick solid white, and solid white to represent fainter-to-brighter emission regions.}
    \label{05}
\end{figure*}
\subsection{A multiwavelength study}\label{multi}
\subsubsection{X-ray, radio, and optical data correlation}
\label{radioHa}
Figure~\ref{05} shows the eRASS:4 mosaic intensity sky map of S147. 
Radio synchrotron data obtained from the Canadian Galactic Plane Survey (CGPS) at $1.42~\mathrm{GHz }$ \citep{2003AJ....125.3145T} are overlaid on the left panel of Fig.~\ref{05} as contours. Optical H$\alpha$ data obtained from the full-sky H$\alpha$ map (6' FWHM resolution)  \citep{2003ApJS..146..407F}, which is a composite of the Virginia Tech Spectral line Survey (VTSS) in the north and the Southern H-Alpha Sky Survey Atlas (SHASSA) in the south, are overlaid on the right panel of Fig.~\ref{05} as contours. The enhanced emission regions and the void-type structure are spatially identical across all three energy bands. The only noticeable difference is detected to the west of the remnant, where both H$\alpha$ and radio-continuum emission seem to extend further to the west in comparison with X-rays. Potential reasons for such a discrepancy are discussed in Sections~\ref{GeVV} and~\ref{eROSEspectra}. For a more detailed study on the nature of this discrepancy, we refer the reader to \citetalias{Spaghettichugai}. At the same time, the aforementioned image clearly demonstrates that the spatial morphology of the SNR in X-rays is nicely confined within the ear-type structure that both H$\alpha$ and radio-continuum data exhibit rather than being matched with typical shell-type shapes. It is noteworthy that in a number of different radio synchrotron surveys operating at different energy bands (i.e., GaLactic and Extragalactic All-sky MWA survey (GLEAM)\footnote{\url{https://www.mwatelescope.org/science/galactic-science/gleam/}} data \citep{2015PASA...32...25W,2017MNRAS.464.1146H,2018MNRAS.480.2743F,2019PASA...36...47H} and/or 4850 MHz radio data obtained from PMN \citep{1993AJ....106.1095C} Southern and tropical surveys, and GB6 \citep{1991AJ....102.2041C,1994AJ....107.1829C}), the "ear-type" structure at the East of the remnant is hardly discernible, if not absent, likely due to its fainter appearance in comparison to radio-continuum emission emanating from the rest of the remnant. However, when employing CGPS data, an arc structure that nicely encapsulates the diffuse X-ray ear-type emission becomes apparent. The latter structure is difficult to display as contours, mainly due to the contamination of the radio-continuum data from nearby regions. Therefore, it is missing on the left panel of Fig.~\ref{05}. However, in Figure 3 of \citetalias{Spaghettichugai}, we show a composite image of X-ray emission from eROSITA and radio emission from CGPS, where this ear-type structure also becomes clearly apparent in the radio data. The SNR's pure thermal nature, as discussed in Section~\ref{section3}, could account for such an interconnection between the different energy bands. 
No universal correlation between H$\alpha$ and X-rays (i.e., warm and hot gas) in SNR environments exists. Nevertheless, it is not unusual that the remnant's X-ray spatial morphology is tightly correlated with the H$\alpha$ emission, which traces the atomic gas, since the X-ray emission mainly stems from hot thermal plasma. The co-existence of X-rays, H$\alpha$, and radio-continuum data has been thoroughly investigated before \citep{1998PASA...15...64C}, and relevant objects have been reported. An example of such a remnant with X-rays, H$\alpha$, and radio fine structure is G332.5-5.6 \citep{2007MNRAS.381..377S}.

\begin{figure*}[]
    \centering
    
    \includegraphics[width=0.495\textwidth,clip=true, trim= 1.0cm 0.1cm 1.0cm 1.0cm]{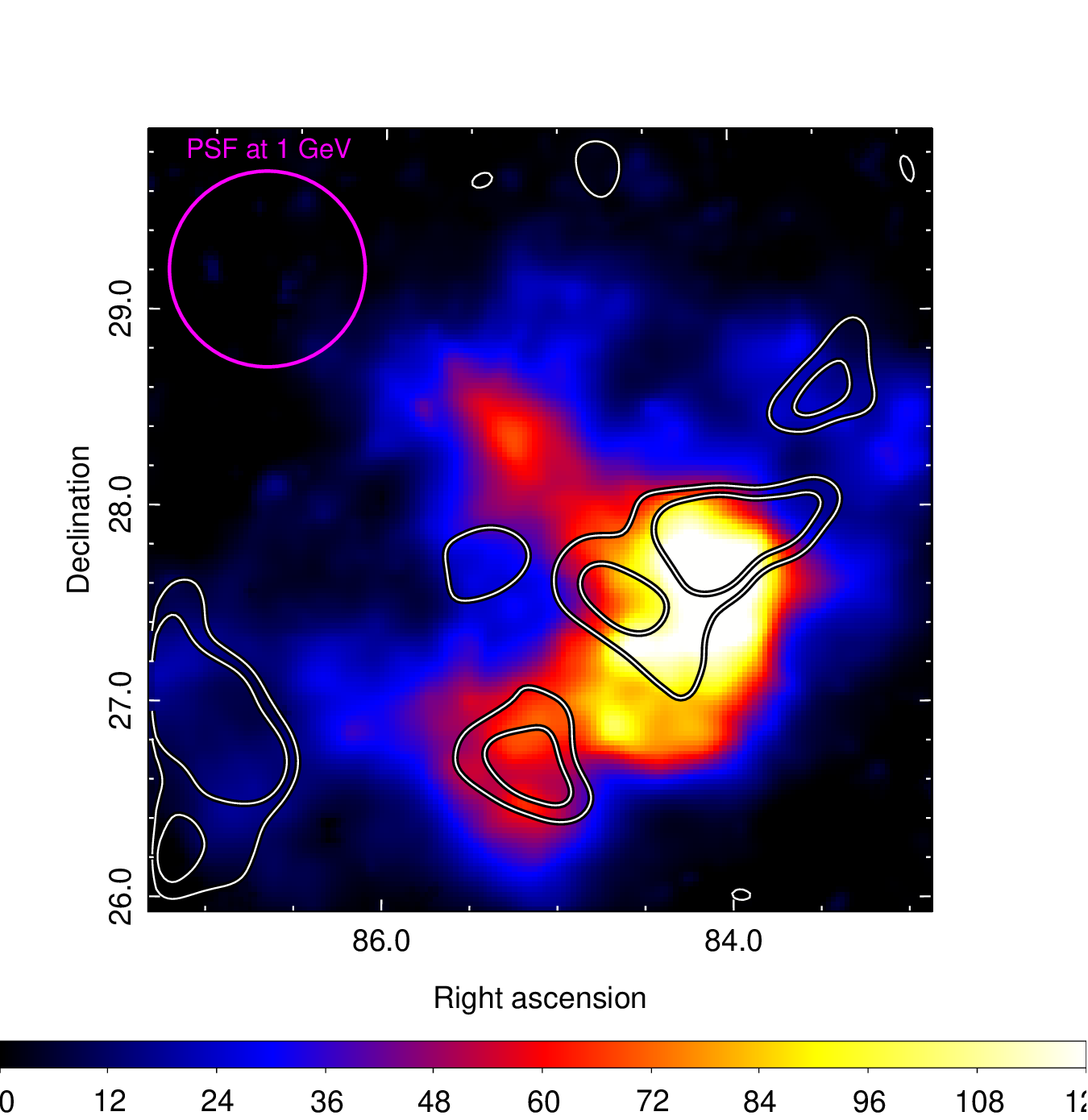}
    \includegraphics[width=0.495\textwidth,clip=true, trim= 1.0cm 0.5cm 2.4cm 2.0cm]{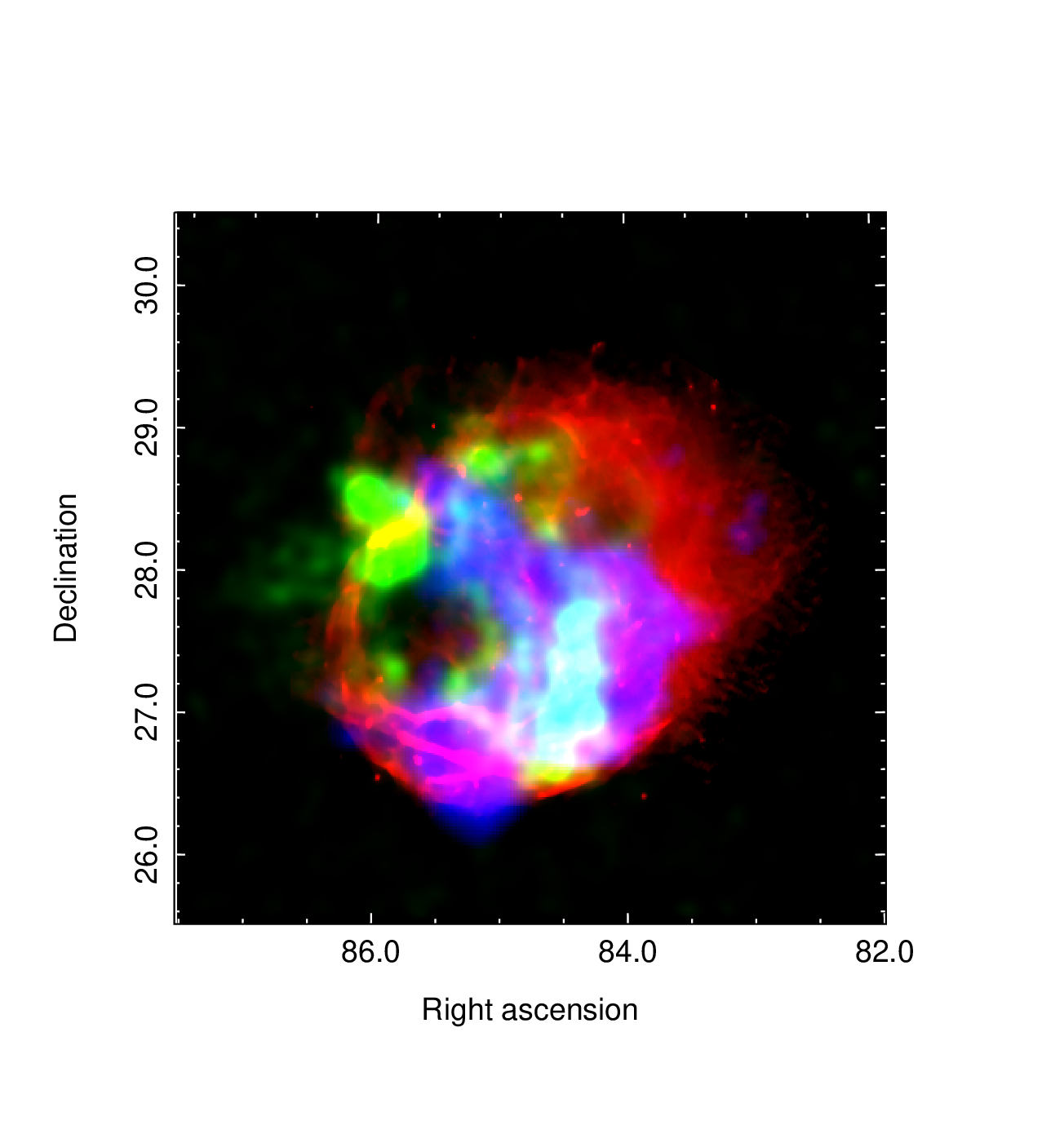}
    \caption{Refined spatial distribution of $\gamma$-ray emission across the remnant's extension and multiwavelength picture. Left panel: $4\degree\times4\degree$ \textit{Fermi}-LAT TS map >1~GeV. The image, of $90''$ pixels, is convolved with a $\sigma=1.5'$ Gaussian. The thick magenta circle represents the $68\%$ containment PSF size, derived at the 1~GeV energy threshold used for the construction of the TS map. The black and white contours represent the location of CO clouds (CO Galactic Plane survey data \citep{2001ApJ...547..792D}) likely interacting with the SNR. Right panel: Combined CGPS data at 1.42~GHz, red, 0.5-1.0~keV eRASS:4 data (green) and \textit{Fermi}-LAT data $>1$~GeV (blue) from the location of the remnant.}
    \label{07}
\end{figure*}

\begin{figure*}[]
    \centering
    \includegraphics[width=0.425\textwidth,clip=true, trim= 0.9cm 0.1cm 1.0cm 0.5cm]{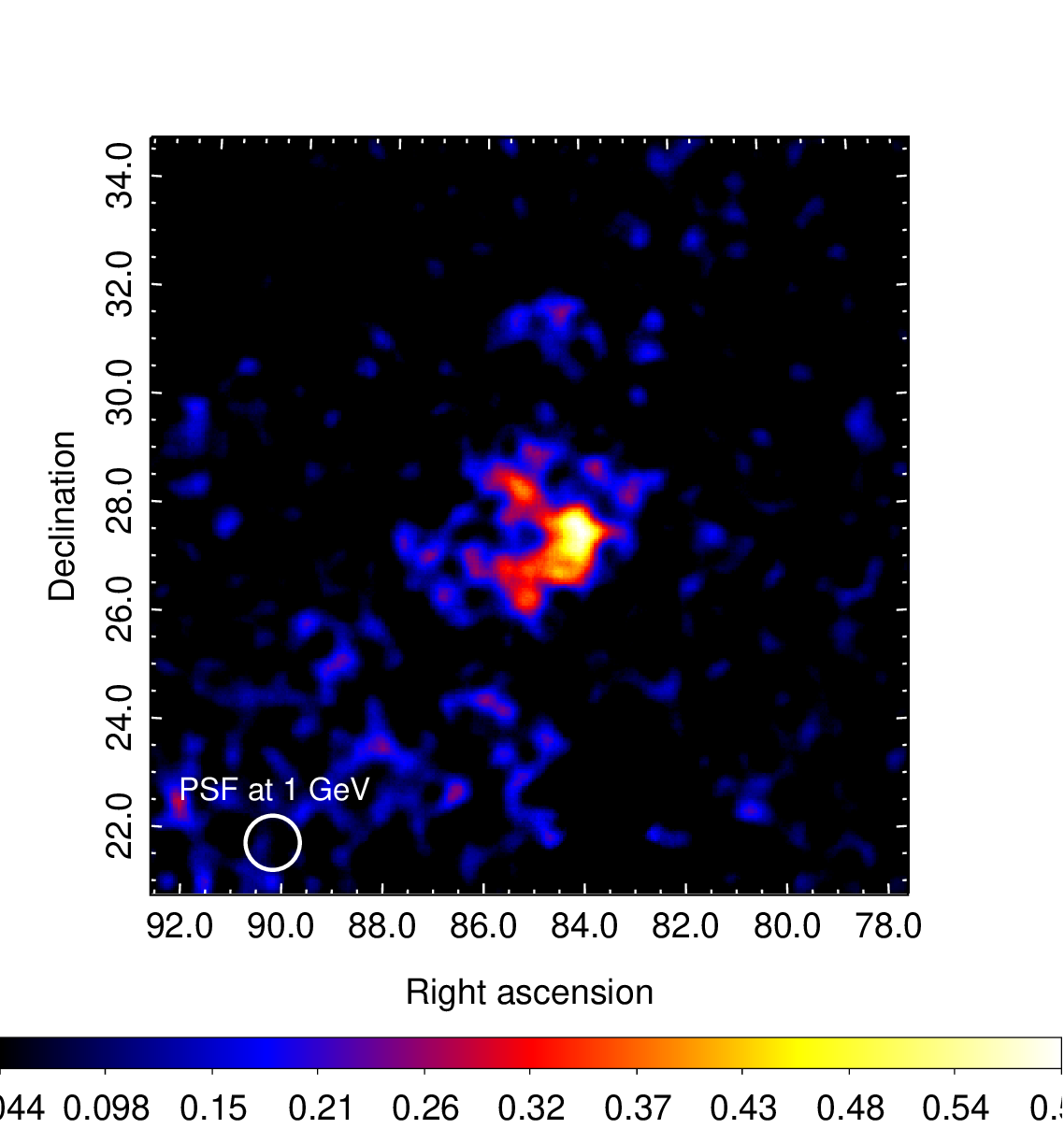}\includegraphics[width=0.62\textwidth]{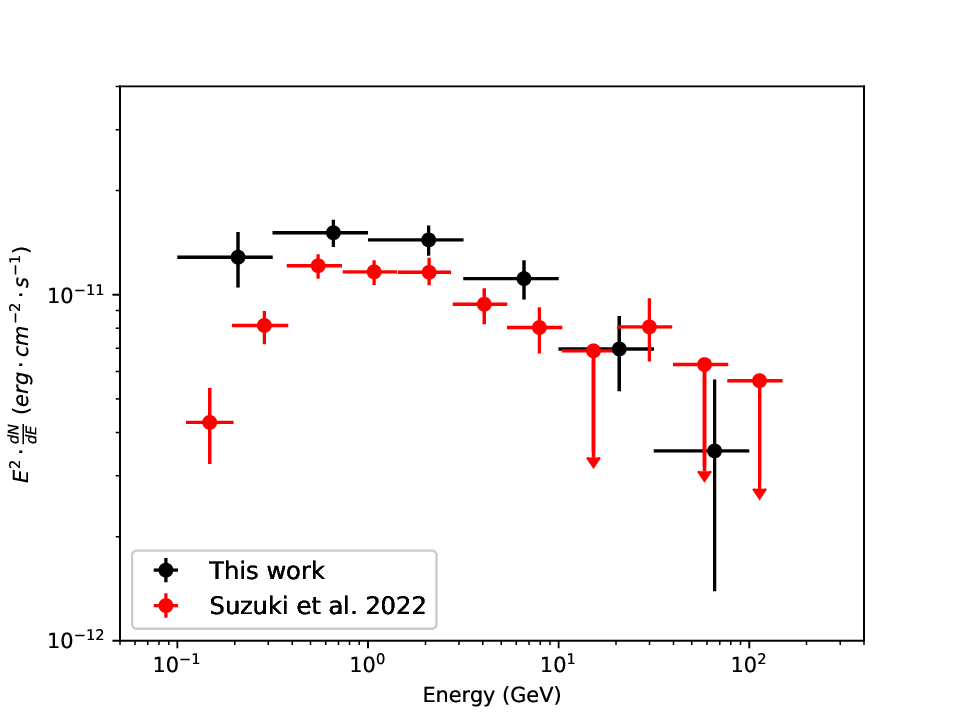}
    \caption{Imaging and spectral analysis results of $\sim15$ years of \textit{Fermi}-LAT data at and around the remnant's location. Left panel: $14\degree\times14\degree$ \textit{Fermi}-LAT residual count map >1~GeV in units of counts per pixel. The image, of $90''$ pixels, is convolved with a $\sigma=10.1'$ Gaussian. The thick white circle represents the $68\%$ containment PSF size derived at the 1~GeV energy threshold used for the construction of the residual count map. Right panel: \textit{Fermi}-LAT giga-electronvolt SED of S147. Black dots correspond to the \textit{Fermi}-LAT spectrum in the 0.1-100~GeV band, obtained in this work. Red dots correspond to giga-electronvolt \textit{Fermi}-LAT results reported in \citet{2022ApJ...924...45S}.}
    \label{08}
\end{figure*}

\subsubsection{\textit{CO \& giga-electronvolt image correlation}}
\label{GeVV}

The giga-electronvolt \textit{Fermi}-LAT data from the location of S147 \citep{2012ApJ...752..135K} and optical and infrared photometry based on the construction of a 3D extinction map \citep{2017MNRAS.472.3924C} suggest that S147 has a giga-electronvolt counterpart, and it might be linked to a dust cloud at a distance of 1 to 1.5 kpc to Earth. According to findings of \citet{2017MNRAS.472.3924C}, the dust cloud that goes by the name "S147 dust cloud" is most likely associated with the SNR itself, and their interaction may have a significant impact on the generation of $\gamma$-rays. 

Compared to the 31 month \textit{Fermi}-LAT data utilized in \citet{2012ApJ...752..135K}, we employed an additional 12 years of data to verify the association of the extended giga-electronvolt source 4FGL J0540.3+2756e or FGES J0537.6+2751 with the remnant and provide updated imaging (refining the remnant's morphology in the giga-electronvolt band) and spectral analysis results. For that, we analyzed Pass 8 LAT data collected from August 4, 2008 to March 28, 2023 using the most recent response function (P8R3) and the $\texttt{fermitools}$ Ver. 2.0.8 analysis tool and by applying recommended cuts. We analyzed a region of $20\degree$ in radius, centered on R.A.: $85.1\degree$, Dec: $27.94\degree$ in the 0.1-200~GeV energy range. We used $\texttt{SOURCE}$ event data, filtered under the $\texttt{evclass}=128$ and $\texttt{evtype}=3$ to employ front and back interactions. Contamination from the Earth's limb is prevented by restricting the zenith angle to a maximum of $90\degree$. We selected a spatial binning size of $0.05\degree$ per pixel, and we set 30 logarithmic energy bins per energy decade for the construction of the exposure map. The likelihood function implemented in \citet{1996ApJ...461..396M} was exploited to
fit the source's spectral data in the most optimal way. Sources were selected from the latest 4FGL catalog alongside the Galactic diffuse-emission component provided in $\texttt{gll\_iem\_v07.fits}$
and the residual background and extragalactic component provided in $\texttt{iso\_P8R3\_SOURCE\_V3\_v1.txt}$. With the aim of refining the giga-electronvolt morphology of S147, we produced both the test statistic (TS) map and the residual count map above 1~GeV as shown on the left panels of Fig.~\ref{07} and Fig.~\ref{08}, respectively. Both panels were produced by fitting the event data, allowing only the normalization of S147 and the normalization of all sources within an area of $5\degree$ radius from the center of the analysis vary (center of the source of interest). The extended giga-electronvolt excess is spatially coincident with the X-ray emission (as illustrated in the right panel of Fig.~\ref{07}) and with the enhanced optical H$\alpha$ regions to the southern part of the remnant, as can be concluded by the comparison of the X-ray and H$\alpha$ data shown in Fig.~\ref{05}, and is consistent with the remnant's size. A robust giga-electronvolt detection is confirmed by both a detection significance greater than 10$\sigma$ and by the absence of significant negative residuals in the corresponding residual count map of Fig.~\ref{08}.

Exploiting CO Galactic Plane survey data \citep{2001ApJ...547..792D}, we investigated a potential spatial correlation of CO clouds and giga-electronvolt \textit{Fermi}-LAT data from the location of the remnant. As shown in the left panel of Fig.~\ref{07}, CO clouds (as plotted with white contours) are spatially coincident with the southern and central regions of the enhanced $\gamma$-ray emission. However, a similar correlation is missing for the north-eastern bright giga-electronvolt blob. Additionally, the three dense clumps, which when combined compose the S147 dust cloud detected in \citet{2017MNRAS.472.3924C}, are in excellent spatial agreement with regions of enhanced CO emission detected in the composite CO survey of the entire MW \citep{2001ApJ...547..792D}. 

It is evident that there is 
excellent spatial correlation between CO clouds and the southern+central rim of S147 that is bright in giga-electronvolt. Examining whether there is a genuine correlation between the SNR-MC interaction and the origin of the SNR's $\gamma$-ray emission or whether the dust cloud (if positioned in the foreground) is illuminated by CRs originating from the SNR and is therefore emitting in the giga-electronvolt band, is beyond the scope of this paper. The scenario that $\gamma$-ray emission could originate from thin filaments observed in the optical and radio-continuum data, mainly due to the good spatial coincidence of the $\gamma$-ray emission, the thin filament, and the giga-electronvolt and H$\alpha$ flux correlation \citep{2012ApJ...752..135K}, cannot be ruled out.

\begin{figure*}[h]
    \centering

    \includegraphics[width=0.51\textwidth]{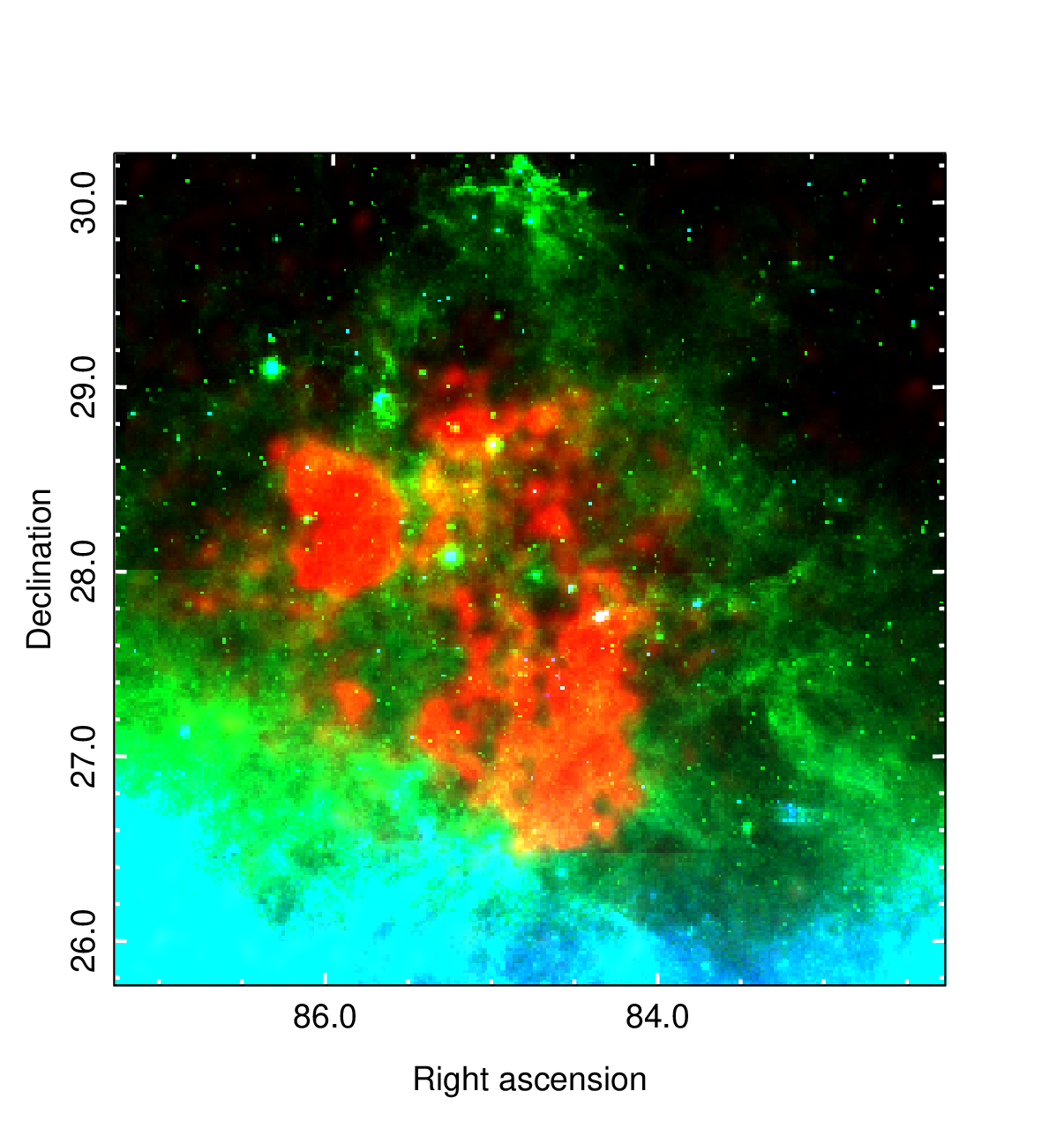}\includegraphics[width=0.51\textwidth]{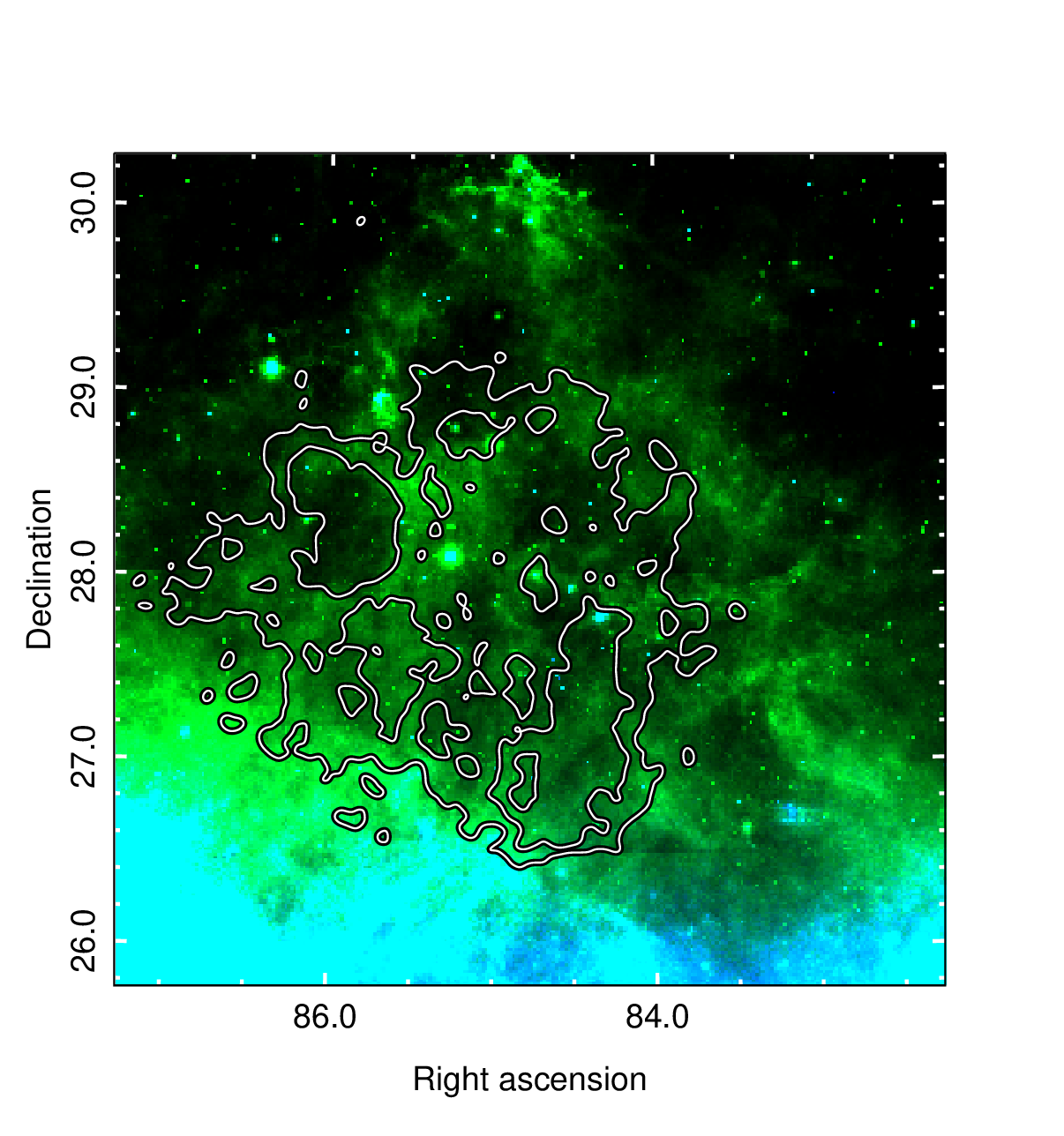}
    \caption{Localized features of spatial anticorrelation between X-ray and infrared emission. Left panel: RGB image, with energy color-coded as follows: eRASS:4 X-ray data in the 0.5-1.0~keV energy band (red), IRAS 25 $\mu$m data (green), and IRAS 100 $\mu$m data (blue) from the location of S147. Right panel: Combined IRAS 25 $\mu$m data (green) and IRAS 100 $\mu$m data (blue) from the same location as in the left panel. The black and white contours represent two levels of eRASS:4 X-ray data in the 0.5-1.0~keV energy band, which we overlaid on the IRAS data, aiming to examine potential anticorrelation features between IR and X-ray emission.}
    \label{06}
\end{figure*}
\begin{figure*}[h]

    \includegraphics[width=0.33\textwidth,clip=true, trim= 1.4cm 0.1cm 1.3cm 1.3cm]{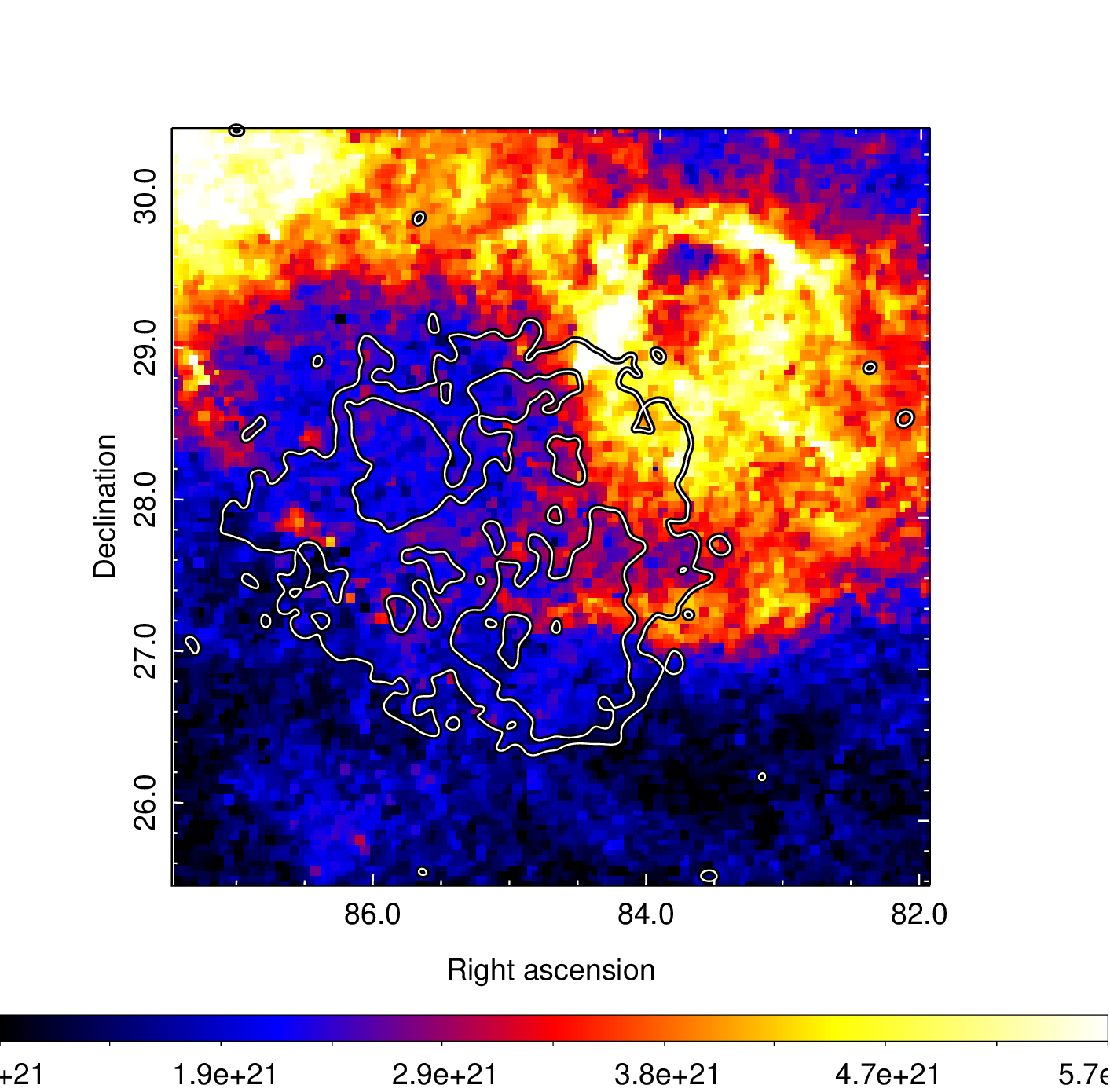}\includegraphics[width=0.33\textwidth,clip=true, trim= 1.4cm 0.1cm 1.3cm 1.2cm]{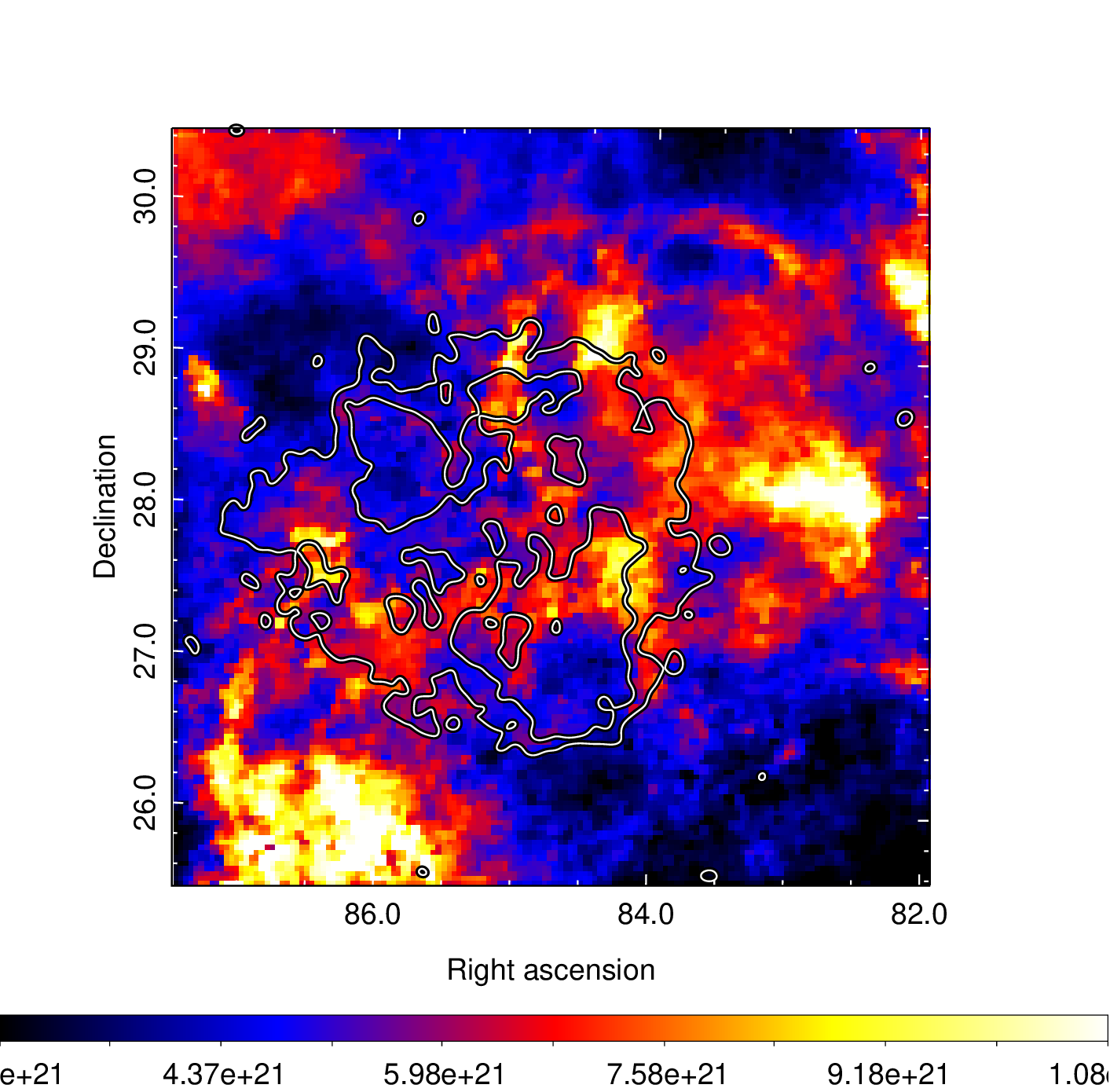}\includegraphics[width=0.33\textwidth,clip=true, trim= 1.4cm 0.1cm 1.3cm 1.2cm]{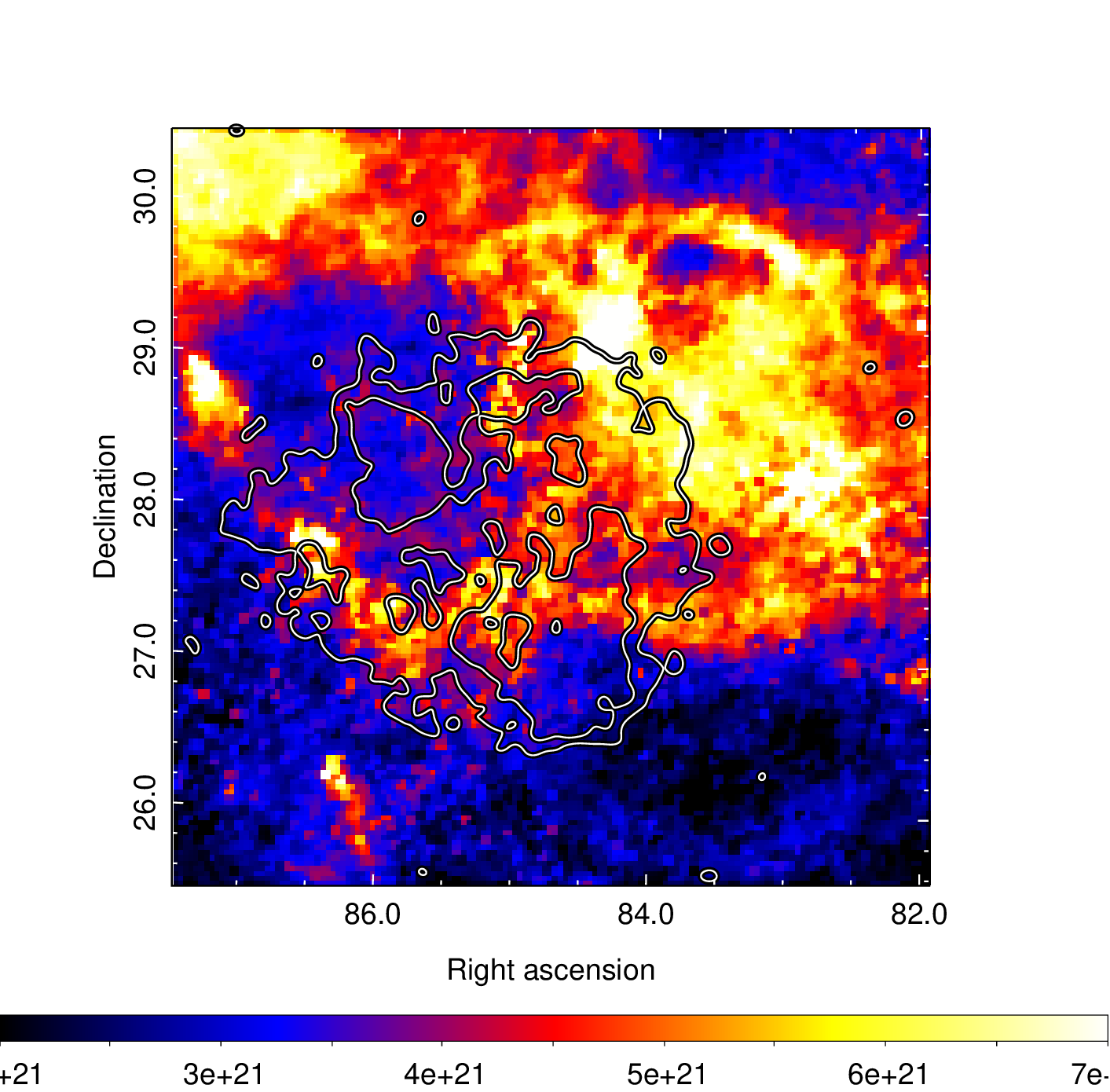}
    
    \caption{Hydrogen column-density maps ($N_\mathrm{H}$, in units of $\mathrm{cm^{-2}}$) derived by utilizing the DUSTMAPS python package \citep{2018JOSS....3..695M}, employing \texttt{bayestar19} data cubes \citep{2019ApJ...887...93G}, and converting the obtained extinction to $N_\mathrm{H}$ according to Eq.~\ref{math} \citep{2016ApJ...826...66F}. Left panel: $N_\mathrm{H}$ map until a 0.6 kpc distance. Middle panel: $N_\mathrm{H}$ map until 1.9 kpc distance. Right panel: $N_\mathrm{H}$ map until 1.33 kpc distance (preferred distance of the SNR given the distance measurements of the associated pulsar and the runaway star-binary companion to the pulsar's progenitor).} 
    \label{09}
\end{figure*}
\subsubsection{\textit{IR data $\&$ dust prevalence}}\label{IR}
Having confirmed the detection of the remnant's giga-electronvolt counterpart and provided the first detailed imaging analysis (with substantially updated results compared to \citet{2012ApJ...752..135K}), we furthermore focus on explaining the nature of apparent interesting structures that the remnant exhibits. Such prominent features are the two strongest (in terms of surface brightness)  diffuse X-ray blobs to the north-east (region B) and south-west (region H) of the remnant, as well as the smaller bright blob at the south-east (region C) of the remnant, and the ear-type structure (region A) and the void structure (region D) as introduced in Section~\ref{eROSEspectra}. The latter structures, besides in X-rays, are also visible in radio synchrotron and H$\alpha$ data, as shown in Fig.~\ref{05}. Fig.~\ref{06} demonstrates the presence of enhanced infrared (IR) emission regions, detected in the infrared band utilizing Infrared Astronomical Satellite (IRAS) data. Examining the two panels of Fig.~\ref{06}, one easily concludes that the two large X-ray blobs and the smaller bright X-ray blob to the south-east of the remnant are spatially coincident with regions where the IR data exhibits the lowest intensity. 
The same applies to the easternmost part of the ear-type structure, which is also the brightest part of the latter structure in X-rays. At and around the void structure, an enhancement of the IR emission in comparison to the aforementioned regions is observed.
Overall, in light of the new X-ray data from eROSITA, even though the image analysis indicates strong anticorrelation features (e.g., at the location of the northern bright X-ray blob) between the X-ray emission and the IRAS data, such an anticorrelation is not seen in the north-western part of the remnant. An X-ray and IR emission anticorrelation was also previously suggested in \citet{2017MNRAS.472.3924C} using ROSAT data. The derived absorption spectral parameters obtained from the fit (see Section~\ref{eROSEspectra} for further analysis details) do not strongly support a clear absorption pattern (except for the void-type structure) either, as shown by the reported values of Table~\ref{TABIS1}. IR emission is strongest in the south-east and north-west of the remnant. The latter assertion is also confirmed by the hydrogen column-density maps of Fig.~\ref{09} and could potentially explain the lack of X-ray emission in the western part of the remnant, assuming that X-ray photons were absorbed due to the prevalence of dust. 
The strongest absorption column density value, obtained from the spectral analysis of the ten individual subregions (see Section~\ref{eROSEspectra} for further details) matches the location of the void structure of the remnant. Such a finding is also supported by the enhanced IR emission at that particular location of the sky. On the contrary, it is also worth noting that a strong anticorrelation between X-ray and IR data (potential nature is dust destruction by X-rays - refer to \citet{1987ApJ...318..674M,2015ApJ...803....7S,2021MNRAS.500.2543P} and references therein, for the efficiency of dust destruction by SNR shockwaves) becomes clearly apparent at the location of the Northern bright X-ray blob of the remnant. This assertion is once again supported by spectral analysis results (that exact region exhibits the lowest absorption column-density value across the entire remnant; Section~\ref{eROSEspectra}). At that part of the remnant, the IR emission seems to respect the X-ray emission creating a "hole" in the IR data of the same size and shape as the northern X-ray bright blob of the remnant. Overall, the comparison of IR and X-ray data from the remnant's location provides evidence for the presence of different physical processes underlined by apparent features (i.e., dust destruction by X-rays and X-ray absorption by dust). Those indications are found to be broadly consistent with the conclusions derived by the spectral analysis of individual parts (subregions) of the remnant.

\begin{figure}[h!]
    
    \includegraphics[width=0.495\textwidth,clip=true, trim= 0.99cm 0.1cm 0.9cm 0.9cm]{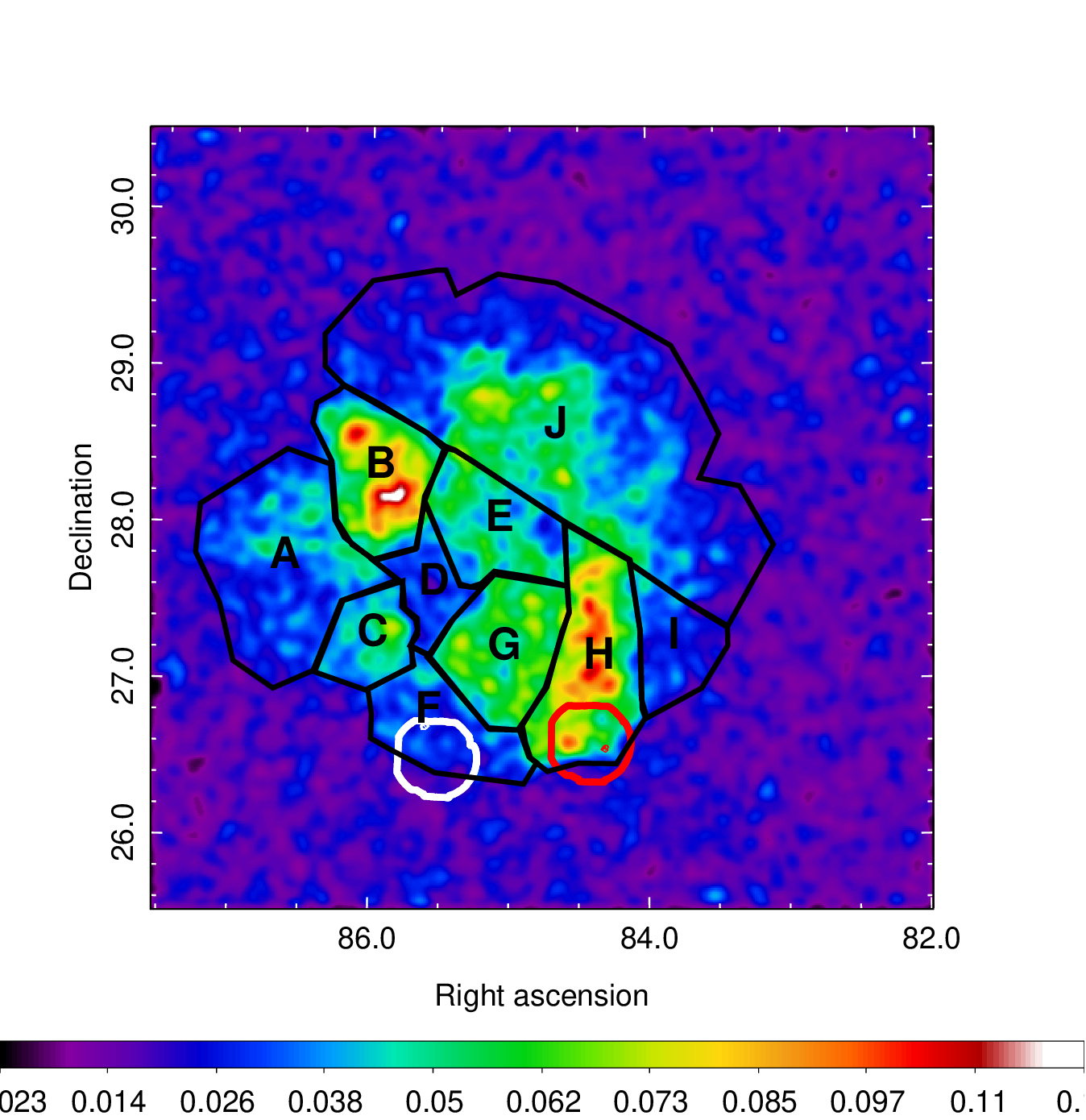}
    \caption{
    Same image as in Fig.~\ref{01} (with identical features in terms of smoothing and point-source removal) depicting the intensity variation across the remnant. 
    The ten distinct subregions selected for further spectral analysis are highlighted as black regions. The entire remnant's spectrum was obtained from the large polygonal shape region formed when combining all ten individual subregions. White and red contours mark the positions of 0693270301 and 0693270401 \textit{XMM-Newton} pointings (shown in Fig.~\ref{04}), respectively.}
    \label{011}
\end{figure}
\section{\textit{Spectral analysis}}
\label{section3}
\subsection{\textit{eROSITA spectra}}
\label{eROSEspectra}
Event data of eRASS:4 were selected from ten distinct spectral extraction regions (optimized based on the surface brightness variations detected across the remnant) of polygon shape, as depicted in Fig.~\ref{011} in black. The selected regions were defined in \texttt{SAOIMAGE DS9} \citep{joye2003new}, which is aimed at inspecting the spectral morphology of the remnant in detail and gaining more insight into interesting individual regions. Among such regions are the ear-type structure, the void region, and the two brightest X-ray blobs at the south-west and north-east of the remnant. The entire remnant's spectrum was also extracted from the area obtained when combining all ten aforementioned subregions and fitted accordingly as described below. A nearby background control region of circular shape 
was chosen to the south-east of the remnant, which is free of emission from the SNR itself (center: R.A.: $89.08\degree$, Dec: $26.31\degree$ radius: $1.58\degree$).
Events from TM5 and TM7 were excluded due to their light leak suffering \citep{2021A&A...647A...1P}. X-ray emitting point sources, detected with a $3\sigma$ excess significance or higher, were excluded using an exclusion circular region of $110''$ radius. Individual bright sources, that is, the HMXB 1A 0535+262 and the associated pulsar (accompanied by its faint PWN) were once again treated independently to avoid likely spectral contamination. The fitting process was conducted in \texttt{Xspec Ver. 12.12.1} \citep{2003HEAD....7.2210D}. A minimum of 30 counts per bin has been set to display the spectrum before fitting. C-statistics \citep{1979ApJ...228..939C} was applied for the fitting procedure due to the limited photon statistics. Simultaneous fitting of the source and background emission from the on-source regions was performed by adopting identical spectral models and an identical fitting approach for the background (astrophysical and instrumental) emission, as described in \citet{G279MILTOS} (Section~4.1).

Even though the majority of the emission from the remnant is confined to the 0.5-1.0~keV energy band, we performed a spectral analysis in the broader range of 0.3-2.3~keV since some faint X-ray emission does exist below 0.5 keV and above 1.0 keV. We restrict the spectral analysis to $< 2.3$~keV since at harder energies the background component becomes dominant. From the spectral analysis of the entire remnant, but also the inspection of individual subregions, we conclude the following. The X-ray emission is found to be solely thermal, with no evidence of a nonthermal component. 
However, from the fit quality it is not clear whether the hot plasma is in collisional ionization equilibrium (CIE) or not. Among the fitted models, both the collisionally ionized diffuse gas in equilibrium (\texttt{VAPEC} model in Xspec notation) and the non-equilibrium collisionally ionized diffuse gas model (\texttt{VNEI} or \texttt{VPSHOCK} model in Xspec notation) provide a good fit to the data. However, it is noteworthy that the obtained absorption column density, derived by optical extinction measurements at the distance of the remnant (as shown in Fig.~\ref{09}), is in good agreement with the corresponding best-fit parameter of nonequilibrium ionization collisional plasma models (NEI). The latter assertion is supported by the lower absorption column density values favored by NEI models compared to CIE model values. Additionally, CIE models require questionably high elemental-abundance values (in particular Mg) to explain spectral characteristics at higher energies (i.e., they underpredict the Mg XI line). 
To take into account the interstellar absorption (of the ISM) at the location of the remnant, all three additive models are modified according to the multiplicative \texttt{TBABS} absorption model \citep{2000ApJ...542..914W} (i.e., we used the following models: \texttt{TBABS(VAPEC/VNEI/VPSHOCK)}). All spectral extraction regions exhibit K-shell oxygen (OVII, OVIII), neon (NeIX, NeX), and magnesium lines (Mg XI, not present in all subregions and with statistical significance much lower compared to O and Ne lines).  
Therefore, in the thermal plasma models described above, elemental abundances were fixed to solar values except for O, Ne, and Mg, which were allowed to vary. It is noteworthy that both models provide poor fits when the aforementioned elemental abundances are fixed to solar values. 
The rest of the source model parameters were left free when fitting the model to the data.

\begin{figure*}[]

    \includegraphics[width=0.49\textwidth]{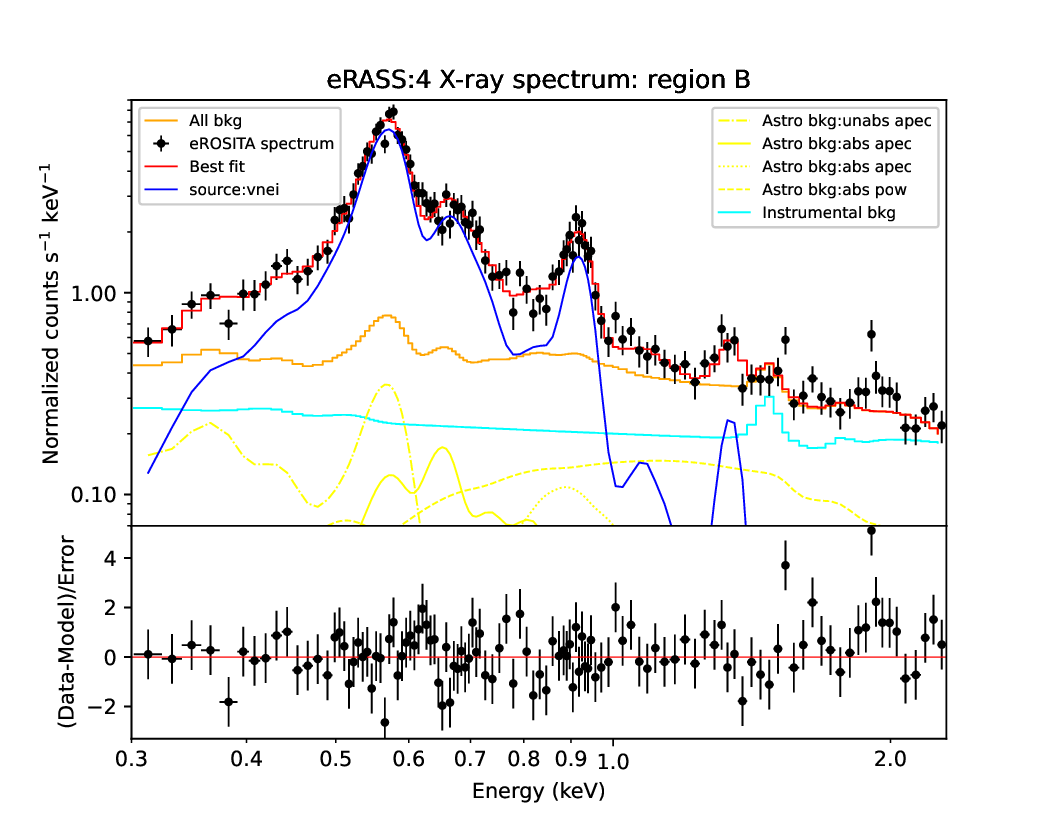}
    \includegraphics[width=0.502\textwidth]{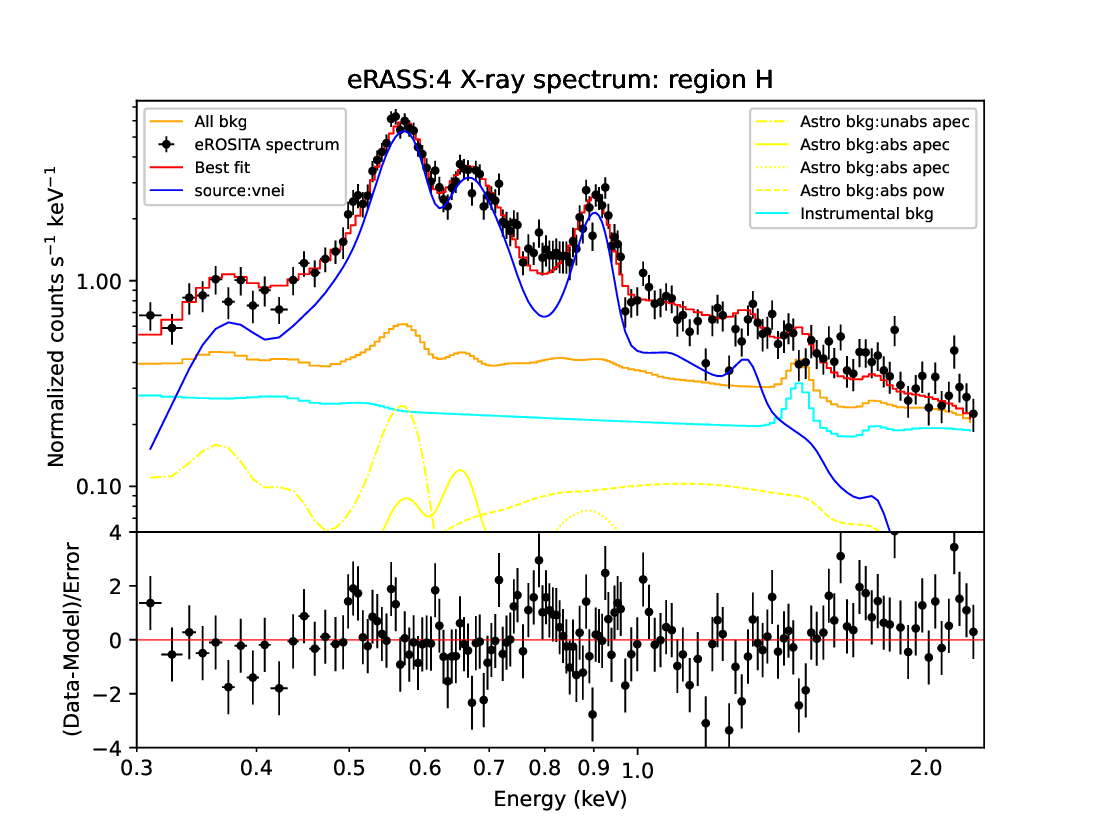}
    \centering
    \includegraphics[width=0.597\textwidth]{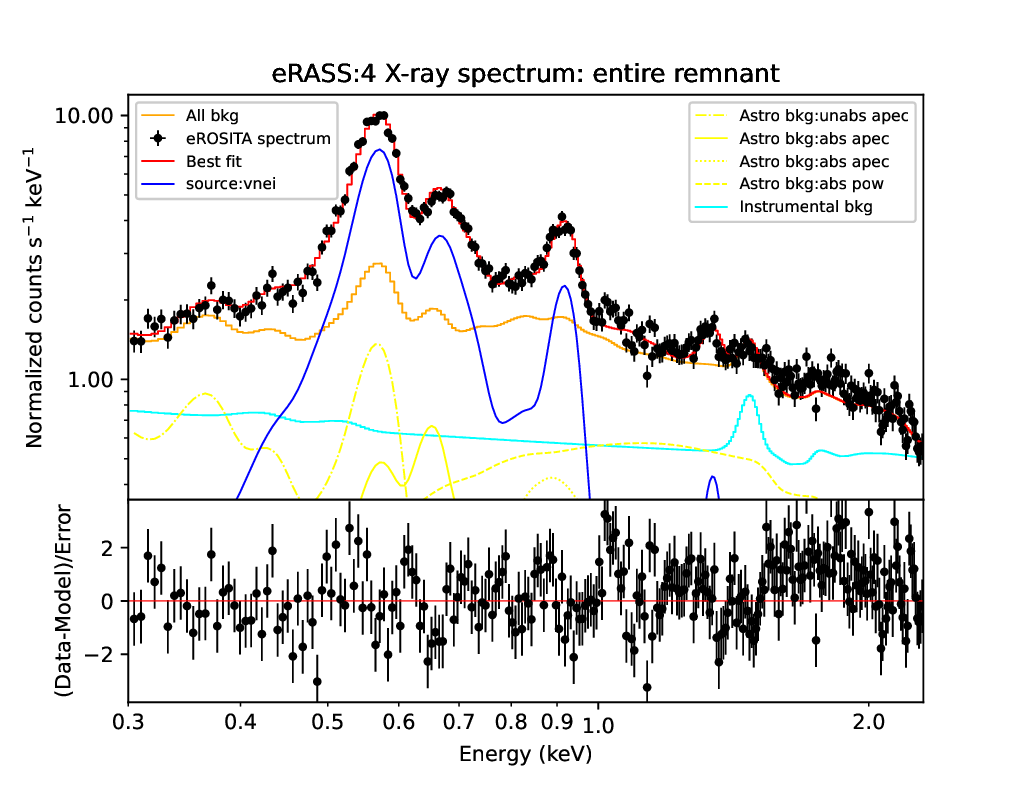}
    \caption{eRASS:4 X-ray spectra in the 0.3-2.3 keV energy band, from the two significantly enhanced regions of X-ray emission (X-rays blobs: region A and region H) and the entire remnant.}
    \label{ALLSUBSPEC}
\end{figure*}

In Fig~\ref{ALLSUBSPEC}, we report the results of the simultaneous fit of the on-source and background emission from the two richest bright X-ray blobs in terms of photon statistics, as well as from the entire remnant, with an absorbed \texttt{VNEI} 
as the optimal model describing the purely thermal S147 spectrum. We do not see significant X-ray spectral shape changes over the remnant's area. The results of the simultaneous fit of the on-source and background emission (using the same source model) from the remaining eight selected subregions are shown in Fig.~\ref{ALLSUBSPEC2} of Appendix A. The best-fit spectral parameters of 
the \texttt{tbabs(vnei)} model, with their $1\sigma$ errors, are reported in Table~\ref{TABIS1} for all ten selected subregions and the entire remnant. The corresponding areas and surface brightness estimates for all regions selected for spectral analysis are summarized in Table~\ref{TABISS1}. Significant temperature variations were detected across the remnant (mainly due to the need to better explain the high energy features of the spectrum of $>1$~keV of some subregions selected for spectral analysis as shown in Table~\ref{TABIS1}, which "hot NEI solutions" provide). The absorption column-density values obtained across the remnant's area neither differ significantly nor exhibit regular patterns. However, particular deviations such as an enhanced absorption column density value at the location of the void structure and a particular low absorption column density value (the lowest value detected across the whole remnant) at the location of the bright X-ray blob to the north-east of the remnant are clearly apparent, as shown in Table~\ref{TABIS1}. Average temperatures of $\sim0.22$ keV and $N_\mathrm{{H}}=0.3~\mathrm{10^{22}cm^{-2}}$ (\texttt{VNEI}), and $\sim0.11$ keV and of $N_\mathrm{{H}}=0.51~\mathrm{10^{22}cm^{-2}}$ (\texttt{VAPEC}) are obtained when exploiting data from the entire remnant area for the two distinct models, respectively. For the best-fit (\texttt{VNEI}) model of the entire remnant reported above, a total flux of $F_{\mathrm{total}}=6.93^{+1.98}_{-1.31}\cdot 10^{-10}~\mathrm{erg~cm^{-2}~s^{-1}}$ is obtained in the 0.3-2.3 keV energy range. However, individual subregions may exhibit much higher plasma temperature values, as reported in Table~\ref{TABIS1}. 
It is worth noting that the obtained elemental abundance values of O, Ne, and Mg (for both tested models) are inextricably linked to the corresponding normalization value, that is to say, the X-ray plasma is characterized by the lack of a strong continuum component and solely exhibits X-ray emission lines. In principle, one could restrict their variation range and draw useful conclusions. We describe such an approach in detail in \citetalias{Spaghettichugai}. In this work, we chose to keep them free and inspect potential variations across the remnant.

\begin{table*}
\centering
\caption{Best-fit spectral parameters.}
\renewcommand{\arraystretch}{1.6}
\setlength{\tabcolsep}{9pt}
\begin{tabular}
{p{2.5cm} p{5.5cm} p{5.5cm} p{5.5cm} p{5.5cm} p{5.5cm} p{5.5cm} p{5.5cm} p{5.5cm} p{5.5cm} }
\hline
Region & \multicolumn{3}{c}{A$\dagger$ region}& \multicolumn{3}{c}{B region} &\multicolumn{3}{c}{C* region} \\ \hline
Model & \multicolumn{9}{c}{vnei}  \\ \hline
kT{\small(keV)}& \multicolumn{3}{c}{$0.41_{-0.05}^{+0.09}$}& \multicolumn{3}{c}{$0.23_{-0.04}^{+0.06}$} & \multicolumn{3}{c}{$0.50_{-0.11}^{+0.21}$}\\ 
$\mathrm{N_{H}}${\small ($\mathrm{10^{22}cm^{-2}}$)}&\multicolumn{3}{c}{$0.23_{-0.04}^{+0.03}$}& \multicolumn{3}{c}{$0.18_{-0.04}^{+0.04}$}  & \multicolumn{3}{c}{$0.32_{-0.04}^{+0.04}$}\\
O&\multicolumn{3}{c}{   $1.30_{-0.28}^{+0.40}$}& \multicolumn{3}{c}{ $1.96_{-0.25}^{+0.32}$} & \multicolumn{3}{c}{$0.47_{-0.13}^{+0.21}$} \\
Ne&\multicolumn{3}{c}{$1.63_{-0.37}^{+0.67}$}& \multicolumn{3}{c}{$3.48_{-0.77}^{+0.96}$}& \multicolumn{3}{c}{$1.70_{-0.54}^{+0.66}$}\\ 
Mg&\multicolumn{3}{c}{$5.95_{-2.07}^{+3.05}$}& \multicolumn{3}{c}{$9.98_{-2.08}^{+4.01}$}& \multicolumn{3}{c}{-}\\
Ionization time (\small{$\mathrm{10^{10}s\cdot cm^{-3}}$)} &\multicolumn{3}{c}{$0.66_{-0.24}^{+0.23}$}& \multicolumn{3}{c}{$4.59_{-2.61}^{+5.74}$}& \multicolumn{3}{c}{$0.09_{-0.02}^{+0.03}$} \\
 \hline
 $\mathrm{\chi^2/dof}$ &\multicolumn{3}{c}{1.04}& \multicolumn{3}{c}{1.09} &  \multicolumn{3}{c}{1.11} \\ 
\hline
Region & \multicolumn{3}{c}{D region}& \multicolumn{3}{c}{E$\dagger$ region} &\multicolumn{3}{c}{F region} \\ \hline
Model & \multicolumn{9}{c}{vnei}  \\ \hline
kT{\small (keV)}& \multicolumn{3}{c}{$0.19_{-0.05}^{+0.18}$}& \multicolumn{3}{c}{$0.42_{-0.06}^{+0.14}$} & \multicolumn{3}{c}{$2.17_{-0.83}^{+1.57}$}\\ 
$\mathrm{N_{H}}${\small ($\mathrm{10^{22}cm^{-2}}$)}&\multicolumn{3}{c}{$0.49_{-0.12}^{+0.11}$}& \multicolumn{3}{c}{$0.28_{-0.06}^{+0.05}$} & \multicolumn{3}{c}{$0.33_{-0.04}^{+0.03}$}  \\
O&\multicolumn{3}{c}{-}& \multicolumn{3}{c}{$2.41_{-0.68}^{+0.62}$} & \multicolumn{3}{c}{$0.80_{-0.23}^{+0.29}$} \\
Ne&\multicolumn{3}{c}{-}& \multicolumn{3}{c}{$2.58_{-0.81}^{+1.66}$}& \multicolumn{3}{c}{$1.19_{-0.42}^{+0.43}$}\\ 
Mg&\multicolumn{3}{c}{-}& \multicolumn{3}{c}{$7.07_{-2.72}^{+4.96}$}& \multicolumn{3}{c}{-}\\
Ionization time (\small{$\mathrm{10^{10}s\cdot cm^{-3}}$)} &\multicolumn{3}{c}{$2.09_{-1.21}^{+6.69}$ }& \multicolumn{3}{c}{$0.64_{-0.19}^{+0.29}$} & \multicolumn{3}{c}{$0.065_{-0.02}^{+0.03}$}\\
\hline
$\mathrm{\chi^2/dof}$ &\multicolumn{3}{c}{1.34}& \multicolumn{3}{c}{0.99} & \multicolumn{3}{c}{1.19}\\ 
 \hline
 Region & \multicolumn{3}{c}{G region}& \multicolumn{3}{c}{H* region} &\multicolumn{3}{c}{I region} \\ \hline
Model & \multicolumn{9}{c}{vnei}  \\ \hline
kT{\small (keV)}& \multicolumn{3}{c}{$0.25_{-0.04}^{+0.12}$}& \multicolumn{3}{c}{ $0.47_{-0.08}^{+0.04}$} & \multicolumn{3}{c}{$0.21_{-0.02}^{+0.06}$} \\ 
$\mathrm{N_{H}}${\small ($\mathrm{10^{22}cm^{-2}}$)}&\multicolumn{3}{c}{$0.29_{-0.08}^{+0.05}$}& \multicolumn{3}{c}{$0.33_{-0.01}^{+0.02}$} & \multicolumn{3}{c}{$0.24_{-0.09}^{+0.10}$}  \\
O&\multicolumn{3}{c}{$2.08_{-0.35}^{+0.43}$}& \multicolumn{3}{c}{$0.33_{-0.04}^{+0.05}$} & \multicolumn{3}{c}{$0.57_{-0.14}^{+0.12}$}  \\
Ne&\multicolumn{3}{c}{$2.62_{-0.65}^{+0.88}$}& \multicolumn{3}{c}{$0.95_{-0.16}^{+0.18}$} & \multicolumn{3}{c}{$0.98_{-0.34}^{+0.35}$}\\ 
Mg&\multicolumn{3}{c}{$9.31_{-2.45}^{+3.31}$}& \multicolumn{3}{c}{-}& \multicolumn{3}{c}{$11.1_{-0.43}^{+0.62}$}  \\ 
Ionization time (\small{$\mathrm{10^{10}s\cdot cm^{-3}}$)} &\multicolumn{3}{c}{$2.32_{-0.87}^{+3.06}$}& \multicolumn{3}{c}{$0.11_{-0.01}^{+0.01}$}& \multicolumn{3}{c}{$11.94_{-7.97}^{+11.69}$} \\
 \hline
$\mathrm{\chi^2/dof}$ &\multicolumn{3}{c}{1.2}& \multicolumn{3}{c}{1.54} & \multicolumn{3}{c}{1.22}  \\ 
\hline
Region & \multicolumn{3}{c}{J* region}& \multicolumn{3}{c}{Entire remnant} &\multicolumn{3}{c}{\textit{XMM-Newton}} \\ \hline
Model &  \multicolumn{9}{c}{vnei} \\ \hline
kT{\small (keV)}& \multicolumn{3}{c}{$0.75_{-0.17}^{+0.11}$}& \multicolumn{3}{c}{$0.22_{-0.03}^{+0.02}$} &\multicolumn{3}{c}{Identical to region H} \\ 
$\mathrm{N_{H}}${\small ($\mathrm{10^{22}cm^{-2}}$)}&\multicolumn{3}{c}{$0.30_{-0.01}^{+0.03}$}& \multicolumn{3}{c}{$0.30_{-0.03}^{+0.04}$}& \multicolumn{3}{c}{}  \\
O&\multicolumn{3}{c}{$0.46_{-0.06}^{+0.06}$}& \multicolumn{3}{c}{ $2.34_{-0.18}^{+0.20}$} &\multicolumn{3}{c}{} \\
Ne&\multicolumn{3}{c}{$1.00_{-0.16}^{+0.18}$}& \multicolumn{3}{c}{$3.13_{-0.34}^{+0.40}$}&\multicolumn{3}{c}{}\\ 
Mg&\multicolumn{3}{c}{-}& \multicolumn{3}{c}{$9.53_{-1.31}^{+1.47}$}&\multicolumn{3}{c}{}\\ 
Ionization time (\small{$\mathrm{10^{10}s\cdot cm^{-3}}$)} &\multicolumn{3}{c}{$0.09_{-0.01}^{+0.01}$}& \multicolumn{3}{c}{$4.27_{-0.99}^{+1.87}$}&\multicolumn{3}{c}{}\\
 \hline
$\mathrm{\chi^2/dof}$ &\multicolumn{3}{c}{1.31}& \multicolumn{3}{c}{1.63} &\multicolumn{3}{c}{}\\ 
\hline
\label{TABIS1}
\end{tabular}
\tablefoot{The best-fit spectral parameters of the regions that have been selected to best represent the spectral variation detected across the remnant are provided with $1\sigma$ statistical errors. Where not defined, elemental abundances are set to solar values. For regions marked with an *, very small N elemental abundance values (essentially equal to zero) were found. For regions marked with a $\dagger$, Si appears to be present, and the corresponding elemental abundance is highly degenerate; thus, we allowed it to vary, aiming to improve the fit quality.}
\end{table*}

\begin{table*}
\centering
\caption{Parameters of spectrally analyzed regions.}
\renewcommand{\arraystretch}{1.6}
\setlength{\tabcolsep}{9pt}
\begin{tabular}
{p{2.5cm} p{5.5cm} p{5.5cm} p{5.5cm} p{5.5cm} p{5.5cm} p{5.5cm} p{5.5cm} p{5.5cm} p{5.5cm} p{5.5cm} p{5.5cm} p{5.5cm} p{5.5cm} p{5.5cm} p{5.5cm} p{5.5cm} p{5.5cm} p{5.5cm}}
\hline
Region & \multicolumn{3}{c}{A region}& \multicolumn{3}{c}{B region} &\multicolumn{3}{c}{C region} & \multicolumn{3}{c}{D region} & \multicolumn{3}{c}{E region} & \multicolumn{3}{c}{F region}\\ \hline
Area ($10^{6}~\mathrm{arcs^2}$) &\multicolumn{3}{c}{14.4}& \multicolumn{3}{c}{7.17} &\multicolumn{3}{c}{3.83} &\multicolumn{3}{c}{2.95}&\multicolumn{3}{c}{5.62}&\multicolumn{3}{c}{6.25}  \\ \hline
$\mathrm{Surf\_bri}$ ($10^{-3}~\mathrm{c\cdot arcs^{-2}}$) &\multicolumn{3}{c}{0.68}& \multicolumn{3}{c}{1.14} &\multicolumn{3}{c}{0.84}& \multicolumn{3}{c}{0.64}&\multicolumn{3}{c}{0.85}&\multicolumn{3}{c}{0.72}   \\ \hline
Region & \multicolumn{3}{c}{G region}& \multicolumn{3}{c}{H region} &\multicolumn{3}{c}{I region} & \multicolumn{3}{c}{J region} & \multicolumn{3}{c}{Entire remnant} & \multicolumn{3}{c}{\textit{XMM-Newton}}\\ \hline
Area ($10^{6}~\mathrm{arcs^2}$) &\multicolumn{3}{c}{7.42}& \multicolumn{3}{c}{7.80} &\multicolumn{3}{c}{4.40} &\multicolumn{3}{c}{38.9}&\multicolumn{3}{c}{105.47}&\multicolumn{3}{c}{1.2}  \\ \hline
$\mathrm{Surf\_bri}$ ($10^{-3}~\mathrm{c\cdot arcs^{-2}}$) &\multicolumn{3}{c}{0.99}& \multicolumn{3}{c}{1.12} &\multicolumn{3}{c}{0.67}& \multicolumn{3}{c}{0.73}&\multicolumn{3}{c}{0.81}&\multicolumn{3}{c}{83.0}   \\ \hline
\label{TABISS1}
\end{tabular}
\tablefoot{Total area and surface brightness of all regions used for spectral analysis in the 0.3-2.3 keV energy range.}
\end{table*}

Overall, the void structure region (region D), the bright blob north-east of the remnant (region B), and the western region (namely region I) that exhibits only weak X-ray diffuse emission appear to be the coolest. 
On the other hand, regions C, F, and the X-ray-bright region H appear to be the hottest across the remnant. 
The absence of X-ray emission in region D, a feature of the remnant denoted as "void structure" in this work and also detected in radio and optical (H$\alpha$) bands, could be (at least partially) attributed to X-ray absorption by dust clouds found at the location of the remnant. The latter conclusion is supported by the significant enhancement of the absorption column density found from the spectral fit of that particular region. However, its nature is not entirely clear. No distinct pattern of absorption column-density variation was found over the remnant, either. The highest $N_\mathrm{H}$ value, derived from the best fit among all seven regions, was obtained at the location of the void structure, denoted as region D, which is well correlated with a significant enhancement of the IR emission at that location, as seen in Fig.~\ref{06}. The lowest $N_\mathrm{H}$ value was derived to the north-west of the remnant, at the precise location of the enhanced X-ray blob (namely region B). The latter region is characterized by the absence of IR emission (potential dust destruction by X-rays) as shown in Fig.~\ref{06}.

The above result is also  supported by Fig.~\ref{09}, which depicts absorption column density sky maps toward the location of S147. The above maps were constructed by employing the \texttt{DUSTMAPS}\footnote{\url{https://dustmaps.readthedocs.io/en/latest/}} python package \citep{2018JOSS....3..695M}. In more detail, we analysed \texttt{bayestar19} data \citep{2019ApJ...887...93G} in the direction of the remnant, making use of the most recently established statistical relation between the mean colour excess-extinction and the absorption column density \citep{2016ApJ...826...66F}, as shown in Eq.~\ref{math1}.
We examined the data in the range of the most probable distance measurements of the remnant, 0.6-1.9~kpc, as derived by multiple works in the past (see Section~\ref{sec:intro} for more details). Comparing the obtained maps with the derived best-fit absorption column density values of Table~\ref{TABIS1}, one concludes that the remnant is placed at a distance greater than 0.6~kpc but moderately smaller than 1.33~kpc (the latter being the distance of both the pulsar associated with the remnant and the runaway star HD 37424, which is considered to be the binary companion of its progenitor).

It is noteworthy that in a number of the ten selected subregions and in the spectrum obtained from the entire remnant, O, Ne, and Mg elemental abundances display sufficiently (for ejecta identification) high values ($>$~2 solar), as shown in Table~\ref{TABIS1}. Regions A and E also appear to be enriched in silicon (Si), indicating X-ray plasma originating from ejecta, despite the evolved state of the remnant (making it perhaps the most evolved SNR that exhibits both swept-up ISM and ejecta components). However, robust conclusions about the presence of Si cannot be obtained due to the limited statistics of the data. The enhanced elemental abundance values obtained by both the analysis of the X-ray spectrum from the entire remnant and the X-ray spectrum from individual subregions suggest that thermal plasma has not yet reached equilibrium; that is to say, an NEI model is favored as the optimal way to describe the remnant's spectral characteristics. In fact, in spite of the common belief that ejecta origin abundances in evolved SNRs are unexpected, recent X-ray observations have revealed an increasing number of SNRs with metal-rich ejecta \citep{10.1093/pasj/51.1.13,Park_2003,2008A&A...485..777T,Hwang_2008}, regardless of their older age. The latter may partly be attributed to the presence of molecular material in the remnant's surroundings. However, S147 does not exhibit strong [OIII] lines in its optical spectrum, and thus it cannot be classified as an O-rich SNR \citep{1981ApJ...248L.105D,1979MNRAS.188..357G,4fbea282-ad61-35c8-8776-164566ece9b7,1980ApJ...242L..73M}, such as the recently discovered X-ray counterpart of G279.0+1.1 SNR that exhibits similar features in its X-ray spectrum \citep{G279MILTOS}.

\begin{figure*}[h]

    \includegraphics[width=0.45\textwidth,clip=true, trim= 0.9cm 0.1cm 1cm 0.cm]{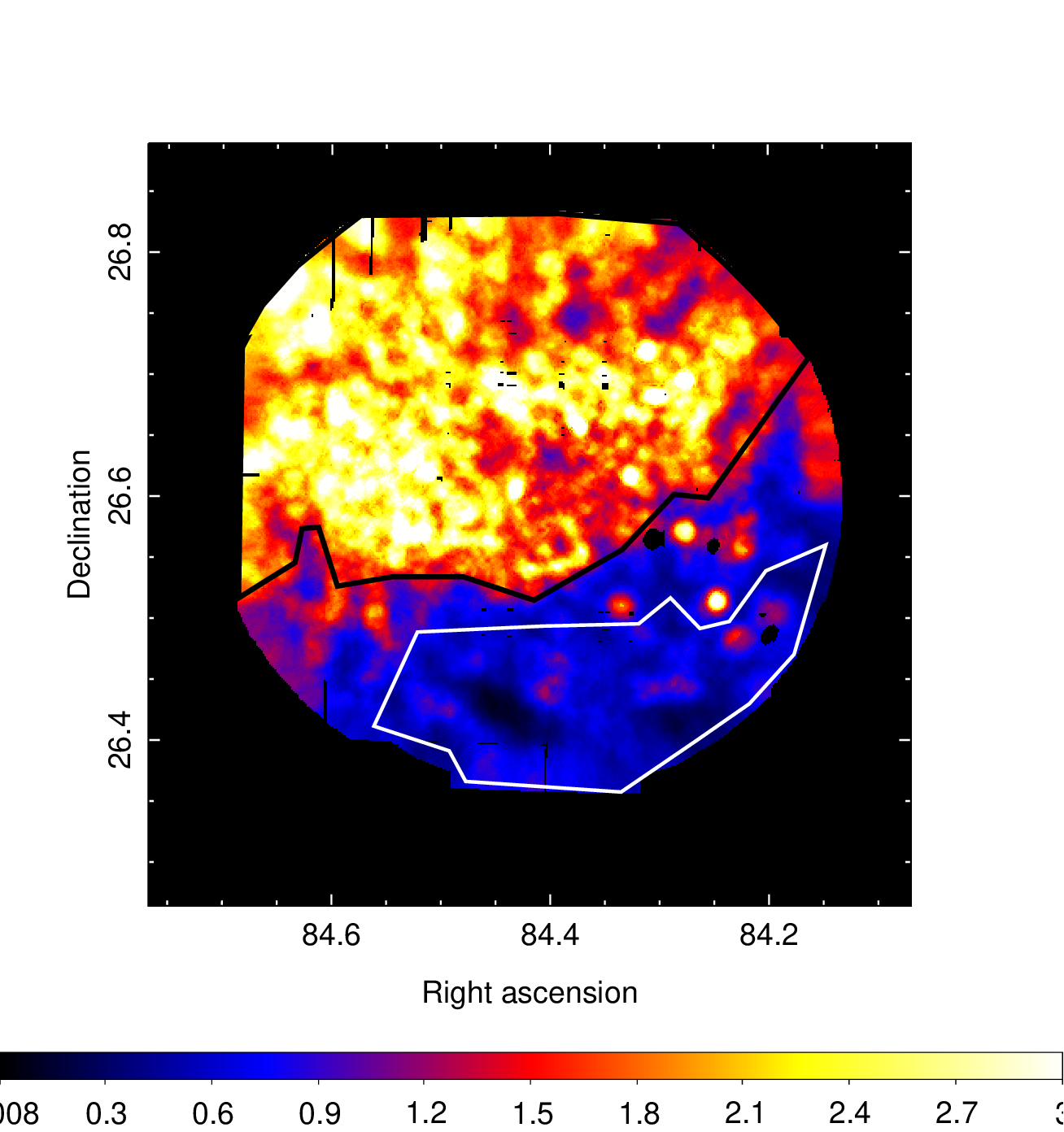}\includegraphics[width=0.63\textwidth]{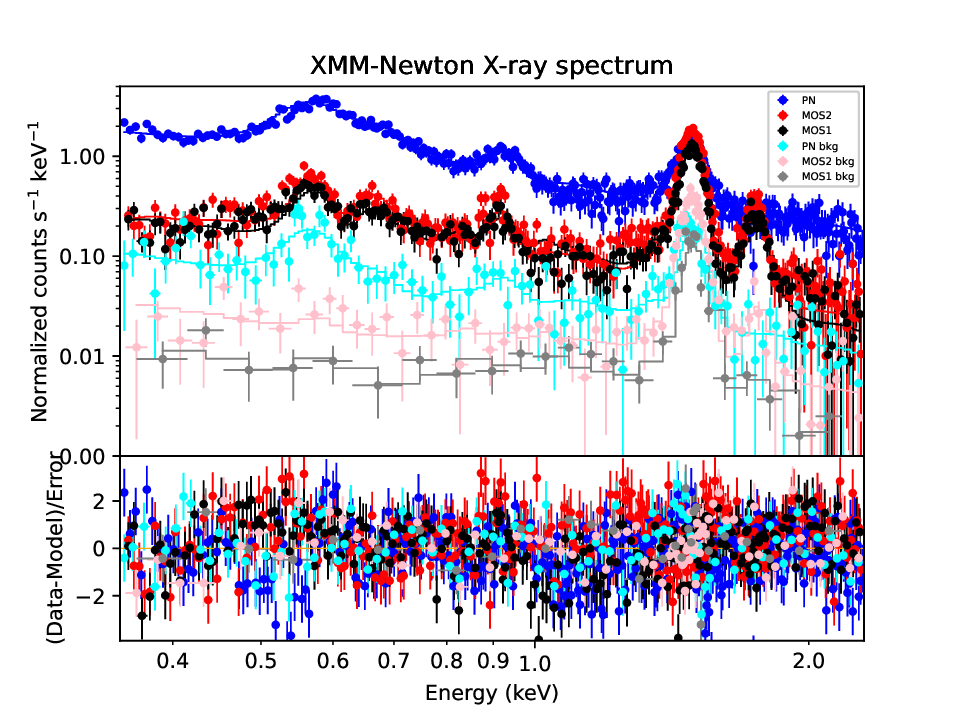}
    \caption{X-ray spectrum from a portion of the remnant that spatially coincides with the FoV of 0693270401 \textit{XMM-Newton} observation. Left panel: 0693270401 \textit{XMM-Newton} observation, identical to the one of Fig.~\ref{04}. The black polygonal region illustrates the on-source region, whereas the white polygonal region is the representative background region selected to be free of the remnant emission.  
    Right panel: pn, mos1, and mos2 \textit{XMM-Newton} spectrum in the 0.35-2.3 keV energy band.}
    \label{XMM2}
\end{figure*}

\begin{figure*}[h]

    \includegraphics[width=0.53\textwidth]{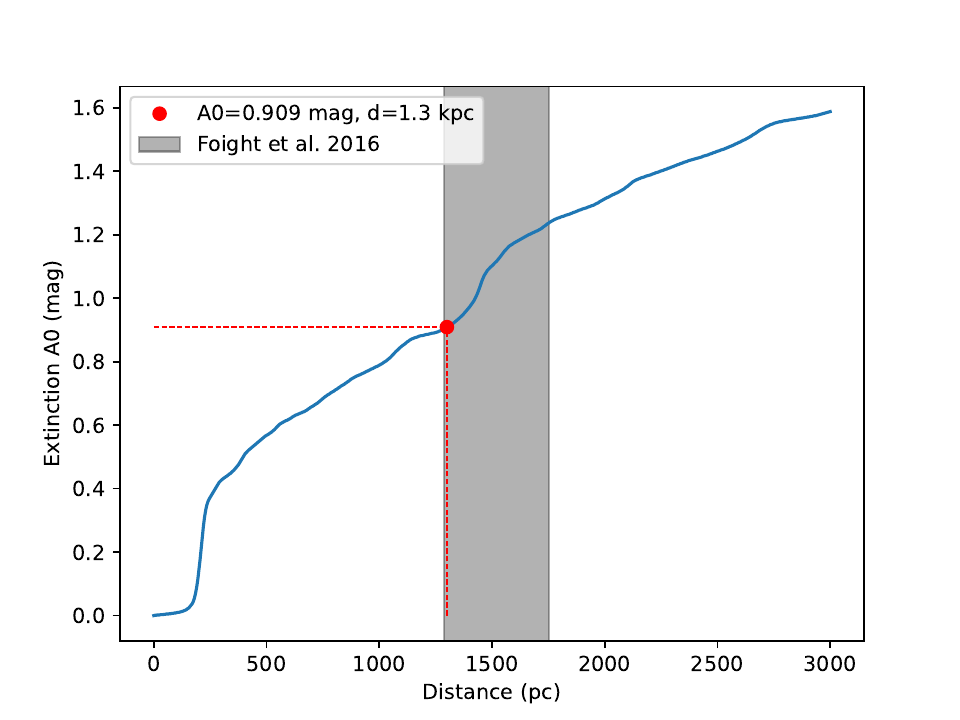}\includegraphics[width=0.53\textwidth]{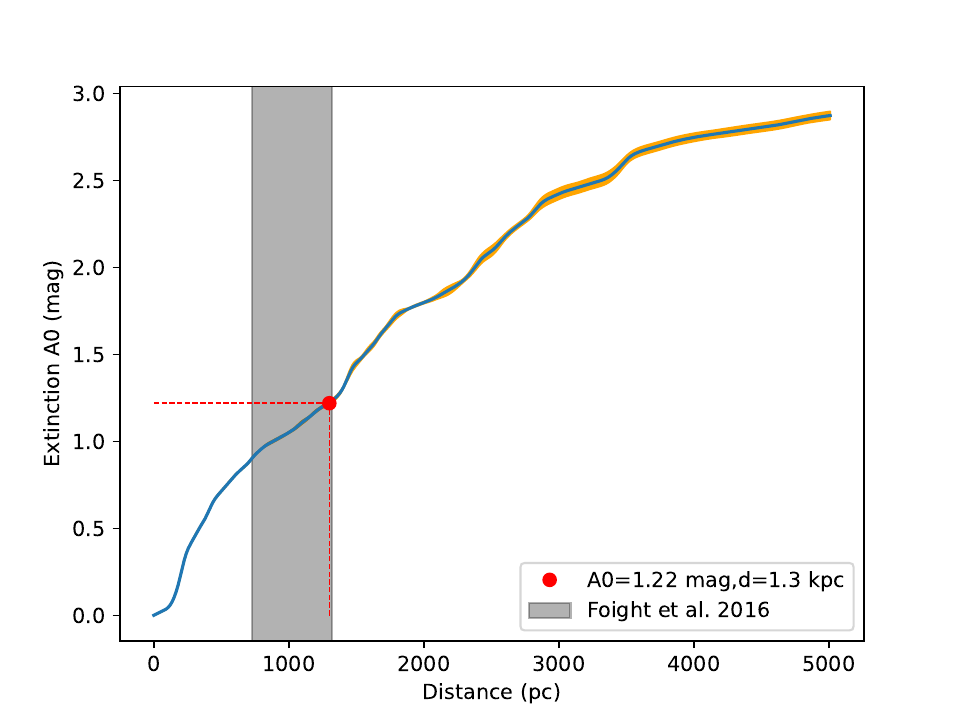}
    \caption{Cumulative extinction as a function of the distance in the remnant's direction. Left panel: One-dimensional cumulative extinction graph as a function of the distance up to $\sim3$kpc (\citet{2019A&A...625A.135L} data sets) toward S147, obtained using the GAIA/2MASS tool for one-dimensional extinction computation\protect\footnote{\url{https://astro.acri-st.fr/gaia_dev/}}. Right panel: One-dimensional cumulative extinction graph as a function of the distance up to $\sim5$kpc (updated \citet{2022A&A...661A.147L} data sets) towards S147, obtained by using the EXPLORE G-Tomo tool for one-dimensional extinction computation\protect\footnote{\url{https://explore-platform.eu/}}. In both panels, the black area corresponds to the distance uncertainty estimation when employing Eq.~\ref{math1}, and the obtained best-fit value of the absorption column density derived from the spectral analysis. The red point represents the obtained extinction when assuming that the remnant is located at a distance of 1.3 kpc.}
    \label{Extinction3}
\end{figure*}

\subsection{\textit{\textit{XMM-Newton} spectra}}\label{XMMsp}

For both available \textit{XMM-Newton} pointings in the direction of S147, a spectral analysis was carried out using \texttt{eSAS} software. Here, we only show results from ObsId 0693270401 since the statistical quality of its data is significantly higher in comparison to ObsId 0693270301. Fig.~\ref{XMM2} delineates the 0693270401 \textit{XMM-Newton} pointing in the soft 0.5-1.0~keV energy band, with the same parameters as the image at the right panel of Fig.~\ref{04}. The polygon-shaped on-source and background regions selected for spectral extraction are defined in \texttt{SAOIMAGE DS9} and marked with blue and white colours, respectively. \texttt{pn\_spectra}, \texttt{mos\_spectra,} and $\texttt{pn\_back}$ $\texttt{mos\_back}$ (for an estimate of the quiescent particle background) eSAS tasks were employed for the image construction and spectral extraction. The independently modeled on-source and background emission spectra of MOS1/MOS2/PN are shown in the right panel of Fig.~\ref{XMM2}. From the spectral analysis, we conclude that both an NEI model (either the $\texttt{VNEI}$ or $\texttt{VPSHOCK}$ model in Xspec notation) and a CIE model $\texttt{(VAPEC}$ in Xspec notation)  can be used to describe the physical conditions of the S147 plasma. However, in the right panel of Fig.~\ref{XMM2} we report  the best-fit \texttt{tbabs(vnei)} model, since it provides fitting results that are well aligned with optical extinction measurements at the remnant's distance (i.e., compatible absorption column density values; the same argument applies for the choice of the eRASS models). Similarly to the eROSITA spectral fitting process, the actual modeling involved the simultaneous fitting of the on-source and background emission, which is in the 0.35-2.3~keV energy band here. The \textit{XMM-Newton} spectral analysis was restricted above 0.35 keV since some anomalous fluctuations were observed in the 0.3-0.35 keV energy band. The Xspec package and C-statistics \citep{1979ApJ...228..939C} (as the measure for the goodness of the fit due to limited data statistics) were used for the spectral analysis. 
Three distinct model components were employed with the aim of describing the X-ray emission spectral features from the portions of the remnant as seen with \textit{XMM-Newton}: $i)$ source emission: \texttt{tbabs$\times$ vnei}; $ii)$ X-ray background: \texttt{const$\times$const$\times$(apec+tbabs$\times$(apec+apec+pow))}; and $iii)$ soft proton events or instrumental background: \texttt{unabsorbed~power~law+gauss+gauss}.
Table~\ref{TABIS1} provides a comprehensive description of the diffuse X-ray emission's spectral parameters of that region. The spectral features that this region exhibits can be well described by adapting the same best-fit model applied to region H (when utilizing eRASS:4 data), which encompasses the emission within the \textit{XMM-Newton} observation 0693270401.

\subsection{\textit{Distance to the SNR}}\label{Di}

A distance consistency check based on the absorption column density observed in X-rays (value obtained from spectral fitting and reported in Tab.~\ref{TABIS1}), the distribution of the mean color excess $\mathrm{E}_{\mathrm{B-V}}$ reported by \citet{1978A&A....64..367L}, and the extinction estimate obtained from the combination of GAIA and 2MASS photometric data \citep{2019A&A...625A.135L, 2022A&A...661A.147L} are applied for the first time in light of the eROSITA X-ray data. In particular, in this work we employed the most recent optical extinction data sets (GAIA-2MASS 3D Galactic Interstellar extinction dust maps \citep{2022A&A...661A.147L}) aimed at calculating the expected
absorption column-density value in the direction of S147. The following statistics relations have been established between the mean color excess and extinction and the X-ray absorption column density \citep{1995A&A...293..889P}:
\begin{equation}
  \begin{array}{lcl}
       N_\mathrm{H}/E_{\mathrm{B-V}}&=&5.3\times10^{21}~\mathrm{cm^{-2}\cdot mag^{-1}}\\N_\mathrm{H}\mathrm{[cm^{-2}}/A_{\mathrm{\nu}}]&=&1.79\times10^{21}
 \label{math}
 \end{array}
\end{equation}

and \citet{2016ApJ...826...66F}
\begin{equation}
  \begin{array}{lcl}
       N_\mathrm{H}/E_{\mathrm{B-V}}&=&8.9\times10^{21}~\mathrm{cm^{-2}\cdot mag^{-1}}\\N_\mathrm{H}\mathrm{[cm^{-2}}/A_{\mathrm{\nu}}]&=&2.87\pm0.12\times10^{21}
 \label{math1}
 \end{array}
.\end{equation}

The estimation of the distance was performed with regard to the absorption column densities obtained from the spectral analysis of the entire remnant, ranging between $2.7$ and $3.4\times10^{21}~\mathrm{cm^{-2}}$. A distance of $0.99_{-0.26}^{+0.34}$~kpc was derived by making use of Eq.~\ref{math1} and employing the \citet{2022A&A...661A.147L} data sets, as shown in Fig.~\ref{Extinction3}. A larger distance of $1.45_{-0.17}^{+0.30}$~kpc is obtained when employing the older data sets \citep{2019A&A...625A.135L}. Overall, the derived distance values of the remnant are in good agreement with previous reports and place the remnant at compatible distances with the associated pulsar and progenitor runaway star. We note that one obtains even larger distance values when using Eq.~\ref{math}. However, we argue that since \citet{1995A&A...293..889P} employed ROSAT data, whereas \citet{2016ApJ...826...66F} utilized higher sensitivity \textit{Chandra} data, a more accurate measurement might be obtained in the latter case (Eq.~\ref{math1}).

The interior of the SNR is dominated by thermal, hot, shocked plasma, which radiates through two-body processes; therefore, a proportionality with the electron and ion densities can be obtained \citep{1976ApJ...204..290R}: $n_\mathrm{e}=1.2n_\mathrm{H}$ (fully ionized plasma). The emission measure (EM) can be expressed by the product of the electron and hydrogen number density integrated over the volume:
\begin{equation}
    \boldsymbol{EM}=\int n_\mathrm{e}\cdot n_\mathrm{H} dV
    \label{likelihood44}
.\end{equation}

Knowing the distance (D) at which the X-ray emitter is positioned and the normalization (norm) parameter (which is obtained from the X-ray-fitting process) one can derive the EM (assuming all units in cgs) as follows:

\begin{equation}
    \boldsymbol{norm}=\frac{10^{-14}}{4\pi D^{2}}EM
    \label{likelihood45}
.\end{equation}

Once again, we consider the aforementioned distances of the associated pulsar and runaway star an accurate distance measurement of the remnant. Adopting a remnant distance of 1.33~kpc and a $3\degree$ angular size of the object in the X-ray band, we compute its real diameter as 70~pc (or 35~pc in radius). Further assuming a spherical distribution of the X-ray emission, a volume of $5.28\cdot 10^{60}~\mathrm{cm^{-3}}$ is derived.
Using the obtained normalization from the X-ray spectral fit of the entire remnant ($norm=0.10_{-0.03}^{+0.06}$) and the above-derived value of the volume (assuming the latter to be spherical, the plasma uniformly distributed, and a filling factor $\eta$=1; that is to say, the X-ray emission fills the entire remnant) and combining Equations~\ref{likelihood44} and \ref{likelihood45}, one obtains a hot plasma density of $n_{\mathrm{H}}=0.018_{-0.003}^{+0.005}~\mathrm{cm^{-3}}$ (or $n_{\mathrm{e}}=0.022_{-0.004}^{+0.006}~\mathrm{cm^{-3}}$) using

\begin{equation}
    n_\mathrm{H}=\sqrt{\frac{norm\times4\pi\cdot D^{2}}{1.2\cdot\eta\cdot10^{-14}\cdot V}}
    \label{likelihood455}
.\end{equation}

\subsection{\textit{An age-old evolutionary scenario}}\label{Ag}
To set the stage for further with more detailed modeling, we used the \citet{Leahy_2017} SNR evolutionary calculator, assuming that the SNR expands in a homogeneous ISM. 
In particular, we used the derived values for the local ISM $n_\mathrm{H}$ as input for the obtained absorption column density of the entire remnant, $0.3_{-0.03}^{+0.04}~\mathrm{10^{22}cm^{-2}}$, under the assumption that it is representative along the entire line of sight, and a distance of $1.33\pm0.1$~kpc (a local $n_\mathrm{H}=0.73_{-0.12}^{+0.16}~\mathrm{cm^{-3}}$ is thus obtained). In addition, we considered an explosion energy of $1-3\times10^{51}$~erg \citep{2012ApJ...752..135K}. Finally, by maintaining the standard values for the remaining parameters of the model, the age of S147 is estimated at $t_{\mathrm{age}}=1.7_{-0.35}^{+0.30}\times10^5~\mathrm{years}$ and $t_{\mathrm{age}}=0.76_{-0.10}^{+0.13}\times10^5~\mathrm{years}$ for $E_0=1\times10^{51}$~erg and $E_0=3\times10^{51}$~erg, respectively. 
For cross-checking purposes, one can compute the age of the remnant utilizing the relation employed in \citet{2009A&A...507..841G}, $t\sim\frac{\tau}{n_\mathrm{e}}$, where $\tau$ is the ionization timescale of the emission plasma. Making use of the derived $n_\mathrm{e}$ value and the ionization timescale of the hot plasma for the entire remnant, as obtained from the spectral analysis procedure (and reported in Table~\ref{TABIS1}), one derives a remnant age of $61.5_{-24.4}^{+46.5}$~kyr. The latter estimate is consistent with the result obtained from the previous methodology; that is to say, the derived age of S147 is of the order of $10^5$ years. 

To conclude, the above results are not broadly consistent -even within uncertainties- with the pulsar's kinematic age of $3\pm0.4\cdot10^{4}$~yrs \citep{2003ApJ...593L..31K}. The above discrepancy can be partially or totally attributed to the lack of knowledge of the actual local density distribution, which, for the purposes of this work, was considered to be homogeneous. The multi-shell appearance of the remnant in H$\alpha$ supports a heavily disturbed local medium. In addition, such a scenario of an old SNR contradicts the X-ray properties of the remnant (i.e., an ionization timescale $<10^{11}~\mathrm{s\cdot cm^{-3}}$ that supports a young age X-ray plasma in nonequilibrium ionization) and the detection of the giga-electronvolt emission originating from the SNR. Thus, in \citetalias{Spaghettichugai} we adopted a significantly lower density medium under the assumption that the remnant's progenitor's stellar winds greatly disturbed the local ISM (i.e., a supernova-in-cavity scenario resulting in a young SNR age). We were attempting to address some of the remnant's observational characteristics that appear to contradict the scenario of an old SNR.

\protect\footnotetext[4]{\url{https://astro.acri-st.fr/gaia_dev/}}
\protect\footnotetext[5]{\url{https://explore-platform.eu/}}
\section{Giga-electronvolt spectra and multiwavelength SED}
\label{sectionGEVPEC}

We conclude the binned likelihood analysis applied to the extended giga-electronvolt source 4FGL J0540.3+2756e, which is likely associated with S147, by evaluating its SED in the 0.1-100~GeV energy band. For the purposes of the spectral analysis we divided the aforementioned energy range into six equally logarithmically spaced energy bins. The best-fit spatial template implemented in \citet{2012ApJ...752..135K} was adopted. The spectral fitting process favors a LogParabola (based on the \texttt{Signif$\_$Curve} task) as the best model to fit the giga-electronvolt data instead of a simple power law reported in \citet{2012ApJ...752..135K,2022ApJ...924...45S}. As shown in the right panel of Fig.~\ref{08}, the newly derived spectrum deviates from previous results. Spectral points are, however, largely insensitive to the adopted spectral model (best-fit log-parabola or best-fit power-law) used to construct the SED. Our results are found to be in good agreement with the 4FGL-DR3 LogParabola model and an updated spectral plot of the remnant \footnote{\url{https://fermi.gsfc.nasa.gov/ssc/data/access/lat/12yr_catalog/}}. 
Overall, we conclude that the discrepancy on the final spectral shape can originate neither from the choice of the spectral model nor from the additional years of \textit{Fermi}-LAT data employed in this work for the construction of the remnant's giga-electronvolt SED. \citet{2022ApJ...924...45S} derived broadly consistent spectral results with \citet{2012ApJ...752..135K} by using ten additional years. Hence, this inconsistency is rather likely attributed to the updated model used in the 4FGL-DR3 data to model the Galactic diffuse component and the isotropic diffuse component ($\texttt{gll\_iem\_v07.fits}$ and $\texttt{iso\_P8R3\_SOURCE\_V3\_v1.txt}$ \textit{Fermi}-LAT files, respectively). Regarding the specifics of the fit, the normalization of all 4FGL-DR3 sources positioned within $5\degree$ of the center of the remnant was left to vary; the same applies to the normalizations of the Galactic diffuse and isotropic backgrounds. 
Additionally, the normalization of the LogParabola model of the S147 counterpart was allowed to vary to obtain the best-fit spectral results.

As part of the remnant's multiwavelength SED study presented in this work, two distinct scenarios can be assumed. $\mathrm{\gamma-ray}$ emission can be produced by either inverse Compton (IC) scattering of relativistic electrons interacting with the Cosmic Microwave background (CMB) (leptonic scenario) and/or $\pi_0$ decay originating from the interaction of relativistic protons (or heavier nuclei) with gas (hadronic scenario). Based on the detailed study of the giga-electronvolt counterpart of the remnant performed in \citet{2012ApJ...752..135K}, a hadronic scenario appears as the most plausible option. In this work, we provide updated spectral results in the giga-electronvolt band that are moderately different compared to previous works mainly due to the updated models implemented. The obtained shape of the giga-electronvolt SED and the likely interaction of the remnant with the nearby molecular clouds still supports the aforementioned finding (i.e., a hadronic production of $\gamma$-rays). Despite the significant change of the remnant's giga-electronvolt SED at lower energies, a hadronic scenario is still favored by the obtained SED shape (see the expectations for evolved proton distributions; e.g., \citet{2018A&A...615A.108Y}). However, a mixture of hadronic and leptonic contributions to the total SED cannot be excluded. No nonthermal X-ray emission component was detected from the remnant, and thus it does not exist, or it is of a subdominant nature and cannot be detected with eROSITA. Therefore, no further constraints were provided utilizing the remnant's multiwavelength SED.

\section{Discussion}
\label{section5}

Using eRASS:4 data of the first four completed SRG/eROSITA all-sky surveys, we report the detection of thermal X-ray emission from most of the $3\degree$-sized angular extension of the SNR S147, as defined in the radio-continuum energy range. A comparison with earlier ROSAT Survey data yields good consistency, as does the comparison with archival, yet-unpublished, \textit{XMM-Newton} pointings toward small portions to the south and south-east of the SNR.

The X-ray spatial morphology in the interior of the remnant is in excellent agreement with the morphology of both its optical (H$\alpha$) and radio-continuum counterparts. The X-ray emission fills almost the entire remnant, except for the void structure east of the center of the remnant, where the emission consistently drops in all three wavebands. An arc-like feature detected to the eastern boundary of the remnant is also present in all three wavebands, giving a unique, ear-type appearance to its shell-type morphology. Thus, we consider it to be a shell-type SNR with some peculiar characteristics. The only noticeable difference between X-rays and lower energy emission is detected to the west of the remnant, where both radio-continuum and H$\alpha$ emission are seemingly extended further to the west. A detailed study on the nature of this discrepancy is reported in \citetalias{Spaghettichugai}.
The purely thermal emission of the SNR in X-rays can be well described by either a nonequilibrium collisional plasma of $\sim0.22$ keV in temperature with $\mathrm{N_{H}}\sim0.3~\mathrm{10^{22}cm^{-2}}$, or a hot plasma in equilibrium of $\sim0.11$ keV in temperature and with $\mathrm{N_{H}}\sim0.51~\mathrm{10^{22}cm^{-2}}$. Its purely thermal nature can also explain the excellent morphological agreement between the X-ray and H$\alpha$ emission (hot and warm gas) and the general interconnection of the emission in all three wavebands as discussed in Section~\ref{radioHa}. Between the two X-ray models, an NEI model appears as the preferable option mainly due to obtained absorption column density values and the indication for ejecta presence. The absorption column density obtained from NEI models is found to be well-aligned with the expected values derived from optical extinction measurements. On the contrary, CIE models require higher absorption column-density values (compared to optical-extinction measurements) and unreasonably high elemental-abundance values to explain the remnant's spectral characteristics. 
Significant temperature variation across the remnant was observed, as shown in Table~\ref{TABIS1}. The X-ray emission is predominantly soft, with a strong detection between 0.5 and 1.0 keV and no detection above 2.3 keV. In all ten selected spectral extraction subregions, the ionization timescale is far from the expected equilibrium values reported in Table~\ref{TABIS1} (full ionization equilibrium is typically reached at $\tau$ values $\geq10^{12}~\mathrm{cm^3\cdot s}$ \citep{1984Ap&SS..98..367M}). The high statistical quality of the X-ray spectra obtained from both eROSITA and \textit{XMM-Newton} observations of the remnant strongly supports the presence of ejecta, which is noteworthy in such a particularly evolved SNR. The remnant is rich in O (OVII, OVIII), Ne (NeIX+X), and Mg (MgXI), whereas it lacks high-Z elements.

Strongly increased X-ray absorption at the location of the void structure (spectral extraction region D) cannot be excluded. The latter hypothesis is supported by the spectral analysis reported in Table~\ref{TABIS1}. Thus, the nature of the SNR's void, found at the center-east of the remnant and consistently appearing in all three wavebands, can be at least partially explained. However, X-ray absorption might not be the primary cause of the lack of X-ray emission from that center-east region of the SNR. 

Overall, no particular pattern of absorption column density variation was observed across the remnant, except for the significant decrease of the absorption column density toward region B -observed as a cavity in IR data as seen in Fig.~\ref{06}. 
The latter feature spatially matches with regions of significantly  enhanced X-ray emission (northern X-ray blob) as seen in images displayed in Fig.~\ref{06}. In addition, the significant increase of the X-ray absorption column density towards region D, the void structure, is found to be spatially
coincident with regions of enhanced IR emission as seen in images displayed in Fig.~\ref{06}. 
An excellent spatial correlation between the southern bright X-ray blob (region H) and CO emission is obtained, as shown in Fig.~\ref{07}. The latter could potentially explain the detected giga-electronvolt emission (i.e., both CO emission and a high EM in X-rays are tracers of high gas densities, which boosts giga-electronvolt emission of relativistic particles (if present)).
However, the bright giga-electronvolt blob positioned to the north of the remnant lacks the presence of CO clouds that could possibly explain the origin of the $\gamma$-ray emission.

The detection of extended $\gamma$-ray emission from the remnant \citep{2012ApJ...752..135K} is confirmed in this study. In particular, we employed $\sim15$ years of \textit{Fermi}-LAT data to successfully identify the nature of the extended diffuse source as the S147 giga-electronvolt counterpart, which is found to be spatially coincident with the brightest parts of the optical and X-ray emission of the remnant. Additionally, our imaging analysis provides a substantial improvement in the morphology details of the $\gamma$-ray emission. A LogParabola fits best the remnant's giga-electronvolt SED, instead of a simple power law as reported in \citet{2012ApJ...752..135K,2022ApJ...924...45S}. The shape of the giga-electronvolt SED is still well consistent with a $\pi^{o}$-decay (hadronic) spectrum, which is the most plausible interpretation given the age of the SNR and the association with CO emission tracing molecular gas.
In particular, the presence of CO clouds at the southern rim of the remnant is found to be inextricably linked to the $\gamma$-ray emission, therefore strengthening the hypothesis of hadronic production of $\gamma$-rays. Nevertheless, a similar association cannot be confirmed for the northern giga-electronvolt blob.

The absorption column ($N_\mathrm{H}$) analysis carried out in this work lends further support to the association of the pulsar or binary companion to the SNR and a distance to the latter of around 1.3 kpc. A physical SNR size of 70~pc in X-rays is obtained when using an SNR distance of 1.33~kpc (as obtained from measurements of the distance of the associated runaway star and pulsar). 
Assuming an evolution of the SNR in undisturbed ISM of $\sim1~\mathrm{cm^{-3}}$, an $\sim0.66-2\times10^5$~yr age estimate for the SNR was obtained using the SNR model calculator reported by \citet{Leahy_2017}. This could place it among the oldest Galactic SNRs, if not the oldest. However, we note that the above estimate is highly uncertain due to the assumption on the remnant's explosion energy, but most importantly due to the assumption of a homogenous local medium.  
An alternative scenario of a younger SNR, evolving inside a pre-existing cavity, is considered in \citetalias{Spaghettichugai}, which yields more consistent results with some of the remnant's observational characteristics and the associated pulsar's kinematic age ($\sim30$~kyr). However, neither scenario takes into account the possibility of inhomogeneity in the surrounding medium, and further refinements of the models are necessary.
Observations of the remnant with future experiments (e.g., XRISM) will shed light on the complexity of its nature.\\

\noindent\textit{Acknowledgements}
This work is based on data from eROSITA, the soft X-ray instrument aboard SRG, a joint Russian-German science mission supported by the Russian Space Agency (Roskosmos), in the interests of the Russian Academy of Sciences represented by its Space Research Institute (IKI), and the Deutsches Zentrum für Luft- und Raumfahrt (DLR). The SRG spacecraft was built by Lavochkin Association (NPOL) and its subcontractors, and is operated by NPOL with support from the Max Planck Institute for Extraterrestrial Physics (MPE).

The development and construction of the eROSITA X-ray instrument was led by MPE, with contributions from the Dr. Karl Remeis Observatory Bamberg $\&$ ECAP (FAU Erlangen-Nuernberg), the University of Hamburg Observatory, the Leibniz Institute for Astrophysics Potsdam (AIP), and the Institute for Astronomy and Astrophysics of the University of Tübingen, with the support of DLR and the Max Planck Society. The Argelander Institute for Astronomy of the University of Bonn and the Ludwig Maximilians Universität Munich also participated in the science preparation for eROSITA.
The eROSITA data shown here were processed using the eSASS/NRTA software system developed by the German eROSITA consortium.

IK acknowledges support by the COMPLEX project from the European Research Council (ERC) under the European Union’s Horizon 2020 research and innovation program grant agreement ERC-2019-AdG 882679. A.M.B. was supported by the RSF grant 21-72-20020. His modeling was performed at the Joint Supercomputer Center JSCC RAS and at the Peter the Great Saint-Petersburg Polytechnic University Supercomputing Center. Rashid Sunyaev acknowledges the support of dvp - N/A pertaining to Member Notification Merge at the Institute for Advanced Study.
This research made use of \texttt{Montage}\footnote{\url{http://montage.ipac.caltech.edu}}.
It is funded by the National Science Foundation under Grant Number ACI-1440620,
 and was previously funded by the National Aeronautics and Space Administration's Earth Science Technology Office,
 Computation Technologies Project, under Cooperative Agreement Number NCC5-626 between NASA and the California Institute of Technology.

We thank the EXPLORE team for providing us access to the G-TOMO tool of the EXPLORE platform \url{https://explore-platform.eu/} and thus allowing us to exploit updated GAIA/2MASS data which we employed to derive the latest estimates of the remnant age. We thank Denys Malyshev, Victor Doroshenko, and Lorenzo Ducci for fruitful discussions.

\bibliography{biblio}

\begin{thebibliography}{92}
\expandafter\ifx\csname natexlab\endcsname\relax\def\natexlab#1{#1}\fi

\bibitem[{{Anderson} {et~al.}(1996){Anderson}, {Cadwell}, {Jacoby},
  {Wolszczan}, {Foster}, \& {Kramer}}]{1996ApJ...468L..55A}
{Anderson}, S.~B., {Cadwell}, B.~J., {Jacoby}, B.~A., {et~al.} 1996, \apjl,
  468, L55

\bibitem[{{Aschenbach}(1996)}]{1996rftu.proc..213A}
{Aschenbach}, B. 1996, in Roentgenstrahlung from the Universe, ed. H.~U.
  {Zimmermann}, J.~{Tr{\"u}mper}, \& H.~{Yorke}, 213--220

\bibitem[{Aschenbach(1996)}]{MPEproceedings}
Aschenbach, B. 1996, ROSAT observations of Supernova Remnants

\bibitem[{{Boubert} {et~al.}(2017){Boubert}, {Fraser}, {Evans}, {Green}, \&
  {Izzard}}]{2017A&A...606A..14B}
{Boubert}, D., {Fraser}, M., {Evans}, N.~W., {Green}, D.~A., \& {Izzard}, R.~G.
  2017, \aap, 606, A14

\bibitem[{{Brunner} {et~al.}(2022){Brunner}, {Liu}, {Lamer}, {Georgakakis},
  {Merloni}, {Brusa}, {Bulbul}, {Dennerl}, {Friedrich}, {Liu}, {Maitra},
  {Nandra}, {Ramos-Ceja}, {Sanders}, {Stewart}, {Boller}, {Buchner}, {Clerc},
  {Comparat}, {Dwelly}, {Eckert}, {Finoguenov}, {Freyberg}, {Ghirardini},
  {Gueguen}, {Haberl}, {Kreykenbohm}, {Krumpe}, {Osterhage}, {Pacaud},
  {Predehl}, {Reiprich}, {Robrade}, {Salvato}, {Santangelo}, {Schrabback},
  {Schwope}, \& {Wilms}}]{2022A&A...661A...1B}
{Brunner}, H., {Liu}, T., {Lamer}, G., {et~al.} 2022, \aap, 661, A1

\bibitem[{{Cash}(1979)}]{1979ApJ...228..939C}
{Cash}, W. 1979, \apj, 228, 939

\bibitem[{Chatterjee {et~al.}(2009)Chatterjee, Brisken, Vlemmings, Goss, Lazio,
  Cordes, Thorsett, Fomalont, Lyne, \& Kramer}]{Chatterjee_2009}
Chatterjee, S., Brisken, W.~F., Vlemmings, W. H.~T., {et~al.} 2009, The
  Astrophysical Journal, 698, 250

\bibitem[{{Chen} {et~al.}(2017){Chen}, {Liu}, {Ren}, {Yuan}, {Huang}, {Yu},
  {Xiang}, {Wang}, {Tian}, \& {Zhang}}]{2017MNRAS.472.3924C}
{Chen}, B.~Q., {Liu}, X.~W., {Ren}, J.~J., {et~al.} 2017, \mnras, 472, 3924

\bibitem[{Chen {et~al.}(2014)Chen, Jiang, Zhou, Su, Zhou, Li, \&
  Zhang}]{ChenB2014}
Chen, Y., Jiang, B., Zhou, P., {et~al.} 2014, Supernova Environmental Impacts,
  ed. A. Ray and R. A. McCray, 170–177

\bibitem[{{Clark} \& {Caswell}(1976)}]{1976MNRAS.174..267C}
{Clark}, D.~H. \& {Caswell}, J.~L. 1976, \mnras, 174, 267

\bibitem[{{Collischon} {et~al.}(2021){Collischon}, {Sasaki}, {Mecke}, {Points},
  \& {Klatt}}]{2021A&A...653A..16C}
{Collischon}, C., {Sasaki}, M., {Mecke}, K., {Points}, S.~D., \& {Klatt}, M.~A.
  2021, \aap, 653, A16

\bibitem[{{Condon} {et~al.}(1991){Condon}, {Broderick}, \&
  {Seielstad}}]{1991AJ....102.2041C}
{Condon}, J.~J., {Broderick}, J.~J., \& {Seielstad}, G.~A. 1991, \aj, 102, 2041

\bibitem[{{Condon} {et~al.}(1994){Condon}, {Broderick}, {Seielstad}, {Douglas},
  \& {Gregory}}]{1994AJ....107.1829C}
{Condon}, J.~J., {Broderick}, J.~J., {Seielstad}, G.~A., {Douglas}, K., \&
  {Gregory}, P.~C. 1994, \aj, 107, 1829

\bibitem[{{Condon} {et~al.}(1993){Condon}, {Griffith}, \&
  {Wright}}]{1993AJ....106.1095C}
{Condon}, J.~J., {Griffith}, M.~R., \& {Wright}, A.~E. 1993, \aj, 106, 1095

\bibitem[{{Cram} {et~al.}(1998){Cram}, {Green}, \&
  {Bock}}]{1998PASA...15...64C}
{Cram}, L.~E., {Green}, A.~J., \& {Bock}, D.~C.~J. 1998, \pasa, 15, 64

\bibitem[{{Dame} {et~al.}(2001){Dame}, {Hartmann}, \&
  {Thaddeus}}]{2001ApJ...547..792D}
{Dame}, T.~M., {Hartmann}, D., \& {Thaddeus}, P. 2001, \apj, 547, 792

\bibitem[{{Dickel} \& {McKinley}(1969)}]{1969ApJ...155...67D}
{Dickel}, J.~R. \& {McKinley}, R.~R. 1969, \apj, 155, 67

\bibitem[{Dinçel {et~al.}(2015)Dinçel, Neuhäuser, Yerli, Ankay, Tetzlaff,
  Torres, \& Mugrauer}]{10.1093/mnras/stv124}
Dinçel, B., Neuhäuser, R., Yerli, S.~K., {et~al.} 2015, Monthly Notices of
  the Royal Astronomical Society, 448, 3196

\bibitem[{{Dopita} {et~al.}(1981){Dopita}, {Tuohy}, \&
  {Mathewson}}]{1981ApJ...248L.105D}
{Dopita}, M.~A., {Tuohy}, I.~R., \& {Mathewson}, D.~S. 1981, \apjl, 248, L105

\bibitem[{{Dorman} {et~al.}(2003){Dorman}, {Arnaud}, \&
  {Gordon}}]{2003HEAD....7.2210D}
{Dorman}, B., {Arnaud}, K.~A., \& {Gordon}, C.~A. 2003, in AAS/High Energy
  Astrophysics Division, Vol.~7, AAS/High Energy Astrophysics Division \#7,
  22.10

\bibitem[{{Ferrand} \& {Safi-Harb}(2012)}]{2012AdSpR..49.1313F}
{Ferrand}, G. \& {Safi-Harb}, S. 2012, Advances in Space Research, 49, 1313

\bibitem[{{Fesen} {et~al.}(1985){Fesen}, {Blair}, \&
  {Kirshner}}]{1985ApJ...292...29F}
{Fesen}, R.~A., {Blair}, W.~P., \& {Kirshner}, R.~P. 1985, \apj, 292, 29

\bibitem[{{Finkbeiner}(2003)}]{2003ApJS..146..407F}
{Finkbeiner}, D.~P. 2003, \apjs, 146, 407

\bibitem[{{Foight} {et~al.}(2016){Foight}, {G{\"u}ver}, {{\"O}zel}, \&
  {Slane}}]{2016ApJ...826...66F}
{Foight}, D.~R., {G{\"u}ver}, T., {{\"O}zel}, F., \& {Slane}, P.~O. 2016, \apj,
  826, 66

\bibitem[{{For} {et~al.}(2018){For}, {Staveley-Smith}, {Hurley-Walker},
  {Franzen}, {Kapi{\'n}ska}, {Filipovi{\'c}}, {Collier}, {Wu}, {Grieve},
  {Callingham}, {Bell}, {Bernardi}, {Bowman}, {Briggs}, {Cappallo},
  {Deshpande}, {Dwarakanath}, {Gaensler}, {Greenhill}, {Hancock}, {Hazelton},
  {Hindson}, {Johnston-Hollitt}, {Kaplan}, {Lenc}, {Lonsdale}, {McKinley},
  {McWhirter}, {Mitchell}, {Morales}, {Morgan}, {Morgan}, {Oberoi}, {Offringa},
  {Ord}, {Prabu}, {Procopio}, {Shankar}, {Srivani}, {Subrahmanyan}, {Tingay},
  {Wayth}, {Webster}, {Williams}, {Williams}, \& {Zheng}}]{2018MNRAS.480.2743F}
{For}, B.~Q., {Staveley-Smith}, L., {Hurley-Walker}, N., {et~al.} 2018, \mnras,
  480, 2743

\bibitem[{{Gaze} \& {Shajn}(1952)}]{1952IzKry...9...52G}
{Gaze}, V.~F. \& {Shajn}, G.~A. 1952, Izvestiya Ordena Trudovogo Krasnogo
  Znameni Krymskoj Astrofizicheskoj Observatorii, 9, 52

\bibitem[{{Giacani} {et~al.}(2009){Giacani}, {Smith}, {Dubner}, {Loiseau},
  {Castelletti}, \& {Paron}}]{2009A&A...507..841G}
{Giacani}, E., {Smith}, M.~J.~S., {Dubner}, G., {et~al.} 2009, \aap, 507, 841

\bibitem[{{Goss} {et~al.}(1979){Goss}, {Shaver}, {Zealey}, {Murdin}, \&
  {Clark}}]{1979MNRAS.188..357G}
{Goss}, W.~M., {Shaver}, P.~A., {Zealey}, W.~J., {Murdin}, P., \& {Clark},
  D.~H. 1979, \mnras, 188, 357

\bibitem[{{Green}(2009)}]{2009yCat.7253....0G}
{Green}, D.~A. 2009, VizieR Online Data Catalog, VII/253

\bibitem[{{Green}(2019)}]{2019JApA...40...36G}
{Green}, D.~A. 2019, Journal of Astrophysics and Astronomy, 40, 36

\bibitem[{{Green}(2018)}]{2018JOSS....3..695M}
{Green}, G. 2018, The Journal of Open Source Software, 3, 695

\bibitem[{{Green} {et~al.}(2019){Green}, {Schlafly}, {Zucker}, {Speagle}, \&
  {Finkbeiner}}]{2019ApJ...887...93G}
{Green}, G.~M., {Schlafly}, E., {Zucker}, C., {Speagle}, J.~S., \&
  {Finkbeiner}, D. 2019, \apj, 887, 93

\bibitem[{{Guseinov} {et~al.}(2004){Guseinov}, {Ankay}, \&
  {Tagieva}}]{2004SerAJ.168...55G}
{Guseinov}, O.~H., {Ankay}, A., \& {Tagieva}, S.~O. 2004, Serbian Astronomical
  Journal, 168, 55

\bibitem[{{Hurley-Walker} {et~al.}(2017){Hurley-Walker}, {Callingham},
  {Hancock}, {Franzen}, {Hindson}, {Kapi{\'n}ska}, {Morgan}, {Offringa},
  {Wayth}, {Wu}, {Zheng}, {Murphy}, {Bell}, {Dwarakanath}, {For}, {Gaensler},
  {Johnston-Hollitt}, {Lenc}, {Procopio}, {Staveley-Smith}, {Ekers}, {Bowman},
  {Briggs}, {Cappallo}, {Deshpande}, {Greenhill}, {Hazelton}, {Kaplan},
  {Lonsdale}, {McWhirter}, {Mitchell}, {Morales}, {Morgan}, {Oberoi}, {Ord},
  {Prabu}, {Shankar}, {Srivani}, {Subrahmanyan}, {Tingay}, {Webster},
  {Williams}, \& {Williams}}]{2017MNRAS.464.1146H}
{Hurley-Walker}, N., {Callingham}, J.~R., {Hancock}, P.~J., {et~al.} 2017,
  \mnras, 464, 1146

\bibitem[{{Hurley-Walker} {et~al.}(2019){Hurley-Walker}, {Hancock}, {Franzen},
  {Callingham}, {Offringa}, {Hindson}, {Wu}, {Bell}, {For}, {Gaensler},
  {Johnston-Hollitt}, {Kapi{\'n}ska}, {Morgan}, {Murphy}, {McKinley},
  {Procopio}, {Staveley-Smith}, {Wayth}, \& {Zheng}}]{2019PASA...36...47H}
{Hurley-Walker}, N., {Hancock}, P.~J., {Franzen}, T.~M.~O., {et~al.} 2019,
  \pasa, 36, e047

\bibitem[{Hwang {et~al.}(2008)Hwang, Petre, \& Flanagan}]{Hwang_2008}
Hwang, U., Petre, R., \& Flanagan, K.~A. 2008, The Astrophysical Journal, 676,
  378

\bibitem[{Jiang {et~al.}(2010)Jiang, Chen, Wang, Su, Zhou, Safi-Harb, \&
  DeLaney}]{Jiang_2010}
Jiang, B., Chen, Y., Wang, J., {et~al.} 2010, The Astrophysical Journal, 712,
  1147

\bibitem[{Joye \& Mandel(2003)}]{joye2003new}
Joye, W.~A. \& Mandel, E. 2003, in Astronomical data analysis software and
  systems XII, Vol. 295, 489

\bibitem[{{Katsuta} {et~al.}(2012){Katsuta}, {Uchiyama}, {Tanaka}, {Tajima},
  {Bechtol}, {Funk}, {Lande}, {Ballet}, {Hanabata}, {Lemoine-Goumard}, \&
  {Takahashi}}]{2012ApJ...752..135K}
{Katsuta}, J., {Uchiyama}, Y., {Tanaka}, T., {et~al.} 2012, \apj, 752, 135

\bibitem[{Keane \& Kramer(2008)}]{10.1111/j.1365-2966.2008.14045.x}
Keane, E.~F. \& Kramer, M. 2008, Monthly Notices of the Royal Astronomical
  Society, 391, 2009

\bibitem[{{Khabibullin} {et~al.}(2024){Khabibullin}, {Churazov}, {Chugai},
  {Bykov}, {Sunyaev}, {Utrobin}, {Zinchenko}, {Michailidis}, {Puehlhofer},
  {Becker}, {Freyberg}, {Merloni}, {Santangelo}, \& {Sasaki}}]{Spaghettichugai}
{Khabibullin}, I.~I., {Churazov}, E.~M., {Chugai}, N.~N., {et~al.} 2024,
  accepted to \aap, arXiv e-prints, arXiv:2401.17261

\bibitem[{{Kirshner} \& {Arnold}(1979)}]{1979ApJ...229..147K}
{Kirshner}, R.~P. \& {Arnold}, C.~N. 1979, \apj, 229, 147

\bibitem[{{Kochanek}(2021)}]{2021MNRAS.507.5832K}
{Kochanek}, C.~S. 2021, \mnras, 507, 5832

\bibitem[{{Kochanek} {et~al.}(2024){Kochanek}, {Raymond}, \&
  {Caldwell}}]{2024arXiv240313892K}
{Kochanek}, C.~S., {Raymond}, J.~C., \& {Caldwell}, N. 2024, submitted to ApJ,
  arXiv e-prints, arXiv:2403.13892

\bibitem[{{Kramer} {et~al.}(2003){Kramer}, {Lyne}, {Hobbs}, {L{\"o}hmer},
  {Carr}, {Jordan}, \& {Wolszczan}}]{2003ApJ...593L..31K}
{Kramer}, M., {Lyne}, A.~G., {Hobbs}, G., {et~al.} 2003, \apjl, 593, L31

\bibitem[{{Kundu} {et~al.}(1980){Kundu}, {Angerhofer}, {Fuerst}, \&
  {Hirth}}]{1980A&A....92..225K}
{Kundu}, M.~R., {Angerhofer}, P.~E., {Fuerst}, E., \& {Hirth}, W. 1980, \aap,
  92, 225

\bibitem[{{Lallement} {et~al.}(2019){Lallement}, {Babusiaux}, {Vergely},
  {Katz}, {Arenou}, {Valette}, {Hottier}, \& {Capitanio}}]{2019A&A...625A.135L}
{Lallement}, R., {Babusiaux}, C., {Vergely}, J.~L., {et~al.} 2019, \aap, 625,
  A135

\bibitem[{{Lallement} {et~al.}(2022){Lallement}, {Vergely}, {Babusiaux}, \&
  {Cox}}]{2022A&A...661A.147L}
{Lallement}, R., {Vergely}, J.~L., {Babusiaux}, C., \& {Cox}, N.~L.~J. 2022,
  \aap, 661, A147

\bibitem[{Lasker(1979)}]{4fbea282-ad61-35c8-8776-164566ece9b7}
Lasker, B.~M. 1979, Publications of the Astronomical Society of the Pacific,
  91, 153

\bibitem[{Leahy \& Williams(2017)}]{Leahy_2017}
Leahy, D.~A. \& Williams, J.~E. 2017, The Astronomical Journal, 153, 239

\bibitem[{{Lucke}(1978)}]{1978A&A....64..367L}
{Lucke}, P.~B. 1978, \aap, 64, 367

\bibitem[{{Masai}(1984)}]{1984Ap&SS..98..367M}
{Masai}, K. 1984, \apss, 98, 367

\bibitem[{{Mathewson} {et~al.}(1980){Mathewson}, {Dopita}, {Tuohy}, \&
  {Ford}}]{1980ApJ...242L..73M}
{Mathewson}, D.~S., {Dopita}, M.~A., {Tuohy}, I.~R., \& {Ford}, V.~L. 1980,
  \apjl, 242, L73

\bibitem[{{Mattox} {et~al.}(1996){Mattox}, {Bertsch}, {Chiang}, {Dingus},
  {Digel}, {Esposito}, {Fierro}, {Hartman}, {Hunter}, {Kanbach}, {Kniffen},
  {Lin}, {Macomb}, {Mayer-Hasselwander}, {Michelson}, {von Montigny},
  {Mukherjee}, {Nolan}, {Ramanamurthy}, {Schneid}, {Sreekumar}, {Thompson}, \&
  {Willis}}]{1996ApJ...461..396M}
{Mattox}, J.~R., {Bertsch}, D.~L., {Chiang}, J., {et~al.} 1996, \apj, 461, 396

\bibitem[{{McKee} {et~al.}(1987){McKee}, {Hollenbach}, {Seab}, \&
  {Tielens}}]{1987ApJ...318..674M}
{McKee}, C.~F., {Hollenbach}, D.~J., {Seab}, G.~C., \& {Tielens}, A.~G.~G.~M.
  1987, \apj, 318, 674

\bibitem[{{Merloni} {et~al.}(2024){Merloni}, {Lamer}, {Liu}, {Ramos-Ceja},
  {Brunner}, {Bulbul}, {Dennerl}, {Doroshenko}, {Freyberg}, {Friedrich},
  {Gatuzz}, {Georgakakis}, {Haberl}, {Igo}, {Kreykenbohm}, {Liu}, {Maitra},
  {Malyali}, {Mayer}, {Nandra}, {Predehl}, {Robrade}, {Salvato}, {Sanders},
  {Stewart}, {Tub{\'\i}n-Arenas}, {Weber}, {Wilms}, {Arcodia}, {Artis},
  {Aschersleben}, {Avakyan}, {Aydar}, {Bahar}, {Balzer}, {Becker}, {Berger},
  {Boller}, {Bornemann}, {Br{\"u}ggen}, {Brusa}, {Buchner}, {Burwitz},
  {Camilloni}, {Clerc}, {Comparat}, {Coutinho}, {Czesla}, {Dannhauer},
  {Dauner}, {Dauser}, {Dietl}, {Dolag}, {Dwelly}, {Egg}, {Ehl}, {Freund},
  {Friedrich}, {Gaida}, {Garrel}, {Ghirardini}, {Gokus}, {Gr{\"u}nwald},
  {Grandis}, {Grotova}, {Gruen}, {Gueguen}, {H{\"a}mmerich}, {Hamaus},
  {Hasinger}, {Haubner}, {Homan}, {Ider Chitham}, {Joseph}, {Joyce},
  {K{\"o}nig}, {Kaltenbrunner}, {Khokhriakova}, {Kink}, {Kirsch}, {Kluge},
  {Knies}, {Krippendorf}, {Krumpe}, {Kurpas}, {Li}, {Liu}, {Locatelli},
  {Lorenz}, {M{\"u}ller}, {Magaudda}, {Mannes}, {McCall}, {Meidinger},
  {Michailidis}, {Migkas}, {Mu{\~n}oz-Giraldo}, {Musiimenta}, {Nguyen-Dang},
  {Ni}, {Olechowska}, {Ota}, {Pacaud}, {Pasini}, {Perinati}, {Pires},
  {Pommranz}, {Ponti}, {Poppenhaeger}, {P{\"u}hlhofer}, {Rau}, {Reh},
  {Reiprich}, {Roster}, {Saeedi}, {Santangelo}, {Sasaki}, {Schmitt},
  {Schneider}, {Schrabback}, {Schuster}, {Schwope}, {Seppi}, {Serim},
  {Shreeram}, {Sokolova-Lapa}, {Starck}, {Stelzer}, {Stierhof}, {Suleimanov},
  {Tenzer}, {Traulsen}, {Tr{\"u}mper}, {Tsuge}, {Urrutia}, {Veronica},
  {Waddell}, {Willer}, {Wolf}, {Yeung}, {Zainab}, {Zangrandi}, {Zhang},
  {Zhang}, \& {Zheng}}]{2023Merloni}
{Merloni}, A., {Lamer}, G., {Liu}, T., {et~al.} 2024, \aap, 682, A34

\bibitem[{{Michailidis} {et~al.}(2024){Michailidis}, {P{\"u}hlhofer},
  {Santangelo}, {Becker}, \& {Sasaki}}]{G279MILTOS}
{Michailidis}, M., {P{\"u}hlhofer}, G., {Santangelo}, A., {Becker}, W., \&
  {Sasaki}, M. 2024, \aap, 685, A23

\bibitem[{{Minkowski}(1958)}]{1958RvMP...30.1048M}
{Minkowski}, R. 1958, Reviews of Modern Physics, 30, 1048

\bibitem[{Ng {et~al.}(2007)Ng, Romani, Brisken, Chatterjee, \&
  Kramer}]{Ng_2007}
Ng, C.-Y., Romani, R.~W., Brisken, W.~F., Chatterjee, S., \& Kramer, M. 2007,
  The Astrophysical Journal, 654, 487

\bibitem[{Park {et~al.}(2003)Park, Hughes, Slane, Burrows, Warren, Garmire, \&
  Nousek}]{Park_2003}
Park, S., Hughes, J.~P., Slane, P.~O., {et~al.} 2003, The Astrophysical
  Journal, 592, L41

\bibitem[{{Pavlinsky} {et~al.}(2021){Pavlinsky}, {Tkachenko}, {Levin},
  {Alexandrovich}, {Arefiev}, {Babyshkin}, {Batanov}, {Bodnar}, {Bogomolov},
  {Bubnov}, {Buntov}, {Burenin}, {Chelovekov}, {Chen}, {Drozdova}, {Ehlert},
  {Filippova}, {Frolov}, {Gamkov}, {Garanin}, {Garin}, {Glushenko}, {Gorelov},
  {Grebenev}, {Grigorovich}, {Gureev}, {Gurova}, {Ilkaev}, {Katasonov},
  {Krivchenko}, {Krivonos}, {Korotkov}, {Kudelin}, {Kuznetsova}, {Lazarchuk},
  {Lomakin}, {Lapshov}, {Lipilin}, {Lutovinov}, {Mereminskiy}, {Molkov},
  {Nazarov}, {Oleinikov}, {Pikalov}, {Ramsey}, {Roiz}, {Rotin}, {Ryadov},
  {Sankin}, {Sazonov}, {Sedov}, {Semena}, {Semena}, {Serbinov}, {Shirshakov},
  {Shtykovsky}, {Shvetsov}, {Sunyaev}, {Swartz}, {Tambov}, {Voron}, \&
  {Yaskovich}}]{2021A&A...650A..42P}
{Pavlinsky}, M., {Tkachenko}, A., {Levin}, V., {et~al.} 2021, \aap, 650, A42

\bibitem[{Phillips {et~al.}(1981)Phillips, Gondhalekar, \&
  Blades}]{10.1093/mnras/195.3.485}
Phillips, A.~P., Gondhalekar, P.~M., \& Blades, J.~C. 1981, Monthly Notices of
  the Royal Astronomical Society, 195, 485

\bibitem[{{Predehl} {et~al.}(2021){Predehl}, {Andritschke}, {Arefiev},
  {Babyshkin}, {Batanov}, {Becker}, {B{\"o}hringer}, {Bogomolov}, {Boller},
  {Borm}, {Bornemann}, {Br{\"a}uninger}, {Br{\"u}ggen}, {Brunner}, {Brusa},
  {Bulbul}, {Buntov}, {Burwitz}, {Burkert}, {Clerc}, {Churazov}, {Coutinho},
  {Dauser}, {Dennerl}, {Doroshenko}, {Eder}, {Emberger}, {Eraerds},
  {Finoguenov}, {Freyberg}, {Friedrich}, {Friedrich}, {F{\"u}rmetz},
  {Georgakakis}, {Gilfanov}, {Granato}, {Grossberger}, {Gueguen}, {Gureev},
  {Haberl}, {H{\"a}lker}, {Hartner}, {Hasinger}, {Huber}, {Ji}, {Kienlin},
  {Kink}, {Korotkov}, {Kreykenbohm}, {Lamer}, {Lomakin}, {Lapshov}, {Liu},
  {Maitra}, {Meidinger}, {Menz}, {Merloni}, {Mernik}, {Mican}, {Mohr},
  {M{\"u}ller}, {Nandra}, {Nazarov}, {Pacaud}, {Pavlinsky}, {Perinati},
  {Pfeffermann}, {Pietschner}, {Ramos-Ceja}, {Rau}, {Reiffers}, {Reiprich},
  {Robrade}, {Salvato}, {Sanders}, {Santangelo}, {Sasaki}, {Scheuerle},
  {Schmid}, {Schmitt}, {Schwope}, {Shirshakov}, {Steinmetz}, {Stewart},
  {Str{\"u}der}, {Sunyaev}, {Tenzer}, {Tiedemann}, {Tr{\"u}mper}, {Voron},
  {Weber}, {Wilms}, \& {Yaroshenko}}]{2021A&A...647A...1P}
{Predehl}, P., {Andritschke}, R., {Arefiev}, V., {et~al.} 2021, \aap, 647, A1

\bibitem[{{Predehl} \& {Schmitt}(1995)}]{1995A&A...293..889P}
{Predehl}, P. \& {Schmitt}, J.~H.~M.~M. 1995, \aap, 293, 889

\bibitem[{{Priestley} {et~al.}(2021){Priestley}, {Chawner}, {Matsuura}, {De
  Looze}, {Barlow}, \& {Gomez}}]{2021MNRAS.500.2543P}
{Priestley}, F.~D., {Chawner}, H., {Matsuura}, M., {et~al.} 2021, \mnras, 500,
  2543

\bibitem[{{Raymond} {et~al.}(1976){Raymond}, {Cox}, \&
  {Smith}}]{1976ApJ...204..290R}
{Raymond}, J.~C., {Cox}, D.~P., \& {Smith}, B.~W. 1976, \apj, 204, 290

\bibitem[{{Reich}(2002)}]{2002nsps.conf....1R}
{Reich}, W. 2002, in Neutron Stars, Pulsars, and Supernova Remnants, ed.
  W.~{Becker}, H.~{Lesch}, \& J.~{Tr{\"u}mper}, 1

\bibitem[{{Reich} {et~al.}(2003){Reich}, {Zhang}, \&
  {F{\"u}rst}}]{2003A&A...408..961R}
{Reich}, W., {Zhang}, X., \& {F{\"u}rst}, E. 2003, \aap, 408, 961

\bibitem[{{Ren} {et~al.}(2018){Ren}, {Liu}, {Chen}, {Xiang}, {Yuan}, {Huang},
  {Zhang}, {Wang}, {Tian}, {Liu}, \& {Wu}}]{2018RAA....18..111R}
{Ren}, J.-J., {Liu}, X.-W., {Chen}, B.-Q., {et~al.} 2018, Research in Astronomy
  and Astrophysics, 18, 111

\bibitem[{{Romani} \& {Ng}(2003)}]{2003ApJ...585L..41R}
{Romani}, R.~W. \& {Ng}, C.~Y. 2003, \apjl, 585, L41

\bibitem[{{Sallmen} \& {Welsh}(2004)}]{2004A&A...426..555S}
{Sallmen}, S. \& {Welsh}, B.~Y. 2004, \aap, 426, 555

\bibitem[{{Sauvageot} {et~al.}(1990){Sauvageot}, {Ballet}, \&
  {Rothenflug}}]{1990A&A...227..183S}
{Sauvageot}, J.~L., {Ballet}, J., \& {Rothenflug}, R. 1990, \aap, 227, 183

\bibitem[{{Slavin} {et~al.}(2015){Slavin}, {Dwek}, \&
  {Jones}}]{2015ApJ...803....7S}
{Slavin}, J.~D., {Dwek}, E., \& {Jones}, A.~P. 2015, \apj, 803, 7

\bibitem[{Smale(2021{\natexlab{a}})}]{bworld}
Smale, D. A.~P. 2021{\natexlab{a}}, \url{http://heasarc.gsfc.nasa.gov/}

\bibitem[{Smale(2021{\natexlab{b}})}]{cworld}
Smale, D. A.~P. 2021{\natexlab{b}},
  \url{http://heasarc.gsfc.nasa.gov/docs/software/floods/ftools_menu.html}

\bibitem[{{Sofue} {et~al.}(1980){Sofue}, {Furst}, \&
  {Hirth}}]{1980PASJ...32....1S}
{Sofue}, Y., {Furst}, E., \& {Hirth}, W. 1980, \pasj, 32, 1

\bibitem[{{Str{\"u}der} {et~al.}(2001){Str{\"u}der}, {Briel}, {Dennerl},
  {Hartmann}, {Kendziorra}, {Meidinger}, {Pfeffermann}, {Reppin}, {Aschenbach},
  {Bornemann}, {Br{\"a}uninger}, {Burkert}, {Elender}, {Freyberg}, {Haberl},
  {Hartner}, {Heuschmann}, {Hippmann}, {Kastelic}, {Kemmer}, {Kettenring},
  {Kink}, {Krause}, {M{\"u}ller}, {Oppitz}, {Pietsch}, {Popp}, {Predehl},
  {Read}, {Stephan}, {St{\"o}tter}, {Tr{\"u}mper}, {Holl}, {Kemmer}, {Soltau},
  {St{\"o}tter}, {Weber}, {Weichert}, {von Zanthier}, {Carathanassis}, {Lutz},
  {Richter}, {Solc}, {B{\"o}ttcher}, {Kuster}, {Staubert}, {Abbey}, {Holland},
  {Turner}, {Balasini}, {Bignami}, {La Palombara}, {Villa}, {Buttler},
  {Gianini}, {Lain{\'e}}, {Lumb}, \& {Dhez}}]{2001A&A...365L..18S}
{Str{\"u}der}, L., {Briel}, U., {Dennerl}, K., {et~al.} 2001, \aap, 365, L18

\bibitem[{{Stupar} {et~al.}(2007){Stupar}, {Parker}, {Filipovi{\'c}}, {Frew},
  {Boji{\v{c}}i{\'c}}, \& {Aschenbach}}]{2007MNRAS.381..377S}
{Stupar}, M., {Parker}, Q.~A., {Filipovi{\'c}}, M.~D., {et~al.} 2007, \mnras,
  381, 377

\bibitem[{{Sun} {et~al.}(1995){Sun}, {Aschenbach}, {Becker}, {Truemper},
  {Anderson}, \& {Wolszczan}}]{1995IAUC.6187....2S}
{Sun}, X., {Aschenbach}, B., {Becker}, W., {et~al.} 1995, \iaucirc, 6187, 2

\bibitem[{{Sunyaev} {et~al.}(2021){Sunyaev}, {Arefiev}, {Babyshkin},
  {Bogomolov}, {Borisov}, {Buntov}, {Brunner}, {Burenin}, {Churazov},
  {Coutinho}, {Eder}, {Eismont}, {Freyberg}, {Gilfanov}, {Gureyev}, {Hasinger},
  {Khabibullin}, {Kolmykov}, {Komovkin}, {Krivonos}, {Lapshov}, {Levin},
  {Lomakin}, {Lutovinov}, {Medvedev}, {Merloni}, {Mernik}, {Mikhailov},
  {Molodtsov}, {Mzhelsky}, {M{\"u}ller}, {Nandra}, {Nazarov}, {Pavlinsky},
  {Poghodin}, {Predehl}, {Robrade}, {Sazonov}, {Scheuerle}, {Shirshakov},
  {Tkachenko}, \& {Voron}}]{2021A&A...656A.132S}
{Sunyaev}, R., {Arefiev}, V., {Babyshkin}, V., {et~al.} 2021, \aap, 656, A132

\bibitem[{{Suzuki} {et~al.}(2022){Suzuki}, {Bamba}, {Yamazaki}, \&
  {Ohira}}]{2022ApJ...924...45S}
{Suzuki}, H., {Bamba}, A., {Yamazaki}, R., \& {Ohira}, Y. 2022, \apj, 924, 45

\bibitem[{{Taylor} {et~al.}(2003){Taylor}, {Gibson}, {Peracaula}, {Martin},
  {Landecker}, {Brunt}, {Dewdney}, {Dougherty}, {Gray}, {Higgs}, {Kerton},
  {Knee}, {Kothes}, {Purton}, {Uyaniker}, {Wallace}, {Willis}, \&
  {Durand}}]{2003AJ....125.3145T}
{Taylor}, A.~R., {Gibson}, S.~J., {Peracaula}, M., {et~al.} 2003, \aj, 125,
  3145

\bibitem[{{Troja} {et~al.}(2008){Troja}, {Bocchino}, {Miceli}, \&
  {Reale}}]{2008A&A...485..777T}
{Troja}, E., {Bocchino}, F., {Miceli}, M., \& {Reale}, F. 2008, \aap, 485, 777

\bibitem[{{Turner} {et~al.}(2001){Turner}, {Abbey}, {Arnaud}, {Balasini},
  {Barbera}, {Belsole}, {Bennie}, {Bernard}, {Bignami}, {Boer}, {Briel},
  {Butler}, {Cara}, {Chabaud}, {Cole}, {Collura}, {Conte}, {Cros}, {Denby},
  {Dhez}, {Di Coco}, {Dowson}, {Ferrando}, {Ghizzardi}, {Gianotti}, {Goodall},
  {Gretton}, {Griffiths}, {Hainaut}, {Hochedez}, {Holland}, {Jourdain},
  {Kendziorra}, {Lagostina}, {Laine}, {La Palombara}, {Lortholary}, {Lumb},
  {Marty}, {Molendi}, {Pigot}, {Poindron}, {Pounds}, {Reeves}, {Reppin},
  {Rothenflug}, {Salvetat}, {Sauvageot}, {Schmitt}, {Sembay}, {Short},
  {Spragg}, {Stephen}, {Str{\"u}der}, {Tiengo}, {Trifoglio}, {Tr{\"u}mper},
  {Vercellone}, {Vigroux}, {Villa}, {Ward}, {Whitehead}, \&
  {Zonca}}]{2001A&A...365L..27T}
{Turner}, M.~J.~L., {Abbey}, A., {Arnaud}, M., {et~al.} 2001, \aap, 365, L27

\bibitem[{{van den Bergh} {et~al.}(1973){van den Bergh}, {Marscher}, \&
  {Terzian}}]{1973ApJS...26...19V}
{van den Bergh}, S., {Marscher}, A.~P., \& {Terzian}, Y. 1973, \apjs, 26, 19

\bibitem[{{Voges} {et~al.}(1999){Voges}, {Aschenbach}, {Boller},
  {Br{\"a}uninger}, {Briel}, {Burkert}, {Dennerl}, {Englhauser}, {Gruber},
  {Haberl}, {Hartner}, {Hasinger}, {K{\"u}rster}, {Pfeffermann}, {Pietsch},
  {Predehl}, {Rosso}, {Schmitt}, {Tr{\"u}mper}, \&
  {Zimmermann}}]{1999A&A...349..389V}
{Voges}, W., {Aschenbach}, B., {Boller}, T., {et~al.} 1999, \aap, 349, 389

\bibitem[{{Voges} {et~al.}(2000){Voges}, {Aschenbach}, {Boller}, {Brauninger},
  {Briel}, {Burkert}, {Dennerl}, {Englhauser}, {Gruber}, {Haberl}, {Hartner},
  {Hasinger}, {Pfeffermann}, {Pietsch}, {Predehl}, {Schmitt}, {Trumper}, \&
  {Zimmermann}}]{2000IAUC.7432....3V}
{Voges}, W., {Aschenbach}, B., {Boller}, T., {et~al.} 2000, \iaucirc, 7432, 3

\bibitem[{{Wayth} {et~al.}(2015){Wayth}, {Lenc}, {Bell}, {Callingham},
  {Dwarakanath}, {Franzen}, {For}, {Gaensler}, {Hancock}, {Hindson},
  {Hurley-Walker}, {Jackson}, {Johnston-Hollitt}, {Kapi{\'n}ska}, {McKinley},
  {Morgan}, {Offringa}, {Procopio}, {Staveley-Smith}, {Wu}, {Zheng}, {Trott},
  {Bernardi}, {Bowman}, {Briggs}, {Cappallo}, {Corey}, {Deshpande}, {Emrich},
  {Goeke}, {Greenhill}, {Hazelton}, {Kaplan}, {Kasper}, {Kratzenberg},
  {Lonsdale}, {Lynch}, {McWhirter}, {Mitchell}, {Morales}, {Morgan}, {Oberoi},
  {Ord}, {Prabu}, {Rogers}, {Roshi}, {Shankar}, {Srivani}, {Subrahmanyan},
  {Tingay}, {Waterson}, {Webster}, {Whitney}, {Williams}, \&
  {Williams}}]{2015PASA...32...25W}
{Wayth}, R.~B., {Lenc}, E., {Bell}, M.~E., {et~al.} 2015, \pasa, 32, e025

\bibitem[{{Wilms} {et~al.}(2000){Wilms}, {Allen}, \&
  {McCray}}]{2000ApJ...542..914W}
{Wilms}, J., {Allen}, A., \& {McCray}, R. 2000, \apj, 542, 914

\bibitem[{Xiao {et~al.}(2008)Xiao, F{\"u}rst, Reich, \& Han}]{xiao2008radio}
Xiao, L., F{\"u}rst, E., Reich, W., \& Han, J. 2008, Astronomy \& Astrophysics,
  482, 783

\bibitem[{Yamauchi {et~al.}(1999)Yamauchi, Koyama, Tomida, Yokogawa, \&
  Tamura}]{10.1093/pasj/51.1.13}
Yamauchi, S., Koyama, K., Tomida, H., Yokogawa, J., \& Tamura, K. 1999,
  Publications of the Astronomical Society of Japan, 51, 13

\bibitem[{{Yang} {et~al.}(2018){Yang}, {Kafexhiu}, \&
  {Aharonian}}]{2018A&A...615A.108Y}
{Yang}, R.-z., {Kafexhiu}, E., \& {Aharonian}, F. 2018, \aap, 615, A108

\end{thebibliography}

\bibliographystyle{aa}

\begin{appendix}

\section{eRASS:4 X-ray spectral plots of individual subregions}\label{subs}

Figure~\ref{ALLSUBSPEC2} shows the results of the simultaneous fit of the on-source and background emission from eight selected subregions, with an absorbed \texttt{VNEI} 
as the optimal model describing the purely thermal S147 spectrum. The corresponding best-fit spectral parameters are summarized in Table~\ref{TABIS1}.

\begin{figure*}
    \includegraphics[width=0.50\textwidth]{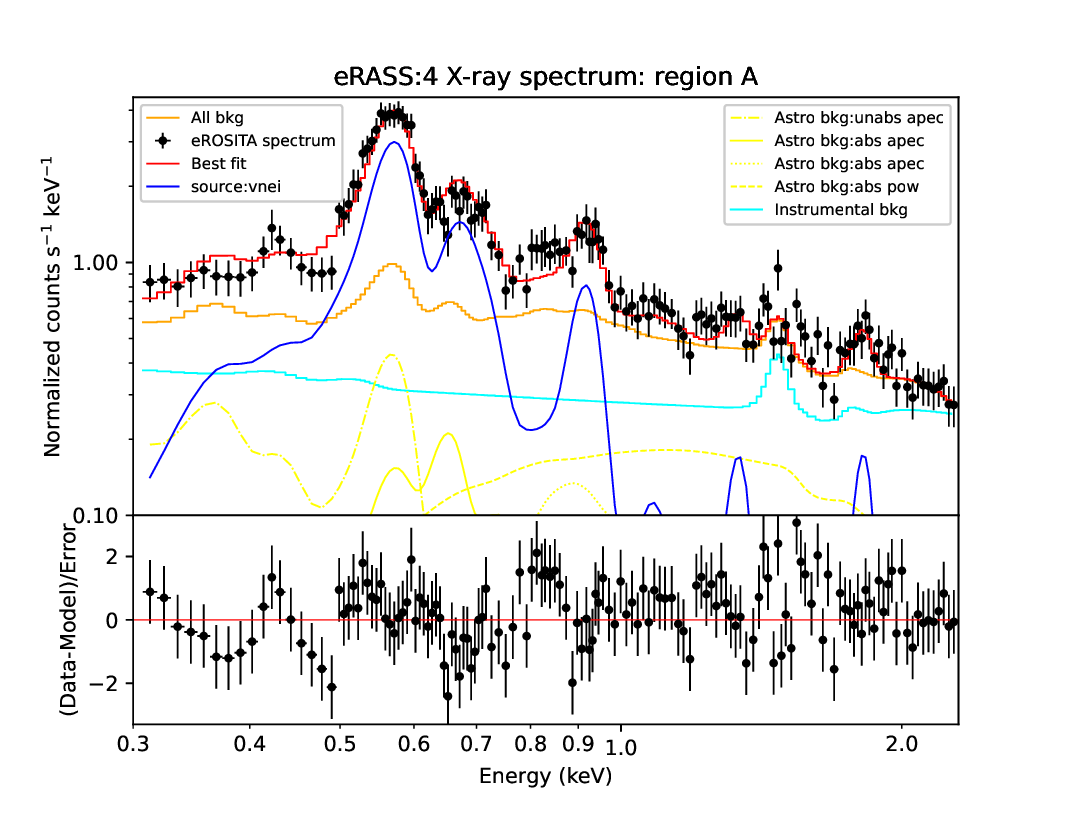}
    \includegraphics[width=0.502\textwidth]{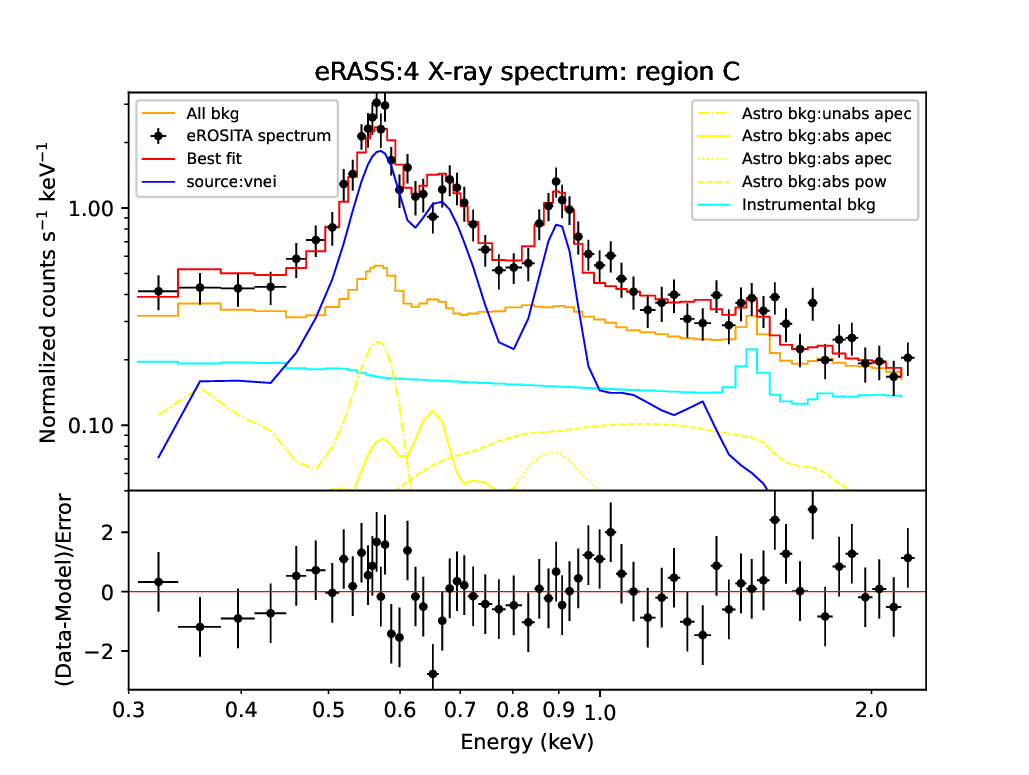}
    \includegraphics[width=0.504\textwidth]{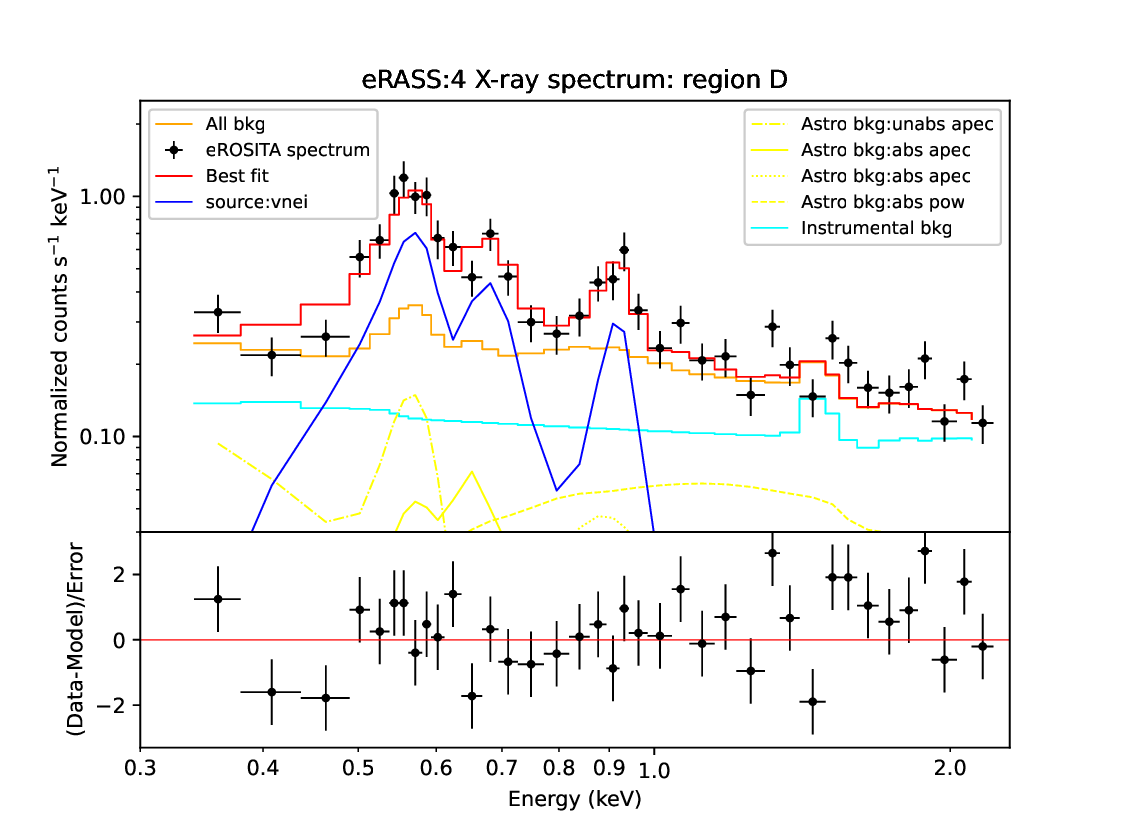}
    \includegraphics[width=0.499\textwidth]{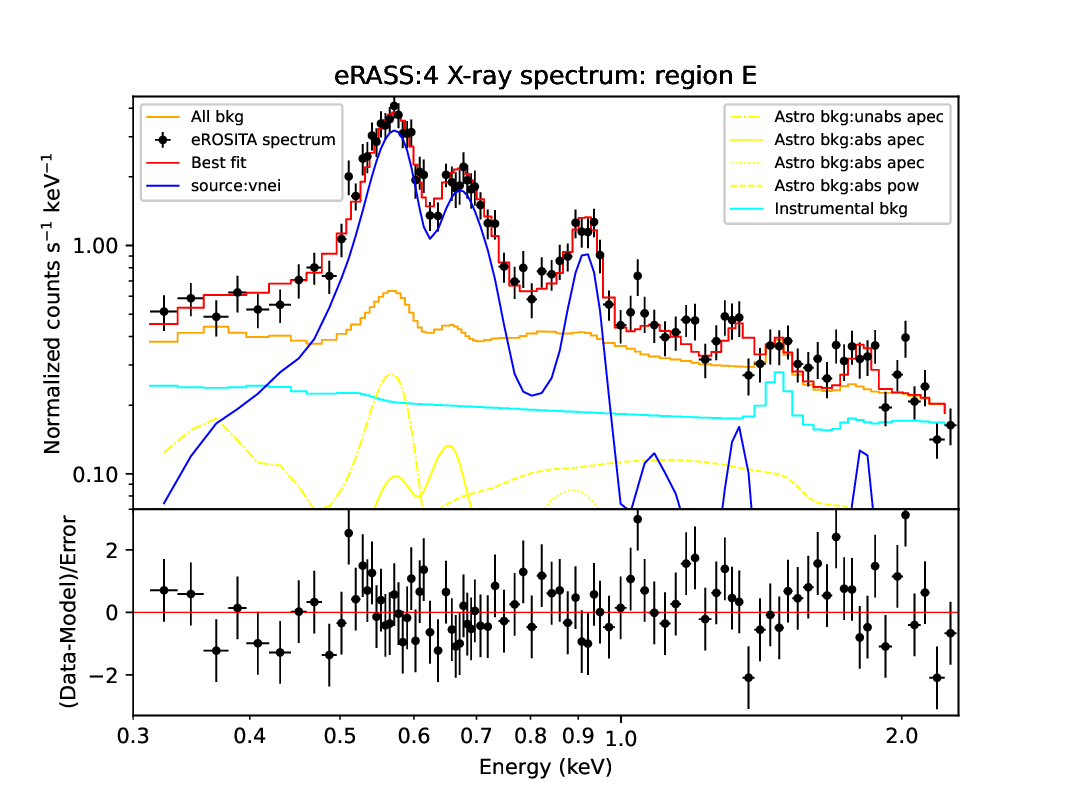}
    \includegraphics[width=0.495\textwidth]{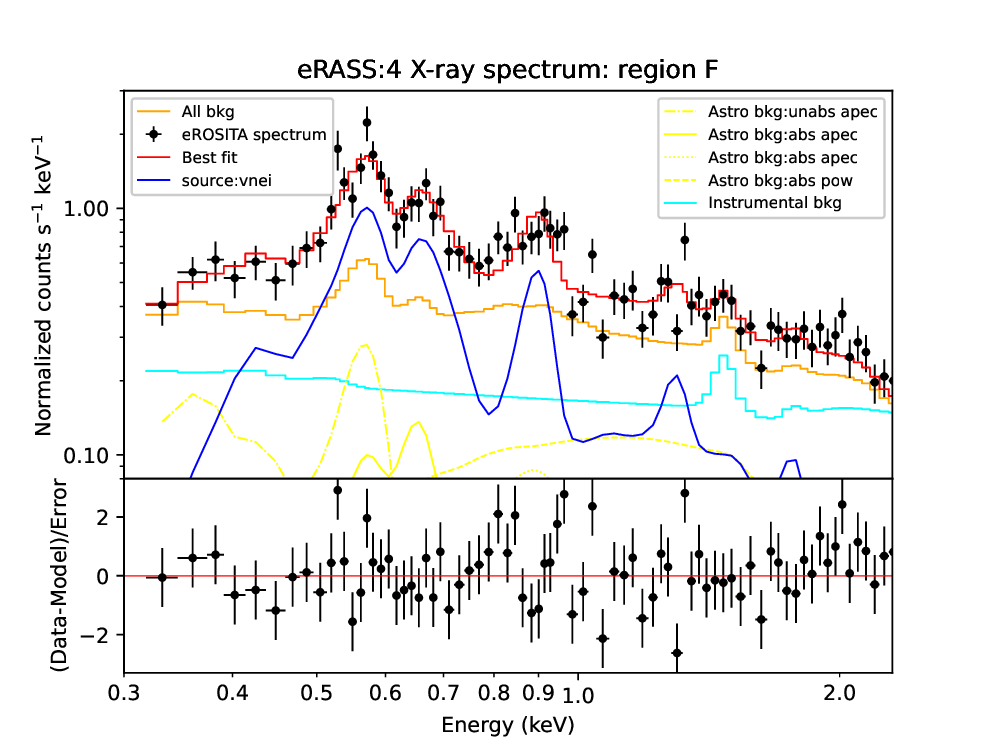}
    \includegraphics[width=0.502\textwidth]{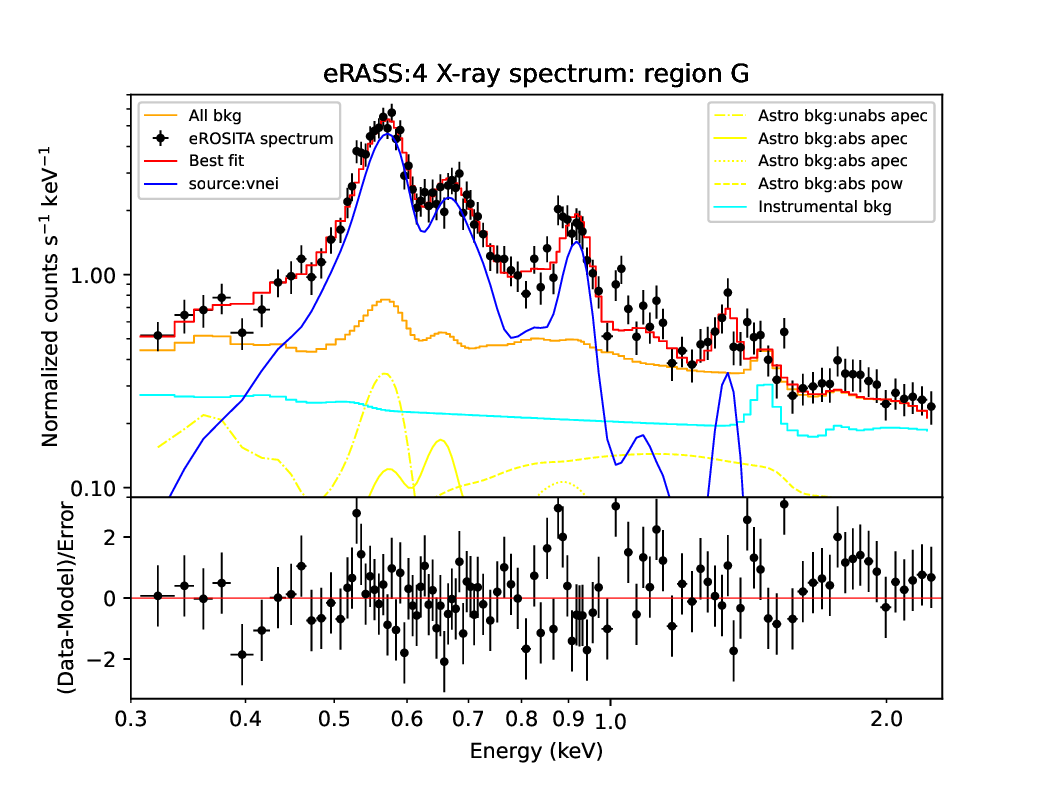}
    \caption{eRASS:4 X-ray spectra in the 0.3-2.3 keV energy band, from eight selected subregions of the remnant.}
\end{figure*}
\begin{figure*}
\ContinuedFloat   
    \includegraphics[width=0.499\textwidth]{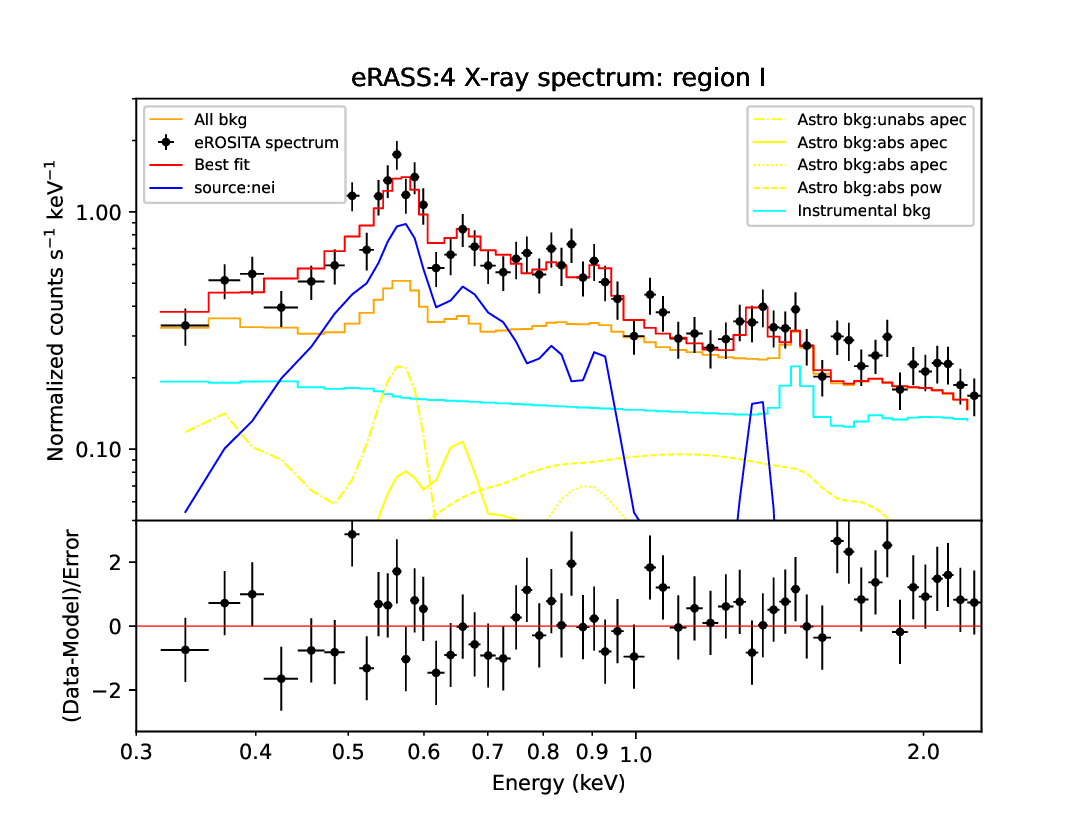}\includegraphics[width=0.495\textwidth]{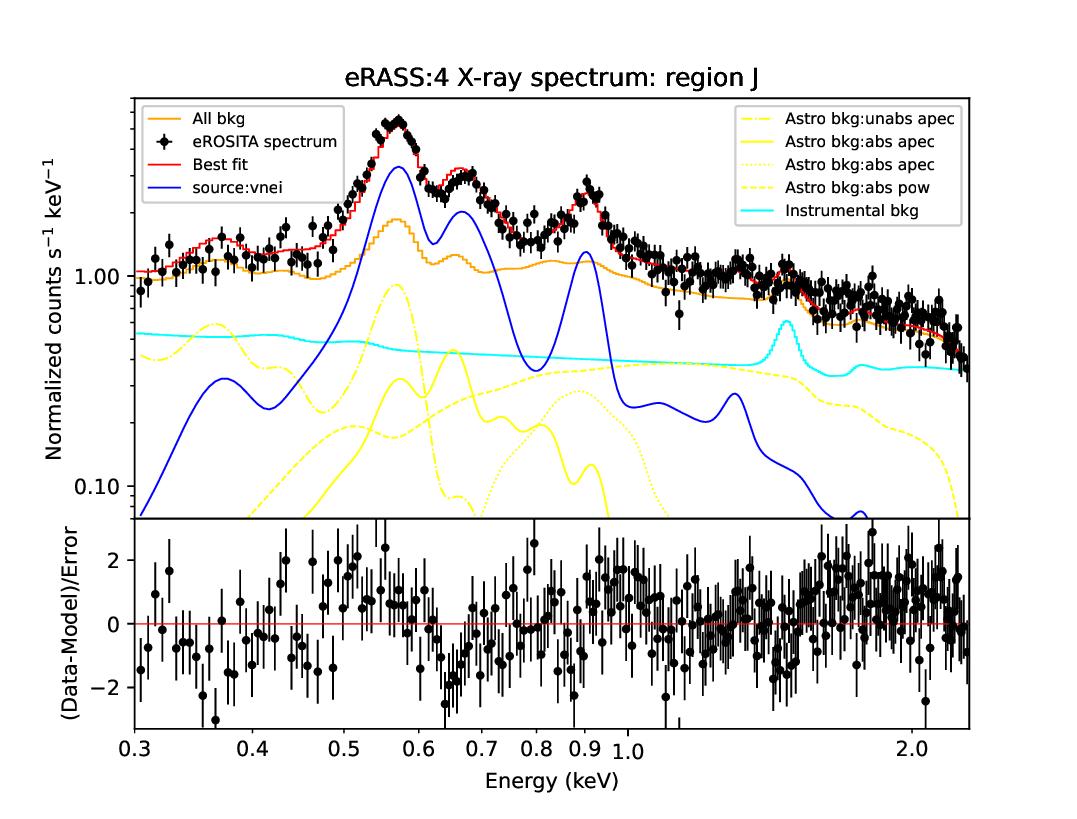}
    \caption{continued.}
    \label{ALLSUBSPEC2}
\end{figure*}

\end{appendix}
\end{document}